\documentclass[10pt]{article}

\usepackage{amsmath}
\usepackage{amssymb}

\usepackage{graphicx}

\usepackage{cite}

\usepackage{color}

\usepackage[labelfont=bf,labelsep=period,justification=raggedright]{caption}

\makeatletter
\renewcommand{\@biblabel}[1]{\quad#1.}
\makeatother

\date{}

\begin{document}

% Title must be 150 characters or less
\begin{flushleft}
{\Large
\textbf{Re-examination of globally flat space-time}
}
% Insert Author names, affiliations and corresponding author email.
\\
Michael R. Feldman$^{1,\ast}$
\\
\bf{1} Michael R. Feldman Private researcher, New York, NY, United States of America
\\
$\ast$ E-mail: mrf44@cornell.edu
\end{flushleft}

% Please keep the abstract between 250 and 300 words
\section*{Abstract}

In the following, we offer a novel approach to modeling the observed effects currently attributed to the theoretical concepts of `dark energy', `dark matter', and `dark flow'. Instead of assuming the existence of these theoretical concepts, we take an alternative route and choose to redefine what we consider to be inertial motion as well as what constitutes an inertial frame of reference in flat space-time. We adopt none of the features of our current cosmological models except for the requirement that special and general relativity be local approximations within our revised definition of inertial systems. Implicit in our ideas is the assumption that at ``large enough" scales one can treat objects within these inertial systems as point-particles having an insignificant effect on the curvature of space-time. We then proceed under the assumption that time and space are fundamentally intertwined such that time- and spatial-translational invariance are not inherent symmetries of flat space-time (i.e. observable clock rates depend upon both relative velocity and spatial position within these inertial systems) and take the geodesics of this theory in the radial Rindler chart as the proper characterization of inertial motion. With this commitment, we are able to model solely with inertial motion the observed effects expected to be the result of `dark energy', `dark matter', and `dark flow'. In addition, we examine the potential observable implications of our theory in a gravitational system located within a confined region of an inertial reference frame, subsequently interpreting the Pioneer anomaly as support for our redefinition of inertial motion. As well, we extend our analysis into quantum mechanics by quantizing for a real scalar field and find a possible explanation for the asymmetry between matter and antimatter within the framework of these redefined inertial systems.

% Please keep the Author Summary between 150 and 200 words
% Use first person. PLoS ONE authors please skip this step. 
% Author Summary not valid for PLoS ONE submissions.

\section*{Introduction}

The purpose of this paper is to present the foundational groundwork for a new metric theory of {\it flat} space-time which takes into account the observed effects currently expected to be the result of `dark energy'\cite{Peebles}, `dark matter'\cite{Trimble}, and `dark flow' \cite{Kashlinsky} without resorting to these theoretical concepts that we have yet to observe in the laboratory. We emphasize above the fact that we are working in flat space-time as this paper is not concerned with reformulating gravity. Meaning, we assume gravity is the consequence of {\it local} curvature in space-time resulting from the energy-momentum content associated with an object as formulated by Einstein in his theory of general relativity. However, for our discussion, we operate under the assumption that at ``large enough" scales we may treat massive objects in our proposed inertial reference frames as point-particles having an insignificant effect on the curvature of space-time for the purpose of examining the motion of said objects within the context of these larger scales. Consequently, we assume that {\it space-time} is essentially flat at these scales, and therefore, the energy density throughout our inertial systems is taken to be approximately zero. Thus, we assume that the deviation away from flat space-time inertial paths due to curvature in space-time is insignificant in our analysis. Furthermore, it is assumed that the observed motion of these large-scale objects about central points (e.g. stars orbiting the center of a galaxy, galaxies orbiting the center of a group/cluster, groups/clusters orbiting the center of a supercluster, etc.) is not due to the presence of gravitational sources at these centers but is instead a manifestation of the way in which objects move when no net external forces are acting upon them. In other words, the following work is concerned with reformulating our understanding of inertial motion. Furthermore, we focus on reformulating the global properties of an inertial reference frame while disregarding the potential local effects that objects moving within this global inertial system may have on the curvature of space-time. To begin with our reformulation, we explicitly state for the reader the assumptions of flat space-time as given by Einstein's special relativity \cite{Einstein_specialrel}:

1. An object will travel in a straight line at a constant speed when no net external forces are acting upon this object (inertial motion adopted from Newton; see section titled ``Definitions" in \cite{Newton}).

2. An observer undergoing inertial motion has the freedom to describe events by ``carrying rulers" in any three arbitrarily chosen spatial directions (perpendicular to one another) and calibrating clocks according to Einstein's prescription for synchronization (an inertial frame of reference). As well, inertial reference frames moving with uniform (constant velocity) rectilinear motion relative to one another are treated equally (i.e. there are no preferred inertial frames of reference in flat space-time).

3. The speed of light remains constant in all of these observer dependent inertial frames.

While operating under these assumptions in addition to those of general relativity \cite{Einstein_genrel}, our cosmological models (e.g. $\Lambda$CDM \cite{Trodden}) then require a `Big-Bang' event\cite{Lemaitre_Bigbang}\cite{Gamow}\cite{Gamow_Alpher}, `inflation' \cite{Guth}, `dark energy', `dark matter', and `dark flow' as explanations for observed phenomena on cosmological scales given our assumed understanding of inertial motion and inertial reference frames as stated above. In contrast, our claims in this paper are that in order to reproduce the observed behavior attributed to the theoretical concepts of `dark energy', `dark matter' and `dark flow', it is not necessary to assume that these supplements must exist. Instead, it is possible to reproduce this behavior by simply incorporating it into a revised understanding of inertial motion and inertial reference frames in empty flat space-time, thereby no longer assuming the three pillars of theoretical physics as listed above and no longer requiring the occurrence of a `Big-Bang', `inflation', and expansion of space. While seemingly rash at first glance, we claim that in what we term as our ``Theory of Inertial Centers", as laid out in the following work, one can reproduce with inertial motion in our redefined inertial reference frames the following observed features:

1. Accelerated redshifts \cite{Perlmutter} and the Hubble relation \cite{Hubblelaw}.

2. Plateauing orbital velocity curves at large distances from a central point about which objects orbit \cite{Rubin}.

3. Consistent velocity ``flow" of objects toward a central point \cite{Kashlinsky} \cite{Watkins}.

4. An orientation associated with a particular frame of reference \cite{Webb} (i.e. we do not take the cosmological principle to be a valid assumption as can be seen from experimental evidence such as \cite{Clowes}).

In our theory of flat space-time, inertial motion remains defined to be the motion of an object when it is subjected to no net external forces. In addition, an inertial reference frame is defined to be a system within which objects move along inertial trajectories when no net external forces are acting upon them. We then make the following assumptions and requirements in our theory:

1. Inertial motion is {\it not} characterized by an object moving in a straight line at a constant speed. Instead, inertial motion is characterized by geodesics about ``inertial center points" in the radial Rindler chart as examined in the following discussion (the radial Rindler chart has been mentioned in other contexts such as \cite{Witten} and \cite{Culetu}). Note that implicit in this assumption is the idea that time and space are fundamentally intertwined such that time-translational invariance and spatial-translational invariance are {\it not} inherent symmetries of flat space-time. Mathematically, this notion reduces to incorporating {\it both} time and spatial distance into the invariant interval associated with our metric. Meaning, the physically observable elapsed time as measured by a clock carried along a given curve, denoted as ``proper time" $\tau$, is {\it not} our affine parameter and thus is not invariant. Therefore, observable clock rates depend upon both spatial position in a particular inertial frame as well as in which inertial frame the observer is observing. Our affine parameter $\chi$ in the theory of inertial centers is then taken as a function of proper time in a particular inertial frame to be
\begin{displaymath}
\chi = \sqrt{\Lambda} \cdot \int r(\tau) d\tau
\end{displaymath}
where $r = r(\tau)$ represents the physical distance to the inertial center about which the observer moves at a particular observable clock time $\tau$ in the inertial system and $\sqrt{\Lambda}$ is taken to be the Hubble constant\cite{Hubblelaw}\cite{Riess}. In addition, these inertial center points define the centers of our inertial reference frames.

2. An observer does {\it not} have the freedom to describe an inertial reference frame in whichever way he/she chooses as in special relativity. We, as observers, are forced to adopt the {\it orientation} of the inertial reference frame that nature provides for us at the particular scale in which we are describing phenomena. As well, the inertial motion of an object must be thought of {\it relative} to the inertial center point about which said object orbits (throughout this paper, we will use the term ``orbit" to refer to the inertial motion of an object about an inertial center point).

3. The speed of light is {\it not} constant throughout these inertial reference frames.

4. Locally within a confined region of each of these newly defined inertial reference frames, our theory reduces to and abides by the axioms of special relativity and general relativity.

Our analysis is organized in the following manner. First, we explore the limiting behavior of our equations of motion with the radial Rindler chart in flat space-time. Out of this, we come upon the ability to model the observed features as listed above. Second, we determine the limit in which our theory reduces to special relativity, while also proposing the form of our invariant interval in terms of both time and distance to an inertial center. We have stated the form of our affine parameter earlier in this introduction as a preface to the logic used in this proposition. Third, we examine the potential observable effects of this theory within our solar system and interpret the Pioneer anomaly \cite{Anderson} as support for our ideas. Fourth, we extend our analysis by quantizing our theory for a real massive scalar field. Within the context of this extension, we find a potential explanation for the asymmetry between matter and antimatter in our observable universe through the possibility of a parallel region to each inertial system embodied mathematically by the ``other" radial Rindler wedge. We conclude by proposing future work including addressing the source of the cosmic microwave background \cite{Penzias} in this theory, attempting to explain other astrometric anomalies within our solar system besides Pioneer \cite{Nieto}, and extending our quantum mechanical analysis to complex fields with spin.

% Results and Discussion can be combined.

\section*{Discussion}

\subsection*{Geodesic paths} \label{geodesic}

Adopting the signature $(-,+,+,+)$ and employing abstract index notation throughout our analysis (see Chapter 2.4 of \cite{Wald_GR}), we work in the following metric:
\begin{equation} \label{eq:metric}
-d \chi^2 = -{\Lambda}r^2 d t^2 + d r^2 + r^2\cosh^2(\sqrt{\Lambda}t) d{\Omega}^2
\end{equation}
where $d{\Omega}^2 = d{\theta}^2 + d{\phi}^2\sin^2{\theta}$; $0 \leq \theta \leq \pi$, $0 \leq \phi < 2\pi$, $-\infty < t < \infty$, $0 < r^2<\infty$, and $\Lambda$ is a positive constant. In a subsequent section, we'll deduce that $\Lambda$ must be the square of the Hubble constant. $d\chi^{2}$ denotes the invariant interval associated with this metric where $d\chi^2 \neq c^2 d\tau^2$ assuming $\tau$ denotes proper time, defined as the physically observable elapsed time between two events as measured by a clock passing through both events carried along a particular curve, and $c$ denotes the constant associated with the speed of light in special relativity. Therefore, in contrast with special relativity, our proper time interval is {\it not} assumed to be invariant, and the speed of light in flat space-time is {\it not} assumed to be constant. However, in subsequent sections, we shall show how special relativity can be treated as a local approximation to our theory of inertial centers. As in special and general relativity, massless particles travel along null geodesics. Thus, with this radial Rindler chart as the description of our inertial frame of reference and our redefinition of the invariant interval associated with the metric, we implicitly assume that time and space are fundamentally intertwined such that time-translational invariance and spatial-translational invariance are {\it not} inherent symmetries of flat space-time. In other words, one cannot progress coordinate time $t$ forward (i.e. replace $t \rightarrow t+ t_0$ where $t_0$ is a constant) without considering the effect of this action on space and vice versa. As well, this concept requires that we incorporate into the invariant interval associated with our metric {\it both} distance to inertial centers as well as proper time. Later in our analysis, we will express $d\chi^2$ for this theory of inertial centers in terms of the proper time interval in a particular inertial frame.

For the affine connection terms, Ricci tensor elements, curvature scalar and square Riemann tensor, we refer to Appendix A. From these calculations it is clear that this space-time geometry is indeed flat. Taking the Rindler transformation equations, $cT = r\sinh(\sqrt{\Lambda}t)$, $R = r\cosh(\sqrt{\Lambda}t)$, we find our metric equation becomes
\begin{displaymath}
-d \chi^2 = -c^2d T^2 + d R^2 + R^2d \Omega^2, \indent \forall R, c^2T^2 < R^2
\end{displaymath}
where $c = $ speed of light in the local Minkowski reference frame \cite{Minkowski}. If one operates under the assumptions of special relativity, $d\chi^2$ would in fact equal $c^2 d \tau^2$, and then the metric in (\ref{eq:metric}) can be used to model uniformly radially accelerated motion with respect to Minkowski space-time confined to either of the Rindler wedges: left wedge for $|T|<-R/c$ and right wedge for $|T|<R/c$ \cite{Rindler}. For the rest of our analysis, however, we no longer assume that special relativity is valid throughout globally flat space-time (again, $d\chi^2 \neq c^2 d\tau^2$) and instead examine the geodesic motion of point-particles in this radial Rindler coordinate system with time and radial distance from our inertial center point corresponding to the coordinate labels $t$ and $r$, respectively. Additionally, as $d\chi^2 \neq c^2 d\tau^2$, we do not assume that the reference frame itself is radially accelerating. Instead, we are re-examining inertial motion under the guidelines presented in our introduction keeping in mind that the form of our invariant interval is different from that of special and general relativity. And since our affine parameter is different from that of special and general relativity, the geodesics of our theory will also be different. Consequently, our employment of the radial Rindler chart in the following analysis is our way of establishing that this coordinate system is the ``natural" one for describing an inertial system in the theory of inertial centers (i.e. coordinate time in the radial Rindler chart progresses at the same rate as the physical clock of a stationary observer in the inertial system). Thus, in the following work, {\it we abandon the idea that Minkowski coordinates can cover all of an inertial system in flat space-time}. Furthermore, we propose that {\it the radial Rindler chart should be our ``natural" coordinate system for describing an inertial frame of reference in the theory of inertial centers}.

Referring to Appendix B, we find for the equations of motion of a particle within a particular inertial system ($U^{a}{\nabla}_{a}U^{b}=0$, where our `proper velocity' in component form is $U^{\mu} = d x^{\mu}/d \sigma$):
\begin{eqnarray}
 0 = \frac{d^2t}{d \sigma^2} + \frac{2}{r}\frac{d t}{d \sigma}\frac{d r}{d \sigma} + \frac{1}{\sqrt{\Lambda}}\cosh(\sqrt{\Lambda}t)\sinh(\sqrt{\Lambda}t) \bigg[\bigg(\frac{d \theta}{d \sigma}\bigg)^2 + \sin^2{\theta}\bigg(\frac{d \phi}{d \sigma}\bigg)^2 \bigg] \label{eq:tmotion}
\\  0 = \frac{d^2r}{d \sigma^2} + \Lambda r \bigg(\frac{d t}{d \sigma}\bigg)^2 - r\cosh^2(\sqrt{\Lambda}t) \bigg[\bigg(\frac{d\theta}{d \sigma}\bigg)^2 + \sin^2{\theta}\bigg(\frac{d\phi}{d \sigma}\bigg)^2\bigg] \label{eq:rmotion}
\\  0 = \frac{d^2\theta}{d \sigma^2} + 2\frac{d \theta}{d \sigma}\bigg[\sqrt{\Lambda}\tanh(\sqrt{\Lambda}t)\frac{d t}{d \sigma} + \frac{1}{r}\frac{d r}{d\sigma}\bigg] - \sin{\theta}\cos{\theta} \bigg(\frac{d \phi}{d \sigma}\bigg)^2 \label{eq:thetamotion}
\\  0 = \frac{d^2\phi}{d \sigma^2} + 2\frac{d \phi}{d \sigma}\bigg[\sqrt{\Lambda}\tanh(\sqrt{\Lambda}t)\frac{d t}{d \sigma} + \frac{1}{r}\frac{d r}{d \sigma} + \cot{\theta}\frac{d\theta}{d \sigma}\bigg] \label{eq:phimotion}
\end{eqnarray}
And our norm for the `four-velocity' is given by
\begin{equation} \label{eq:fourvelocity}
 -k = g_{ab}U^a U^b = -\Lambda r^2 \bigg(\frac{d t}{d \sigma}\bigg)^2 + \bigg(\frac{d r}{d \sigma}\bigg)^2 + r^2 \cosh^2(\sqrt{\Lambda}t)\bigg[\bigg(\frac{d\theta}{d \sigma}\bigg)^2 + \sin^2{\theta}\bigg(\frac{d\phi}{d \sigma}\bigg)^2\bigg]
\end{equation}
where
\begin{displaymath}
k =\left\{\begin{array}{l l} 0 & \quad \textrm{massless particle} \\ 1 & \quad \textrm{massive particle} \\ \end{array}\right.
\end{displaymath}
and $\sigma = \chi$ for massive particles. Notice that our `four-velocity' $U^{a}$ in this theory is dimensionless for spatial components and has units of [time]/[distance] for our time component since $\chi$ (and therefore $\sigma$) has units of [distance]. Multiplying each term in our radial equation of motion by $r$ and plugging in (\ref{eq:fourvelocity}),
\begin{equation} \label{eq:radialaccel}
0 = r\frac{d^2r}{d \sigma^2}  + \bigg(\frac{d r}{d \sigma}\bigg)^2 + k
\end{equation}
But to remain at a constant radial distance away from our inertial center: $d^2r/d \sigma^2$, $d r/d \sigma=0$. Therefore, only massless particles can have circular orbits. 

Possible geodesic paths obey the relation $U^{a}U_{a} \leq 0$ from (\ref{eq:fourvelocity}), and solving for $d\theta/d t$ and $d\phi/d t$, we find that
\begin{displaymath}
\frac{1}{r^2 \cosh^2(\sqrt{\Lambda}t)}\bigg[ \Lambda r^2 - \bigg(\frac{d r}{d t}\bigg)^2 \bigg] \geq  \bigg(\frac{d\theta}{d t}\bigg)^2 + \sin^2{\theta}\bigg(\frac{d \phi}{d t}\bigg)^2
\end{displaymath}
Examining our $\theta$ equation of motion (\ref{eq:thetamotion}), we see that a particle remains at a constant value of $\theta$ for non-zero angular velocity in $\phi$ if and only if $d\theta/d\sigma = 0$ and $\theta = 0$, $\pi/2$, $\pi$. Consequently, the angular velocity of a particle traveling in the equatorial plane ($\theta = \pi/2$) of this inertial reference frame is bound by the range:
\begin{equation}
-\frac{\sqrt{\Lambda}}{\cosh(\sqrt{\Lambda}t)} \leq \frac{d\phi}{d t}\bigg|_{\theta = \pi/2} \leq +\frac{\sqrt{\Lambda}}{\cosh(\sqrt{\Lambda}t)}
\end{equation}
Then, for a photon traveling in a circular orbit in the equatorial plane, we find
\begin{equation} \label{eq:spincircular}
\frac{d\phi}{d t}\bigg|_{k=0, \theta = \pi/2, {\rm circular}}=\pm \frac{\sqrt{\Lambda}}{\cosh(\sqrt{\Lambda}t)}
\end{equation}
Later, we'll see that a massless particle can have circular orbits only for $\theta = \pi/2$ (orbits with $\phi = \phi_0$ cannot be circular).

For massive particles nearly at rest with respect to the center of this inertial system (i.e. spatial `velocity' terms are much smaller than our `velocity' term in time so that these spatial terms can be taken as nearly zero), these four equations of motion (\ref{eq:tmotion}), (\ref{eq:rmotion}), (\ref{eq:thetamotion}), and (\ref{eq:phimotion}) reduce to two:
\begin{displaymath}
0 = \frac{d^2t}{d \chi^2} \indent {\rm and} \indent 0 =  \frac{d^2r}{d \chi^2} + \Lambda r \bigg(\frac{d t}{d \chi}\bigg)^2
\end{displaymath}
And solving for the radial acceleration, we find that
\begin{equation} \label{eq:stationary_force}
\frac{d^2r}{d t^2} = -\Lambda r
\end{equation}
In this limit, the inertial motion of our point-particle is described by a spatial acceleration in $r$ pulling inward toward the center of this particular reference frame scaled by the square of the time-scale constant. Thus, slowly moving objects at large radial distances experience a large radial acceleration pulling inward toward the center of the inertial system about which the objects orbit.

Then, let us examine the case where the motion of particles far from an inertial center (large $r$) is dominated by angular velocities with approximately circular radial motion ($r \approx {\rm constant}$). Our equations of motion reduce to
\begin{eqnarray}
0 = \frac{d^2t}{d \sigma^2} + \frac{1}{\sqrt{\Lambda}}\cosh(\sqrt{\Lambda} t)\sinh(\sqrt{\Lambda} t) \bigg[\bigg(\frac{d\theta}{d \sigma}\bigg)^2 + \sin^2{\theta}\bigg(\frac{d\phi}{d \sigma}\bigg)^2 \bigg] \nonumber
\\ 0 = \Lambda r \bigg(\frac{d t}{d \sigma}\bigg)^2 - r\cosh^2(\sqrt{\Lambda} t) \bigg[\bigg(\frac{d\theta}{d \sigma}\bigg)^2 + \sin^2{\theta}\bigg(\frac{d\phi}{d \sigma}\bigg)^2 \bigg] \nonumber
\\ 0 = \frac{d^2\theta}{d \sigma^2}  + 2\sqrt{\Lambda} \tanh(\sqrt{\Lambda} t)\frac{d\theta}{d \sigma}\frac{d t}{d \sigma} - \sin{\theta}\cos{\theta}\bigg(\frac{d\phi}{d \sigma}\bigg)^2 \nonumber
\\ 0 = \frac{d^2\phi}{d \sigma^2} + 2\frac{d\phi}{d \sigma}\bigg[\sqrt{\Lambda}\tanh(\sqrt{\Lambda} t)\frac{d t}{d \sigma} + \cot{\theta}\frac{d\theta}{d \sigma}\bigg] \nonumber
\end{eqnarray}
Plugging $d^2t/d \sigma^2$ and $d t/d \sigma$ into our expressions for $d^2\phi/d \sigma^2$ and $d^2\theta/d \sigma^2$:
\begin{eqnarray}
0 = \frac{d^2\phi}{d t^2} + \sqrt{\Lambda} \tanh(\sqrt{\Lambda} t)\frac{d\phi}{d t} + 2\cot{\theta}\frac{d\theta}{d t}\frac{d\phi}{d t} \nonumber
\\ 0 = \frac{d^2\theta}{d t^2} + \sqrt{\Lambda} \tanh(\sqrt{\Lambda} t)\frac{d\theta}{d t} -\sin{\theta}\cos{\theta}\bigg(\frac{d\phi}{d t}\bigg)^2 \nonumber
\end{eqnarray}
Now, if we assume $d\phi/d t \gg d\theta/d t$ and integrate:
\begin{displaymath}
\frac{d\phi}{d t} = \frac{\phi_0}{\cosh(\sqrt{\Lambda} t)}
\end{displaymath}
where $\phi_0 \rightarrow \pm \sqrt{\Lambda}$ for light taking a circular orbit in the equatorial plane and thus $|\phi_0| \leq \sqrt{\Lambda}$ for $\theta = \pi/2$. Plugging in for $d\phi/d t$, we have
\begin{equation}
\frac{d^2\theta}{d t^2} = \sin{\theta}\cos{\theta}\bigg(\frac{\phi_0}{\cosh(\sqrt{\Lambda} t)}\bigg)^2
\end{equation}
where $\phi_0$ is a constant. In the large $d\phi/d \sigma$ limit:
\begin{eqnarray}
\frac{d^2\theta}{d t^2} > 0 \indent \textrm{for}\indent 0<\theta<\frac{\pi}{2}
\\ \frac{d^2\theta}{d t^2} = 0 \indent \textrm{for}\indent \theta = \frac{\pi}{2}
\\ \frac{d^2\theta}{d t^2} < 0 \indent \textrm{for}\indent \frac{\pi}{2}<\theta<\pi
\end{eqnarray}
As long as the particle is not located at either of the poles ($\theta \neq 0$, $\pi$), we see a sinusoidal spatial angular acceleration that decreases with $t$ and moves the object toward $\theta = \pi/2$. One can then picture spiral galaxy formation resulting from objects orbiting an inertial center with large angular velocity in $\phi$.

If we refer back to our expression for $d \phi/d t$, we find for the orbital velocity ($v = r \cdot d\phi/dt$) of a particle in this limit:
\begin{equation}\label{eq:orbvel}
v = \frac{\phi_0}{\cosh(\sqrt{\Lambda} t)}r
\end{equation}
And for $\sqrt{\Lambda}t \approx 0$, our particle's speed is linearly proportional to its radial distance away from the inertial center about which it orbits. In this limit at large $r$, the relationship between orbital velocity and radial distance mimics the relationship between orbital velocity and radial distance found in our observed galaxy rotation curves\cite{Rubin} for comparably small values of $\sqrt{\Lambda}$ and therefore $\phi_0$. However, the analysis above will apply to the classical (in the sense that we are not taking into account quantum mechanics) inertial motion of an object in any particular inertial system (e.g. galaxies, groups, clusters, etc.). Later in our analysis, we'll provide an experimental scale for the time-scale constant $\sqrt{\Lambda}$ by analyzing the inherent redshift that occurs in these inertial frames (i.e. we'll take $\sqrt{\Lambda}$ to be the Hubble constant). Since $|\phi_0| \leq \sqrt{\Lambda}$, this value will also give us an upper limit for the slope of our orbital velocity curves at large $r$. Thus, we claim that the linear relationship found in (\ref{eq:orbvel}) models the experimental relationship found from our observed orbital velocity curves for objects far from the center of the galaxy within which they orbit. We base this claim off of the idea that the plateauing nature of our experimental curves would be interpreted in our model to be the result of the small scale of $\sqrt{\Lambda}$ relative to galactic distance and orbital velocity scales.

\subsection*{Conservation laws} \label{conserve}

Since this metric is just a coordinate transformation away from Minkowski, we expect to find ten linearly independent Killing vector fields as vector fields are geometric objects independent of our coordinate parametrization. One could obtain these using the radial Rindler transformation equations, but we find it helpful to explicitly derive them. We refer to Appendix C for more detail as well as a full list of all Killing vector fields given in the radial Rindler chart. Rewriting here for reference the three we will be using:
\begin{eqnarray}
\rho^{\mu} \rightarrow \langle \frac{1}{\sqrt{\Lambda}r}\cosh(\sqrt{\Lambda}t), -\sinh(\sqrt{\Lambda}t), 0, 0 \rangle \label{eq:Killingradial}
\\ \Theta^{\mu} \rightarrow \langle \frac{1}{\sqrt{\Lambda}}\cos{\theta}, 0, -\sin{\theta}\tanh(\sqrt{\Lambda}t), 0 \rangle \label{eq:Killingtheta}
\\ \psi^{\mu} \rightarrow \langle 0,0,0,1 \rangle
\end{eqnarray}
Applying Noether's theorem ($U^{a}\xi_{a} = {\rm constant}$),
\begin{eqnarray}
E = \sqrt{\Lambda}r\cosh(\sqrt{\Lambda}t)\frac{d t}{d \sigma} + \sinh(\sqrt{\Lambda}t)\frac{d r}{d \sigma} \label{eq:Energy}
\\ \Omega = \sqrt{\Lambda}r^2\cos{\theta}\frac{d t}{d \sigma} + r^2\sin{\theta}\sinh(\sqrt{\Lambda}t)\cosh(\sqrt{\Lambda}t)\frac{d\theta}{d \sigma} \label{eq:AngularParity}
\\ L = r^2\cosh^2(\sqrt{\Lambda}t)\sin^2{\theta}\frac{d\phi}{d \sigma} \label{eq:AngularMomentum}
\end{eqnarray}
Plugging into (\ref{eq:fourvelocity}) and solving for $d t/d \sigma$, we find that
\begin{equation} \label{eq:Clock}
\frac{d t}{d \sigma} = \frac{1}{\sqrt{\Lambda}}\bigg[\frac{E}{r}\sin^2{\theta}\cosh(\sqrt{\Lambda}t) + \frac{\Omega}{r^2}\cos{\theta} \pm \sqrt{}|_{d t/d \sigma}\bigg]
\end{equation}
where
\begin{eqnarray}
\sqrt{}|_{d t/d \sigma}=\bigg\{(\frac{E}{r}\sin^2{\theta}\cosh(\sqrt{\Lambda}t)+\frac{\Omega}{r^2}\cos{\theta})^2 - \bigg[\sin^2{\theta}\bigg(\bigg(\frac{E}{r}\bigg)^2 + k\bigg(\frac{\sinh(\sqrt{\Lambda}t)}{r}\bigg)^2\bigg)
\nonumber \\ + \bigg(\frac{\Omega}{r^2}\bigg)^2 + \bigg(\frac{L\tanh(\sqrt{\Lambda}t)}{r^2}\bigg)^2 \bigg]\bigg\}^{1/2} \label{eq:timeroot}
\end{eqnarray}
requiring
\begin{displaymath}
 (\frac{E}{r}\sin^2{\theta}\cosh(\sqrt{\Lambda}t)+\frac{\Omega}{r^2}\cos{\theta})^2 \geq \sin^2{\theta}\bigg[\bigg(\frac{E}{r}\bigg)^2 + k\bigg(\frac{\sinh(\sqrt{\Lambda}t)}{r}\bigg)^2\bigg] + \bigg(\frac{\Omega}{r^2}\bigg)^2 + \bigg(\frac{L\tanh(\sqrt{\Lambda}t)}{r^2}\bigg)^2
\end{displaymath}
Notice, for massive particles moving radially in the equatorial plane ($k = 1$, $\theta = \pi/2$, $\Omega = 0$, and $L = 0$), this constraint reduces to:
\begin{displaymath}
E^2 \geq 1
\end{displaymath}
which is just our analogue of the statement in special relativity that the energy of an object must be greater than or equal to its rest mass \cite{Einstein_inertia_energy} since in special relativity one would assume this constant $E$ would equal the energy of the particle divided by its rest mass (i.e. in special relativity, $E$ would be equal to $\tilde{E}/mc^2$ where $\tilde{E}$ is the energy of the particle). Using (\ref{eq:Energy}), (\ref{eq:AngularParity}), (\ref{eq:AngularMomentum}), and (\ref{eq:Clock}), we find

\begin{eqnarray}
 \frac{d r}{d t} = \frac{\sqrt{\Lambda}r}{\sinh(\sqrt{\Lambda}t)}\bigg[\frac{1}{\sin^2{\theta}\cosh(\sqrt{\Lambda}t)+\frac{\Omega}{Er}\cos{\theta} \pm \sqrt{}|_{d r/d t}} - \cosh(\sqrt{\Lambda}t) \bigg] \label{eq:radialvel}
\\  \frac{d\theta}{d t} = \frac{\sqrt{\Lambda}}{\sin{\theta}\cosh(\sqrt{\Lambda}t)\sinh(\sqrt{\Lambda}t)}\bigg[\frac{1}{\frac{Er}{\Omega}\sin^2{\theta}\cosh(\sqrt{\Lambda}t) + \cos{\theta} \pm \sqrt{}|_{d\theta/d t}} - \cos{\theta} \bigg] \label{eq:thetavel}
\\  \frac{d\phi}{d t} = \frac{\sqrt{\Lambda} L}{\sin^2{\theta}\cosh^2(\sqrt{\Lambda}t)}\bigg[\frac{1}{Er\sin^2{\theta}\cosh(\sqrt{\Lambda}t) + \Omega\cos{\theta} \pm \sqrt{}|_{d\phi/d t}} \bigg]\label{eq:phivel}
\end{eqnarray}
where
\begin{eqnarray}
\sqrt{}|_{d r/d t} = \bigg\{(\sin^2{\theta}\cosh(\sqrt{\Lambda}t)+\frac{\Omega}{Er}\cos{\theta})^2 - \bigg[\sin^2{\theta}\bigg(1 + k\bigg(\frac{\sinh(\sqrt{\Lambda}t)}{E}\bigg)^2\bigg) + \bigg(\frac{\Omega}{Er}\bigg)^2
 \nonumber \\ + \bigg(\frac{L\tanh(\sqrt{\Lambda}t)}{Er}\bigg)^2 \bigg]\bigg\}^{1/2} \label{eq:radialroot}
\\ \sqrt{}|_{d\theta/d t} = \bigg\{(\frac{Er}{\Omega}\sin^2{\theta}\cosh(\sqrt{\Lambda}t)+\cos{\theta})^2 - \bigg[\sin^2{\theta}\bigg(\bigg(\frac{Er}{\Omega}\bigg)^2 + k\bigg(\frac{r\sinh(\sqrt{\Lambda}t)}{\Omega}\bigg)^2\bigg)
 \nonumber \\ + 1 + \bigg(\frac{L\tanh(\sqrt{\Lambda}t)}{\Omega}\bigg)^2 \bigg]\bigg\}^{1/2}\label{eq:thetaroot}
\\ \sqrt{}|_{d\phi/d t} = \bigg\{(Er\sin^2{\theta}\cosh(\sqrt{\Lambda}t)+\Omega\cos{\theta})^2 - \bigg[\sin^2{\theta}\bigg((Er)^2 + k(r\sinh(\sqrt{\Lambda}t))^2\bigg)
 \nonumber \\ + {\Omega}^{2} + (L\tanh(\sqrt{\Lambda}t))^2 \bigg]\bigg\}^{1/2} \label{eq:phiroot}
\end{eqnarray}
For light traveling radially in the equatorial plane, $\Omega$, $L = 0$ and (\ref{eq:radialvel}) reduces to
\begin{equation}
\frac{d r}{d t}\bigg|_{k=0, \Omega = 0, L = 0} = \pm \sqrt{\Lambda} r
\end{equation}
giving us
\begin{equation}\label{eq:radiallight}
r(t)|_{k=0,\Omega=0,L=0} = r_0 \textrm{exp}({\pm \sqrt{\Lambda}t})
\end{equation}
where $r_0$ is a constant signifying the radial position of the photon at $t=0$. One could have arrived at this expression for the general case of light traveling radially even outside of the equatorial plane simply from (\ref{eq:fourvelocity}) for null geodesics. Let us now pose the question of whether or not it is possible for light to travel from the $r>0$ region of our inertial system to $r=0$ which we regard as our inertial center point. Integrating $d r/d t$ from $0$ to $\Delta t$,
\begin{displaymath}
\Delta r = \pm \sqrt{\Lambda}r_0\int^{\Delta t}_{0}{\textrm{exp}(\pm \sqrt{\Lambda}t)d t} = r_0 [\textrm{exp}(\pm \sqrt{\Lambda}\Delta t) - 1]
\end{displaymath}
Solving for $\Delta t$,
\begin{equation} \label{eq:radialphoton_time}
\Delta t = \ln{\bigg(\frac{\Delta r + r_0}{r_0}\bigg)^{\pm 1/\sqrt{\Lambda}}}
\end{equation}
For a photon traveling radially inward, the sign of the root is negative, and it reaches $r = 0$ in
\begin{displaymath}
\lim_{r_{{\rm final}}\rightarrow0}\Delta t = \ln{\bigg(\frac{(0-r_0) + r_0}{r_0}\bigg)^{- 1/\sqrt{\Lambda}}}=\ln{\bigg(\frac{r_0}{-r_0 + r_0}\bigg)^{1/\sqrt{\Lambda}}} \rightarrow \infty
\end{displaymath}
Consequently, not even light can reach $r=0$ in a finite amount of time. But what about the inertial behavior of massive particles in these systems? At first glance, (\ref{eq:radialvel}) and (\ref{eq:thetavel}) appear to be divergent for $t=0$. However, to evaluate all of these velocity expressions for $t=0$, we return to symmetry equations (\ref{eq:Energy}) and (\ref{eq:AngularParity}):
\begin{displaymath}
E = \sqrt{\Lambda} r \frac{dt}{d\sigma}\bigg|_{t = 0} \indent {\rm and} \indent \Omega = \sqrt{\Lambda}r^2 \cos{\theta}\frac{dt}{d\sigma}\bigg|_{t=0}
\end{displaymath}
Therefore,
\begin{equation} \label{eq:tzeroconstants}
\frac{E}{r} \cos{\theta} = \frac{\Omega}{r^2}\bigg|_{t=0}
\end{equation}
Plugging into (\ref{eq:timeroot}),
\begin{eqnarray}
\sqrt{}|_{d t/d \sigma}|_{t=0} = \bigg\{\bigg(\frac{E}{r}\sin^2{\theta} + \frac{E}{r}\cos^2{\theta}\bigg)^2 - \bigg[\sin^2{\theta}\bigg(\frac{E}{r}\bigg)^2+\bigg(\frac{E}{r}\cos^2{\theta} \bigg)^2 \bigg] \bigg\}^{1/2} \nonumber
\\ = \sqrt{\bigg(\frac{E}{r}\bigg)^2 - \bigg(\frac{E}{r} \bigg)^2} = 0 \nonumber
\end{eqnarray}
which one can plug back into (\ref{eq:Clock}) to find consistency with our expressions for $dt/d\sigma|_{t=0}$ above. Yet we see from (\ref{eq:radialroot}), (\ref{eq:thetaroot}), and (\ref{eq:phiroot}) that
\begin{eqnarray}
\sqrt{}|_{dr/dt} = \frac{r}{E} \cdot \sqrt{}|_{dt/d\sigma}
\\ \sqrt{}|_{d\theta/dt} = \frac{r^2}{\Omega} \cdot \sqrt{}|_{dt/d\sigma}
\\ \sqrt{}|_{d\phi/dt} = r^2 \cdot \sqrt{}|_{dt/d\sigma}
\end{eqnarray}
which implies for $t=0$ that all of these terms vanish. Then our spatial velocity terms for $t=0$ become
\begin{eqnarray}
\frac{d r}{d t}\bigg|_{t=0} = \frac{\sqrt{\Lambda}r}{\sinh(\sqrt{\Lambda}t)}\bigg[\frac{1}{\sin^2{\theta}+\cos^2{\theta}} - 1 \bigg] = \frac{0}{0} \nonumber
\\ \frac{d\theta}{d t}\bigg|_{t=0} = \frac{\sqrt{\Lambda}}{\sin{\theta}\sinh(\sqrt{\Lambda}t)}\bigg[\frac{\cos{\theta}}{\sin^2{\theta} + \cos^2{\theta}} - \cos{\theta} \bigg] = \frac{0}{0} \nonumber
\\  \frac{d\phi}{d t}\bigg|_{t=0} = \frac{\sqrt{\Lambda} L}{\sin^2{\theta}}\bigg[\frac{1}{Er\sin^2{\theta} + Er\cos^2{\theta}} \bigg] = \frac{\sqrt{\Lambda}L}{Er \sin^2{\theta}} \nonumber
\end{eqnarray}
where we have used (\ref{eq:tzeroconstants}) in these limit expressions. So we see that our velocity terms are not necessarily divergent for $t=0$. However, we'll address the issue of motion for small $\sqrt{\Lambda}t$ later when we relate Einstein's special relativity to our theory of inertial centers. One must also keep in mind that expressions (\ref{eq:radialvel}), (\ref{eq:thetavel}), and (\ref{eq:phivel}) represent a set of complex differential equations that we unfortunately will not be able to solve in this paper. The purpose of the following portion of this section is in fact to evaluate the large $t$ behavior of all spatial velocities where it is not explicitly apparent how to evaluate this limit if one were to work in Minkowski coordinates while keeping in mind the notion that he/she must relate back to the radial Rindler chart for inertial time as $d\chi^2 \neq c^2 d\tau^2$ (we'll elaborate further on the term ``inertial time" in our next section). It does appear easier to proceed in this manner of working in Minkowski and relating back to radial Rindler for solely radial motion as we shall do later in this section.

Yet, we return to our velocity expressions from Noether's theorem in order to examine the general expression for $d r/d t$ as $t \rightarrow \infty$. First, we determine the limiting value of $\sqrt{}|_{d r/d t}$:
\begin{displaymath}
\lim_{t\rightarrow \infty}{\sqrt{}|_{d r/d t}} = \sin^2{\theta}\cosh(\sqrt{\Lambda}t)\sqrt{1-\frac{k}{E^2 \sin^2{\theta}}}
\end{displaymath}
Then for massive particles ($k = 1$) assuming $\theta \neq 0$ or $\pi$,
\begin{equation} \label{eq:radiallim}
\lim_{t\rightarrow \infty}{\frac{d r}{d t}} = -\sqrt{\Lambda}r
\end{equation}
which is just the equation for a massless particle traveling radially inward. When $\theta = 0$ or $\pi$, we must return to conservation equations (\ref{eq:AngularParity}) and (\ref{eq:AngularMomentum}). We find $L=0$ and
\begin{displaymath}
\frac{d t}{d \sigma}\bigg|_{\theta = 0, \pi} = \frac{\Omega}{\sqrt{\Lambda}r^2 \cos{\theta}}
\end{displaymath}
Plugging in (\ref{eq:Energy}) and solving when $t \rightarrow \infty$, we again find (\ref{eq:radiallim}). We now understand that eventually all massive particles move toward $r=0$. Yet as the object approaches the center, its speed decreases as well and will only stop moving inward when it reaches this inertial center point in an infinite amount of time. Thus, with this large $t$ behavior, we apparently inherit the ability to model the observed anomalous effects of `dark flow' \cite{Kashlinsky}. In our next section, we will provide an interpretation for the physical significance of our coordinate time in our theory of inertial centers, relating $t$ back to the rate at which physical clocks are observed to tick.

However, we progress onward and look at the large $t$ limits for both $d\theta/d t$ and $d\phi/d t$. Beginning with the former, we find
\begin{displaymath}
\lim_{t \rightarrow \infty}{\sqrt{}|_{d\theta/d t}} = \frac{Er}{\Omega}\sin^2{\theta}\cosh(\sqrt{\Lambda}t)\sqrt{1 - \frac{k}{E^2\sin^2{\theta}}}
\end{displaymath}
Plugging into $d\theta/d t$ and examining for massive particles,
\begin{equation}
\lim_{t \rightarrow \infty}{\frac{d\theta}{d t}} = \frac{-\sqrt{\Lambda}}{\cosh(\sqrt{\Lambda}t)\sinh(\sqrt{\Lambda}t)}\cot{\theta}
\end{equation}
And for $\theta \neq 0$ or $\pi$,
\begin{equation}
\lim_{t \rightarrow \infty}{\frac{d\theta}{dt}}\bigg|_{\theta \neq 0, \pi} = 0
\end{equation}
Solving for $d\theta/d t$ when $\theta = 0$ or $\pi$ using (\ref{eq:thetavel}) and $L=0$,
\begin{displaymath}
\frac{d\theta}{d t}\bigg|_{\theta = 0, \pi} = \frac{\sqrt{\Lambda}}{\cosh(\sqrt{\Lambda}t)\sinh(\sqrt{\Lambda}t)}\tan{\theta}=0
\end{displaymath}
At the poles, particles have no angular velocity in $\theta$ nor angular momentum in $\phi$ ($d\theta/d \sigma, L = 0$). Lastly for our large $t$ limits, we have $d\phi/d t$:
\begin{displaymath}
\lim_{t \rightarrow \infty}{\sqrt{}|_{d\phi/d t}} =  Er\sin^2{\theta}\cosh(\sqrt{\Lambda}t)\sqrt{1 - \frac{k}{E^2\sin^2{\theta}}}
\end{displaymath}
Plugging into $d\phi/d t$,
\begin{displaymath}
 \lim_{t \rightarrow \infty}\frac{d\phi}{d t}  = \frac{\sqrt{\Lambda} L}{\sin^2{\theta}\cosh^2(\sqrt{\Lambda}t)}\bigg[\frac{1}{Er\sin^2{\theta}\cosh(\sqrt{\Lambda}t)(1 \pm \sqrt{1 - \frac{k}{E^2\sin^2{\theta}}})} \bigg]
\end{displaymath}
For massive particles and assuming $\theta \neq 0$ or $\pi$,
\begin{equation}
\lim_{t \rightarrow \infty}\frac{d\phi}{d t} = 0
\end{equation}
We have a clearer picture of the inertial trajectories of massive particles over time in the context of our redefined inertial reference frames. As time progresses, massive objects will eventually move radially inward losing angular velocity in $\theta$ and angular momentum in $\phi$, slowing down in radial velocity as they approach the center point about which they orbit.

Looking back at our expression for $d r/d t$, we ask ourselves the question: for what values of $\theta$ is $d r/d t$ most positive? For positive $d r/d t$, we have particles moving radially outward, and maximizing this expression with respect to $\theta$ provides us with the easiest possible path to be ejected away from our inertial center. Examining particles with large radial `proper velocities' relative to their own angular `proper velocities' which from (\ref{eq:Energy}), (\ref{eq:AngularParity}), (\ref{eq:AngularMomentum}) implies $Er \gg \Omega, L$ since $dr/d\sigma \gg r\cosh(\sqrt{\Lambda}t) \cdot d\theta/d\sigma$ and $dr/d\sigma \gg r\cosh(\sqrt{\Lambda}t) \cdot d\phi/d\sigma$ in this limit:
\begin{eqnarray}
 \frac{d r}{d t}\bigg|_{Er \gg \Omega, L} = \frac{\sqrt{\Lambda}r}{\sinh(\sqrt{\Lambda}t)\cosh(\sqrt{\Lambda}t)} \bigg[\frac{1}{\sin^2{\theta} \pm \sqrt{\sin^4{\theta}-\frac{\sin^2{\theta}}{\cosh^2(\sqrt{\Lambda}t)}(1+k\frac{\sinh^2(\sqrt{\Lambda}t)}{E^2})}} - \cosh^2(\sqrt{\Lambda}t)\bigg] \nonumber
\end{eqnarray}
But the largest positive value of $d r/d t |_{Er \gg \Omega, L}$ occurs if we minimize the denominator of the first term in brackets with respect to $\theta$. Clearly, this term needs to be re-evaluated when $\theta = 0$ or $\pi$. Returning to conservation equations (\ref{eq:Energy}) and (\ref{eq:AngularParity}), we solve for the radial motion of a particle through the poles by plugging into (\ref{eq:fourvelocity}) ($\theta = 0$ or $\pi$ and $d \theta/d \sigma = 0$):
\begin{eqnarray}
 \frac{d r}{d t}\bigg|_{\theta = 0,\pi} = \frac{\sqrt{\Lambda}r}{1 + \frac{k}{E^2}\sinh^2(\sqrt{\Lambda}t)} \cdot \bigg\{-\frac{k}{E^2}\cosh(\sqrt{\Lambda}t)\sinh(\sqrt{\Lambda}t)
\pm \bigg[\bigg(\frac{k}{E^2}\bigg)^2 \cosh^2(\sqrt{\Lambda}t)\sinh^2(\sqrt{\Lambda}t)
\nonumber \\ + \bigg(1+\frac{k}{E^2}\sinh^2(\sqrt{\Lambda}t)\bigg) \bigg(1- \frac{k}{E^2}\cosh^2(\sqrt{\Lambda}t)\bigg)\bigg]^{1/2} \bigg\} \nonumber
\end{eqnarray}
For large proper radial motion, we assume $E^2 \gg k\cosh^2(\sqrt{\Lambda}t)$ (as $E^2 \geq 1$ is our analogue of the rest mass condition from Einstein's special relativity). Then, our expression for radial motion through the poles reduces to
\begin{equation}
\frac{d r}{d t}\bigg|_{\theta = 0, \pi} \approx \pm \sqrt{\Lambda} r
\end{equation}
We see that massive particles can travel at speeds near that of photons through the poles, and therefore it appears that the easiest way for particles to be ejected radially outward away from an inertial center would be through the poles of the inertial system. If we imagine a supernova occurring near the center point of an inertial system, we find that a simple potential scenario for the occurrence of relativistic jets \cite{Meier} in this reference frame would be the expulsion of stellar remnants through the poles. Consequently, if we use this logic to provide an alternative for relativistic jet production, we must then require that each of our inertial frames have a particular orientation governed by the location of these poles and embodied mathematically by the spatial positions for which particular metric components vanish. In other words, when describing a particular inertial frame, these are the $\theta$ values for which $\sin{\theta}=0$ previously referred to as ``coordinate singularities" (e.g. see Chapter 5.1 of \cite{Ellis}) but taken here as a physical attribute of the inertial system reflecting the idea that the radial Rindler chart is the ``natural" coordinate system for an inertial reference frame in flat space-time. Thus we must ask ourselves the following question. How is this orientation established in the theory of inertial centers? As we shall mention later in our paper, this is an open question which we will have to address in future work.

Back to our circular orbit analysis, we solve for the radius at which light can have circular paths in a particular inertial system for possibly both $d\theta/d t = 0$, $d\phi/d t = \pm \sqrt{\Lambda}/\cosh(\sqrt{\Lambda}t)$ and $d\theta/d t = \pm \sqrt{\Lambda}/\cosh(\sqrt{\Lambda}t)$, $d\phi/d t = 0$. For the two, we obtain from (\ref{eq:Energy})
\begin{equation} \label{eq:t_circle}
\frac{d t}{d \sigma}\bigg|_{k=0, {\rm circular}} = \frac{E}{\sqrt{\Lambda} r \cosh(\sqrt{\Lambda}t)}
\end{equation}
In the former situation ($\theta = \pi/2$, $\Omega = 0$), we substitute (\ref{eq:t_circle}) and (\ref{eq:AngularMomentum}) into (\ref{eq:fourvelocity}) and arrive at
\begin{equation} \label{eq:radiuscircular}
r_{\theta = \pi/2, {\rm circular}} = \frac{L}{E}
\end{equation}
Whereas for the latter case ($\phi = \phi_0$, $L = 0$), we plug (\ref{eq:t_circle}) and (\ref{eq:AngularParity}) into (\ref{eq:fourvelocity}) to find
\begin{displaymath}
r = \frac{\Omega \cosh(\sqrt{\Lambda}t)}{E[\cos^2{\theta} \pm \sin{\theta}\sinh(\sqrt{\Lambda}t)]}
\end{displaymath}
which is not constant. Consequently, in our inertial systems, light can travel in circular orbits only in the equatorial plane with angular velocity given by (\ref{eq:spincircular}) at a radius given by (\ref{eq:radiuscircular}). The type of lensing expected from a black hole or `dark matter'\cite{Massey} is evidently reproduced in a similar manner by light traveling with angular velocity about an inertial center point. Although in the analysis above, we studied circular orbits where light remains at a constant $r$, the logic applies similarly for the case where the photon has both radial and angular velocity components.

We come to the redshift factor for light traveling radially. Before we begin with this analysis, we must refer back to our procedure for determining the observed wavelength of a photon when operating under the assumptions of special and general relativity. In general relativity, the observed frequency $f$ of a photon with momentum $p^a$ ($p^a = \hbar k^a$) emitted/received by an observer traveling with proper velocity in component form given by $u^{\mu} = dx^{\mu}/d\tau$ is (see Chapter 6.3 of \cite{Wald_GR})
\begin{displaymath}
-2\pi f = k_a u^a\bigg|_{P}
\end{displaymath}
where $P$ is the location in space-time at which the event in question occurs (i.e. emission/absorption). Dividing through by the Minkowski constant for the speed of light $c$, we have
\begin{displaymath}
-\frac{2\pi}{\lambda} = \sum_{\mu} k_{\mu} \cdot \frac{1}{c}\frac{dx^{\mu}}{d\tau}
\end{displaymath}
where $c = \lambda f$ and $\lambda$ is the wavelength of the photon emitted/received by our observer. Since we require that our theory in flat space-time reduce to special relativity within a localized region of our respective inertial system (i.e. $d\chi^2 \rightarrow c^2 d\tau^2$ in this localized region), it appears necessary to assume that, in our theory of inertial centers, the wavelength of a photon with wave-vector $k^a$ emitted/received by an observer with `proper velocity' $U^a$ is given by
\begin{displaymath}
-\frac{2 \pi}{\lambda} = k_a U^a\bigg|_{P}
\end{displaymath}
where we emphasize to the reader that in our theory the component form of the `four-velocity' for our observer is affinely parametrized by $\chi$ (i.e. $U^{\mu} = dx^{\mu}/d\chi$), in direct contrast to special and general relativity for which the four-velocity of an observer would be affinely parametrized by proper time $\tau$. Proceeding with our radial treatment, the wave-vector for this photon is of the form,
\begin{displaymath}
k^{\mu} \rightarrow \langle k^t, k^r, 0, 0 \rangle
\end{displaymath}
And the wavelength observed by a radially traveling individual is given by
\begin{displaymath}
-\frac{2\pi}{\lambda} = k^{a}U_{a} = -\Lambda r^2 k^t \frac{d t}{d \chi} + k^r \frac{d r}{d \chi}
\end{displaymath}
Using the Killing vector field in (\ref{eq:Killingradial}), we obtain the conservation law ($-\rho_0=k^{a}\rho_{a}$):
\begin{displaymath}
\rho_0 = \sqrt{\Lambda}r \cosh(\sqrt{\Lambda}t)k^t + \sinh(\sqrt{\Lambda}t)k^r
\end{displaymath}
But for a photon, $0 = k^{a}k_{a} \Longrightarrow  k^r = \pm \sqrt{\Lambda} r k^t$. So,
\begin{displaymath}
k^t = \frac{\rho_0}{\sqrt{\Lambda} r [\cosh(\sqrt{\Lambda}t) \pm \sinh(\sqrt{\Lambda}t)]}
\end{displaymath}
where the positive root corresponds to light traveling away from $r = 0$ and negative to light traveling inward. Solving for the motion of the observer in this particular inertial reference frame,
\begin{displaymath}
-1 = -\Lambda r^2 \bigg(\frac{d t}{d \chi} \bigg)^2 + \bigg(\frac{d r}{d \chi} \bigg)^2
\end{displaymath}
And for an observer nearly at rest with respect to the inertial center about which he/she orbits,
\begin{displaymath}
\frac{d t}{d \chi} = \pm \frac{1}{\sqrt{\Lambda}r}
\end{displaymath}
Taking time to move forward, we find that
\begin{displaymath}
\frac{2\pi}{\lambda} = \frac{\rho_0}{\cosh(\sqrt{\Lambda}t) \pm \sinh(\sqrt{\Lambda}t)}
\end{displaymath}
But from our earlier analysis, we found that a radially traveling photon abides by the equation, $r(t) = r_0 \textrm{exp}(\pm \sqrt{\Lambda}t)=r_0[\cosh(\sqrt{\Lambda}t) \pm \sinh(\sqrt{\Lambda}t)]$. Thus,
\begin{equation}
\lambda \propto r
\end{equation}
Then for a light signal sent between two observers at rest in this inertial frame, the redshift factor $z$ is given by the expression:
\begin{equation}
z = \frac{\lambda_{\rm absorber} - \lambda_{\rm emitter}}{\lambda_{\rm emitter}}  = \frac{r(t_{{\rm absorber}})}{r(t_{{\rm emitter}})}  - 1
\end{equation}
Consequently, we see large shifts from emitters much closer to the center of the system (assuming the absorber position remains the same).

Suppose, within the framework of this theory, we examine light propagating at the scale of the inertial reference frame associated with our observable universe. Then analogous to the manner in which the expression for the Hubble parameter \cite{Hubblelaw} is derived in the Friedmann-Lema\^{i}tre-Robertson-Walker (FLRW) metric\cite{Friedmann}\cite{Friedmann_negative_energy}\cite{Lemaitre}, we set the Hubble constant $H_{0}$ equal to
\begin{equation}\label{eq:Hubble}
H_0 = \bigg|\frac{\dot{r}}{r}\bigg| = \sqrt{\Lambda}
\end{equation}
where $\dot{r} = dr/dt$. We notice that
\begin{equation}
\frac{\ddot{r}}{r} = \Lambda > 0
\end{equation}
producing a positive value for the acceleration of cosmological redshift and therefore replicating the observed effects assumed to be the result of `dark energy' \cite{Perlmutter}. Thus, for any particular inertial reference frame, we should see a shift in wavelength similar to the Hubble constant for the radial motion of photons. We'll use this conclusion later when we take the Pioneer anomaly as support for our theory in the context of the inertial system associated with the Milky Way. However, we should be concerned with our expression for $\dot{r}/r$,
\begin{displaymath}
\frac{\dot{r}}{r} = \pm \sqrt{\Lambda}
\end{displaymath}
as this theory then requires that we also have blueshifted objects if the absorber is in fact closer than the emitter to the center of the inertial frame within which the light signal in question propagates (i.e. negative values for $\dot{r}/r$ and $z$). Nevertheless, if we apply our analysis to objects at the scale of the Local Group \cite{vandenBergh_LG}\cite{vandenBergh_modLG} as in Table \ref{tab:gen_redshift_LG}, we would require an alternative interpretation for the observed significant blueshifts. Whereas in current models, this blueshift would be interpreted as the Doppler effect and thus for example as Andromeda (Messier 31) moving with velocity toward the Milky Way \cite{Cox}, in our theory of inertial centers one could interpret a portion of this blueshift (we say portion as the motion of our observers within an inertial system also affects wavelength) as the possibility that Andromeda is farther away from the inertial center associated with the Local Group than we are. In support of these observations, we refer to Table \ref{tab:red_shift_Dec_pattern} where there appears to be an orientation associated with our redshift values. For similar values of right ascension ($\pm 2$ h), we see a steady change in wavelength shift from blue ($-$) to red ($+$) as one proceeds from large positive values of declination to large negative values of declination. In our theory, we would still need to consider differences in radial distance associated with these objects and not just spatial orientation. However, given that our distance modulus values are very much similar for most of these entries ($\approx 24$ mag), it seems that this interpretation for an orientation to the Local Group should be taken into consideration. On the other hand, even if there does appear to be an orientation associated with the Local Group, we must question why we have not seen significant blueshifts at much larger scales. We will come back to these ideas later in our work.

\begin{figure}[!ht]
\begin{center}
\includegraphics[width=8.3cm]{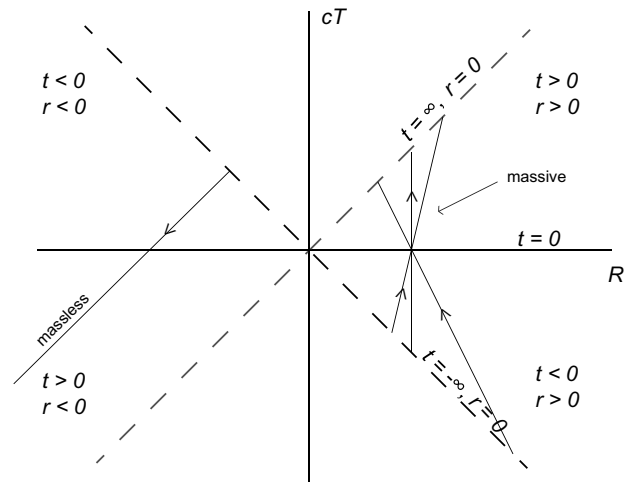}
\end{center}
\caption{{\bf Geodesic paths in Minkowski coordinates}}
\label{fig1}
\end{figure}

Until now, we have assumed that our coordinate time can take values between $-\infty < t < \infty$ without explicitly examining the motion of particles in the $t<0$ region. Reducing our analysis to solely radial motion away from the poles, we analyze geodesic paths in Minkowski coordinates ($T,R$) first for simplicity. However, we must be very clear that under our assumptions $T$ does not represent inertial time as previously stated and in our theory of inertial centers corresponds to an ``unnatural" time coordinate for flat space-time combining both physically observable clock time and spatial distance as $cT = r\sinh(\sqrt{\Lambda}t)$. Then our equations of motion reduce to

\begin{displaymath}
0 = \frac{d^2 T}{d{\sigma}^2} \indent \indent {\rm and} \indent \indent 0 = \frac{d^2 R}{d{\sigma}^2}
\end{displaymath}
leading to the straight lines that we expect in Minkowski coordinates:
\begin{equation}\label{eq:Minkgeo}
R = v \cdot (cT) + R_0
\end{equation}
where $v$ is a constant bounded by $|v| \leq 1$. We leave the physical interpretation of the Minkowski constant $c$ in this theory of inertial centers for the next section. However, using our transformation equations, we find in radial Rindler coordinates
\begin{equation}
r(t) = \frac{R_0}{\cosh(\sqrt{\Lambda}t) - v\sinh(\sqrt{\Lambda}t)}
\end{equation}
One immediately notices that for massive particles ($|v| < 1$), both limiting cases of $t\rightarrow \pm \infty$ result in the particle heading inward toward the $r = 0$ center point of the inertial system. This produces a scenario for inertial motion of massive objects beginning at a center point in the far past, coming to a maximum radial distance away at a later time, and then heading back inward to eventually return to the same center point. In other words, classically, all particles must also originate from the $r =0$ center point of the particular inertial frame in question (see Figure \ref{fig1}).

\subsection* {Reduction to special relativity}

Taking the differential of both Rindler transformation equations:
\begin{displaymath}
c d T = d r \sinh(\sqrt{\Lambda}t) + \sqrt{\Lambda} r \cosh(\sqrt{\Lambda}t)d t 
\end{displaymath}
\begin{displaymath}
d R = d r \cosh(\sqrt{\Lambda}t) + \sqrt{\Lambda} r \sinh(\sqrt{\Lambda}t)d t
\end{displaymath}
where
\begin{displaymath}
\sinh(\sqrt{\Lambda}t) = \sqrt{\Lambda}t + \frac{(\sqrt{\Lambda}t)^3}{3!} + \frac{(\sqrt{\Lambda}t)^5}{5!} + \ldots
\end{displaymath}
\begin{displaymath}
\cosh(\sqrt{\Lambda}t) = 1 + \frac{(\sqrt{\Lambda}t)^2}{2!} + \frac{(\sqrt{\Lambda}t)^4}{4!} + \ldots
\end{displaymath}
Plugging in these expressions above, we find that
\begin{displaymath}
c d T = d r \bigg(\sqrt{\Lambda}t + \frac{(\sqrt{\Lambda}t)^3}{3!} + \ldots \bigg) + \sqrt{\Lambda} r d t \bigg(1 +  \frac{(\sqrt{\Lambda}t)^2}{2!}  + \ldots \bigg)
\end{displaymath}
\begin{displaymath}
d R = d r \bigg(1 + \frac{(\sqrt{\Lambda}t)^2}{2!} + \ldots \bigg) + \sqrt{\Lambda} r d t \bigg(\sqrt{\Lambda}t + \frac{(\sqrt{\Lambda}t)^3}{3!} + \ldots \bigg)
\end{displaymath}
If we localize our view of space-time such that all differential terms
\begin{equation} \label{eq:loclambda}
\mathcal{O}(\Lambda) = 0
\end{equation}
we will then have
\begin{displaymath}
cd T \approx \sqrt{\Lambda} \bigg(td r + rd t \bigg)
\end{displaymath}
\begin{displaymath}
d R \approx d r
\end{displaymath}
Further, we require for this local patch of space-time that
\begin{equation} \label{eq:locradius}
t d r \ll r d t
\end{equation}
and our transformation equations reduce to
\begin{eqnarray}
cT \approx \sqrt{\Lambda}r t
\\ R \approx r
\end{eqnarray}
with differential expressions
\begin{eqnarray}
cd T \approx \sqrt{\Lambda} r d t \\
d R \approx d r
\end{eqnarray}

For the observer remaining a radial distance $r=r_0$ away from the center of his/her reference frame, the radial Rindler chart will be accurately approximated by Minkowski coordinates under conditions (\ref{eq:loclambda}) and (\ref{eq:locradius}) as $R=r$ and $T \propto t$ in this small $\sqrt{\Lambda}t$ limit. If one takes $\sqrt{\Lambda}$ to be a fundamental property of each inertial system in question, it must be that the measured Minkowski value for the speed of light constant $c$ is a byproduct of the reference frame we wish to locally approximate. In other words, in the Minkowski approximation for the radial Rindler chart, an observer, located a radial distance $r = r_0$ away from the inertial center point about which he/she orbits at $t = 0$, will find:
\begin{equation} \label{eq:speedoflight}
c = \sqrt{\Lambda}r_0
\end{equation}
If we treat $t=0$ as the point at which we determine the initial conditions for the particle that we are observing (i.e. boundary conditions for position and velocity), then our object will appear to move along straight line geodesics for small values of $\sqrt{\Lambda}t$, but as we continue to observe for longer periods of time, the properties of the radial Rindler chart which we are approximating become more and more relevant.

\begin{figure}[!ht]
\begin{center}
\includegraphics[width = 8.3cm]{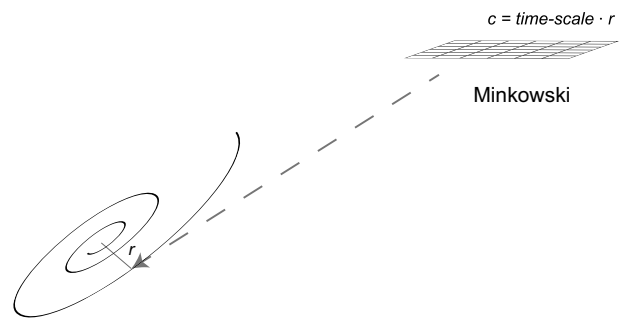}
\end{center}
\caption{{\bf Local approximation of the inertial frame of reference}}
\label{fig2}
\end{figure}

In order for us to relate our theory of inertial centers to special relativity, we must require that coordinate time in the radial Rindler chart progress at the same rate as the proper time of an observer stationary relative to the center of the inertial system within which we are analyzing events (i.e. inertial time). In other words, $dt/d\tau = 1$ for stationary observers located at any particular radial distance $r=r_0$ away from an inertial center. However, keep in mind that stationary observers do not follow along geodesic paths from equation (\ref{eq:radialaccel}). Then, for observers which we can consider as nearly stationary relative to the center of a particular inertial system (i.e. $r = {\rm constant}$), we have $d\chi^2 \rightarrow c^2 d\tau^2$ where $c$ is given by (\ref{eq:speedoflight}), effectively ensuring that our coordinate time $t$ progresses at the same rate as the proper time $\tau$ of a stationary observer. Consequently, we find under (\ref{eq:loclambda}) and (\ref{eq:locradius}) in addition to our stationary observer assumption that our line element can be treated approximately as
\begin{displaymath}
-c^2 d\tau^2 = -c^2 dT^2 + dR^2 + R^2 d\Omega^2
\end{displaymath}
And therefore in this ``stationary" limit (relative to the inertial center), when not operating about the poles, we come upon time- and spatial-translational invariance within our local region where the origin of our coordinate system is located at the inertial center of this reference frame. Because it appears that we have now recovered time- and spatial-translational invariance in this limit, we can naively assume that we have the ability to translate our coordinate system in any way we prefer (e.g. moving the center of our reference frame away from the inertial center). In other words, we can approximate when our motion is not near the poles of our global inertial system with the metric:
\begin{displaymath}
-c^2 d\tau^2 = -c^2 dT^2 + dX^2 + dY^2 + dZ^2
\end{displaymath}
where $X = R\sin{\theta}\cos{\phi} + X_0$, $Y = R\sin{\theta}\sin{\phi} + Y_0$, and $Z = R\cos{\theta} + Z_0$ for $R>0$ and $X_0$, $Y_0$, and $Z_0$ are constants (see Figure \ref{fig2}). Thus, this local stationary approximation reduces our theory to special relativity (see Chapter 4.2 of \cite{Wald_GR}). Additionally, we see from these transformation equations that the Minkowski chart is not able to cover all of space-time in our theory of inertial centers (i.e. $r<0$ values are neglected by the Minkowski chart). We will come back to this idea later in our work. Physically, our localization conditions require that the time-scale constant $\sqrt{\Lambda}$ for our inertial systems be small enough such that we as observers here on Earth would observe only the stationary limit in our ``everyday lives". Of course, this statement also assumes that we are nearly stationary relative to the inertial center about which we orbit taken in our next section to be the center of the Milky Way. Yet, given our redshift analysis, it appears that the Hubble constant provides the necessary scale \cite{Riess} for this requirement from (\ref{eq:Hubble}).

Furthermore, it is clear that in order to express $d\chi^2$ in terms of the proper time of the observer whose motion we wish to analyze in the relevant inertial frame (i.e. the particular system in which the observer can be treated as a point-particle orbiting an inertial center point) and still have our invariant interval reduce to $c^2 d\tau^2$ in our stationary limit where the observer's distance to the inertial center point about which he/she orbits is very nearly constant, we must have
\begin{equation}\label{eq:invariant}
d\chi^2 = \Lambda r^2 d\tau^2
\end{equation}
where $r$ represents the physical distance to the center point of the inertial system in question. In addition, according to our theory of inertial centers, the value that we use in special and general relativity for the constant $c$ in our massive geodesic equations relies on the particular inertial reference frame in which we can regard the object whose local behavior we wish to examine as a point-particle orbiting an inertial center point (in special and general relativity, $g_{ab}u^a u^b = -c^2$ where $u^{\nu} = dx^{\nu}/d\tau$). Therefore, the local Minkowski constant that we measure for the speed of light is dependent upon our position in our most local inertial reference frame (i.e. the frame in which we can be treated as a point-particle orbiting an inertial center). As well, for two different stationary observers orbiting about the same inertial center point, the clock of the observer located {\it closer} to their shared inertial center will appear to {\it run faster} when examined from the perspective of the more distant observer. Meaning, not only do observable clock rates differ due to the relative velocity of individuals as in special relativity, but they also differ due to the difference in distance of each individual from the inertial center about which each orbits. Thus, initially synchronized clocks that are stationary relative to the shared inertial center about which both orbit {\it do not remain synchronized} if they are located at different distances from this inertial center.

\subsection* {Application to a local gravitational system}

Before we present the approximations of this section, it seems necessary to provide remarks as to how gravitation fits into the theory of inertial centers. The formulation of our theory of inertial centers detailed in previous sections deals with the structure of flat space-time ignoring possible issues with curvature. So what we are really asking is the following. How does an object move in flat space-time when absolutely no external forces, fields, etc. are present to affect said object? Nevertheless, we still assume within our model that all objects cause curvature in space-time due to their intrinsic energy-momentum content, but this curvature we take to be a local effect within the far larger inertial system that we are attempting to redefine in this work. However, as presented in the previous section ``Reduction to special relativity", we claim that locally the structure of flat space-time within our redefined inertial reference frames reduces to the flat space-time of Einstein's theory of special relativity as long as the observer remains at very nearly the same distance away from the inertial center about which he/she orbits. As discussed above, we term this the ``local stationary approximation", and in this approximation the observer sees space-time locally within a region located at the same radial distance as the observer away from the inertial center about which he/she orbits as approximately special relativistic, where the speed of light in this confined region of space-time is given by (\ref{eq:speedoflight}) and our affine parameter reduces to $\chi = c \cdot \tau$. If one then considers the influence of an object on the structure of space-time in this local region where special relativity approximately holds, we assume in the theory of inertial centers that this object will bend space-time locally according to Einstein's general relativity. Meaning, gravity remains a consequence of {\it local} curvature in space-time in the theory of inertial centers. However, when we take a perspective far away from our massive object so that the curvature this object induces in space-time looks approximately insignificant for the purpose of examining motion at these larger scales, we claim that we can treat the object very nearly as a particle in a flat space-time inertial reference frame as formulated above, where the inertial motion of the object is dictated by the geodesics of our model. But again, if we focus our attention on the local region around the massive object while disregarding the existence of the larger inertial system, we will still observe the effects resulting from the curvature the object induces in space-time and thus the gravitational effects it has on other objects around it (i.e. general relativity holds locally).

In the following, we give an approximate method under our stationary localization conditions as described in the previous section for determining a potential implication of our theory of inertial centers with regard to the observables of a local gravitational system. We take the view that the Schwarzschild metric \cite{Schwarz_a} applies in the small $\sqrt{\Lambda}t$ limit within confined regions of our inertial reference frame for observers nearly stationary with respect to the inertial center about which they orbit as one would expect from the well-established accuracy of general relativity\cite{C_Will_tests}. Below, our ``mixing" of the Schwarzschild metric with the radial Rindler chart is an approximate way of expressing the fact that locally in the inertial reference frame of our theory the observer can treat the speed of light as nearly constant if one were to remove the massive object and work in flat space-time (i.e. set $M=0$ in the Schwarschild metric) as well as the idea that general relativity holds locally. But the observer must always keep in mind that the inertial frame of Einstein's special relativity is actually {\it not} an inertial frame of the theory of inertial centers, and thus this local system is located in the more globally relevant inertial system where the speed of light is not constant. Then if we no longer take $M=0$ (i.e. return the massive object to the local system), we should still expect gravitation as formulated by Einstein when we examine locally and disregard the larger inertial system from our model. In other words, the Schwarzschild metric still applies locally in the theory of inertial centers when examining motion about an uncharged non-rotating spherically symmetric massive object. However, if we move our observer farther and farther away from the gravitational source, the local limit will no longer apply since we have to take into consideration the structure and properties of the larger inertial system as well as the fact that in our theory objects move inertially along geodesics different from those of Einstein's theories of special and general relativity, even though locally these different geodesics appear to be very similar (i.e. $d\chi^2 \rightarrow c^2 d\tau^2$ in the local stationary limit).

Additionally, we need some approximate way to take into account the fact that the speed of light is not constant throughout the inertial system while still keeping in mind that locally the observer may experience gravitational effects from a massive object nearby. We admit that the methods in this section are rough at best, but it is our hope that in future work we will be able to model far more accurately this transition from the local approximation of general relativity to the more global application of the theory of inertial centers.

Our metric equation takes the form of the Schwarzschild solution:
\begin{displaymath}
-d\chi^2 = -B c^2 dT^2 + \frac{1}{B}dR^2 + R^2 d\Omega_l^2
\end{displaymath}
where
\begin{displaymath}
B(R) = 1- \frac{2MG}{c^2 R} \indent \indent {\rm and} \indent \indent d{\Omega_l}^2 = d{\theta_{l}}^2 + d{\phi_l}^2 \sin^2{\theta_l}
\end{displaymath}
($T,R,\theta_l,\phi_l$) describe our local gravitational system and ($t,r,\theta,\phi$) refer to the global inertial reference frame within which the local system is located. In other words, we assume that the observer takes the coordinate transformations away from the inertial center to cover local space-time in the same manner as outlined in our previous section. Meaning, ignoring the existence of the massive object
\begin{eqnarray}
cT = r \sinh(\sqrt{\Lambda}t) \nonumber
\\ X = r \cosh(\sqrt{\Lambda}t) \sin{\theta}\cos{\phi} + X_0 \nonumber
\\ Y = r\cosh(\sqrt{\Lambda}t) \sin{\theta}\sin{\phi} + Y_0 \nonumber
\\ Z = r\cosh(\sqrt{\Lambda}t) \cos{\theta} + Z_0 \nonumber
\end{eqnarray}
where $X_0$, $Y_0$, and $Z_0$ are constants, $\sqrt{\Lambda}t$ is taken to be small, and we only examine the $r>0$ region of the inertial system. Thus, $T \approx t$ where we employ equation (\ref{eq:speedoflight}) for the local stationary limit. Then taking into account the existence of this massive object in the local region with $M=0$ flat space-time Minkowski coordinates given by ($T,X,Y,Z$) in the local stationary limit, we employ the Schwarzschild metric noting that our affine parameter is approximately given by $\chi = c \cdot \tau$. As well, $c$ refers to the speed of light in the local system at the point in the global inertial frame where the observer and photon meet, and $M$ is the mass of the object. Then we will proceed through a standard treatment of the gravitational redshift for the Schwarzschild metric (see Chapter 6.3 of \cite{Wald_GR}). However, we keep the Minkowski constant $c$ in all of our expressions as we intend to investigate the implications of the variable nature of the speed of light in flat space-time from our theory of inertial centers. For an observer and photon both traveling radially in this local system ($U^{\mu} = dx^{\mu}/d\chi\rightarrow \langle U^T, U^R, 0,0 \rangle$, $k^{\mu} \rightarrow \langle k^T, k^R, 0, 0 \rangle$), we have
\begin{displaymath}
- \frac{2 \pi}{\lambda} = -Bc^2U^Tk^T + \frac{1}{B}U^Rk^R
\end{displaymath}
where $\lambda$ is the wavelength measured by our observer. Applying conservation laws for $U^{a}$ and $k^{a}$ using the time-translationally invariant Killing vector field for the Schwarzschild metric, $\xi^{a} = (\partial/\partial T)^a$:
\begin{displaymath}
U^T = \frac{E}{Bc^2} \indent \indent {\rm and} \indent \indent k^T = \frac{\rho_0}{Bc^2}
\end{displaymath}
And taking into account the motion of the observer and photon ($0 = k^{a}k_{a}$ and $-1 = U^{a}U_{a}$)
\begin{displaymath}
U^R = \pm \sqrt{\bigg(\frac{E}{c}\bigg)^2 - B} \indent \indent {\rm and} \indent \indent k^R = \pm \frac{\rho_0}{c}
\end{displaymath}
Plugging into our expression for the observed wavelength of the photon,
\begin{equation}\label{eq: prop}
\frac{2\pi}{\lambda} = \frac{E \rho_0}{B}\bigg[\frac{1}{c^2} - \bigg(\pm \bigg)\bigg|_{{\rm photon}} \cdot \bigg(\pm \frac{1}{E c}\sqrt{\bigg(\frac{E}{c}\bigg)^2 - B}\bigg)\bigg|_{{\rm observer}} \bigg]
\end{equation}
where $(\pm)|_{{\rm photon}}$ and $(\pm)|_{{\rm observer}}$ refer to the photon/observer traveling radially outward/inward ($+ / -$) in the local system (in $R$). If we assume the observer to be nearly at rest in the local frame ($U^T \gg U^R$), then $B = (E/c)^2$ and expression (\ref{eq: prop}) reduces to
\begin{displaymath}
\lambda \propto c \sqrt{B}
\end{displaymath}
where in the following we approximate in the small $\sqrt{\Lambda}t$ limit with our equation for the local speed of light in the inertial reference frame (\ref{eq:speedoflight}). We employ this ``trick" as the Schwarzschild metric is just an approximation in our model valid under confined regions of the particular inertial system within which the gravitational source is located. However, one should be able to experimentally detect with an apparatus of the necessary sensitivity that these photons progress along the geodesics of our theory of inertial centers (and not straight lines) bent locally due to the curvature in space-time caused by our massive object $M$. Therefore, we find a slight modification to the Schwarzschild redshift factor:
\begin{equation}\label{eq: shift at rest}
z = \frac{\lambda_{\rm absorber} - \lambda_{\rm emitter}}{\lambda_{\rm emitter}} =\frac{r_{{\rm absorber}}}{r_{{\rm emitter}}} \sqrt{\frac{1 - \frac{2M G}{R_{{\rm absorber}}} \cdot \frac{1}{\Lambda r^{2}_{{\rm absorber}}}}{1 - \frac{2MG}{R_{{\rm emitter}}} \cdot \frac{1}{\Lambda r^{2}_{{\rm emitter}}}}} - 1
\end{equation}
where $r_{{\rm absorber/emitter}}$ refers to the radial position of the absorber/emitter in the inertial reference frame (i.e. relative to the inertial center) and $R_{{\rm absorber/emitter}}$ to the radial position relative to the center of our massive object $M$ in the local gravitational system. Consequently, we should see a modified redshift factor consisting of the Schwarzschild expression (Chapter 6.3 of \cite{Wald_GR}) scaled by the solution found in our flat space-time vacuum analysis.

Let us then apply this analysis to the case of a space probe traveling out of our solar system where the $r_{\rm absorber}/r_{\rm emitter}$ factor should have a larger impact on our observations. In our crude example, we treat both the probe and the absorber as essentially stationary. Referring to expression (\ref{eq: shift at rest}) for observers at rest, the absorber wavelength in terms of the emitter is
\begin{displaymath}
\lambda_{{\rm absorber}}|_{{\rm Modified}} = \lambda_{{\rm emitter}} \cdot \frac{r_{{\rm absorber}}}{r_{{\rm emitter}}} \sqrt{\frac{1 - \frac{2MG}{R_{{\rm absorber}}} \cdot \frac{1}{\Lambda r^{2}_{{\rm absorber}}}}{1 - \frac{2MG}{R_{{\rm emitter}}} \cdot \frac{1}{\Lambda r^{2}_{{\rm emitter}}}}}
\end{displaymath}
where our $R$ values in this example refer to local radial distances away from the center of the Sun and $r$ to distances away from the center of the Milky Way. For the ratio between the Schwarzschild wavelength and the modified value above assuming the term under the square root remains approximately the same for small changes in $r$ relative to changes in $R$, we have
\begin{displaymath}
\frac{\lambda_{{\rm absorber}}|_{{\rm Schw}}}{\lambda_{{\rm absorber}}|_{{\rm Modified}}} \approx \frac{r_{{\rm emitter}}}{r_{{\rm absorber}}}
\end{displaymath}
where
\begin{displaymath}
\lambda_{\rm absorber}|_{\rm Schw} = \lambda_{\rm emitter} \cdot \sqrt{\frac{1 - \frac{2MG}{R_{\rm absorber}} \cdot \frac{1}{c^2_{\rm absorber}}}{1 - \frac{2MG}{R_{\rm emitter}} \cdot \frac{1}{c^2_{\rm emitter}}}}
\end{displaymath}
Since Pioneer 10 was on course to travel away from the center of the Milky Way in the general direction of Aldebaran\cite{Anderson}, we can approximate the path of our photon as nearly a radial one in our galactic inertial reference frame. Therefore, if we naively ignore the two-way nature of the Doppler residuals, $r_{\rm absorber} \approx r_{\rm emitter} \cdot e^{-\sqrt{\Lambda}\Delta t}$ where $\Delta t$ is the time it takes the massless particle to travel from the emitter to the absorber, assuming time measured by the emitter progresses at nearly the same rate as that measured by the absorber in this short distance calculation (i.e. $\tau_a \approx \tau_e = t$). Notice, these photons travelled inward for Pioneer 10, so the root is negative. Plugging into our expression above, the fractional difference in wavelength predicted here on Earth is approximately
\begin{displaymath}
\frac{\lambda_{{\rm absorber}}|_{{\rm Schw}}}{\lambda_{{\rm absorber}}|_{{\rm Modified}}} = \frac{1}{e^{-\sqrt{\Lambda}t}} \approx 1+ \sqrt{\Lambda}t
\end{displaymath}
to first order where we assume that our modified expression coincides with our experimental values. Then the observed ``time acceleration" reported in\cite{Turyshev} and \cite{Anderson_Nieto_flyby} provides an estimate for the time-scale of our galaxy of $\sqrt{\Lambda}|_{{\rm MW}} = 2.92 \times 10^{-18}$~s$^{-1}$. The consistency of this value with that of the Hubble constant\cite{Riess} lends support to the argument that the time-scale $\sqrt{\Lambda}$ is universal for all inertial reference frames as we had implicitly assumed from our proposed form of the affine parameter presented in our introduction. However, further experiment is necessary in order to verify this claim.

Clearly, the two-way nature of the Doppler residuals of the Pioneer experiments as well as the difference in clock rates for varying positions within an inertial system in our theory will complicate our analysis further. However, the purpose of this section is to illuminate to the reader the idea that we may have evidence from experiments within our own solar system that support the relevance of this theory of inertial centers and suggest that possibly all inertial reference frames as defined within this theory abide by the same fundamental time-scale constant $\sqrt{\Lambda}$. Nevertheless, others have argued as in \cite{ThermalPioneer} that the Pioneer anomaly is a consequence of the mechanics of the spacecrafts themselves instead of evidence of ``new physics". Therefore, to gain more support for the theory of inertial centers, we must address in future work not only the two-way nature of the Doppler residuals as both Pioneer 10 {\it and} Pioneer 11 appear to report blueshifted wavelengths even when they traveled in opposite directions with respect to the galactic inertial frame of reference but also the possibility that our theory can succinctly explain the other astrometric Solar System anomalies outlined in \cite{Nieto} and \cite{Anderson_Nieto_flyby}.

\subsection* {Quantization of a real scalar field}

We begin our extension into quantum field theory from the covariant form of the Klein-Gordon equation \cite{Fulling}:
\begin{displaymath}
\nabla_{a} \nabla^{a} \phi - \mu^2 \phi = 0
\end{displaymath}
where $\nabla_{a}$ is the derivative operator compatible with the metric $g_{ab}$ (i.e. $\nabla_a g_{bc} = 0$), $\mu = mc/\hbar$, $m$ is the mass associated with our field, and $\hbar$ is the reduced Planck constant. First, we explore how one can intuitively arrive at this equation of motion given our classical assumptions. In special relativity, we have
\begin{displaymath}
-m^2 c^2 = p^{a}p_{a} = \sum_{\nu, \beta} \eta_{\nu \beta} p^{\nu} p^{\beta}
\end{displaymath}
where $p^{\nu} = m \cdot dx^{\nu}/d\tau$ and $\eta_{\nu \beta}$ refers to the Minkowski metric components. Making the substitution $p^{a} \rightarrow - i \hbar \nabla^{a}$, we come upon the Klein-Gordon equation above for a scalar field.

However, in our theory of inertial centers, the equation of motion in terms of `momentum' is given by
\begin{displaymath}
-m^2 = p^{a} p_{a}
\end{displaymath}
where now $p^{\nu} = m \cdot dx^{\nu}/d\chi$ and so we have a major difference in our `momentum' terms. In contrast with our experience in relativity, the `four-velocity' for massive particles in our theory is parametrized by $\chi$ and {\it not} by proper time $\tau$. Unfortunately, there does not appear to be a natural operator substitution for $dx^{\nu}/d\chi$. Yet, if we use expression (\ref{eq:invariant}), we have a potential extension of the Klein-Gordon equation when analyzing motion {\it at a particular scale}. It appears that one should substitute $c \rightarrow \sqrt{\Lambda} r$ to find
\begin{equation}\label{eq:KG extend}
\Box \phi - \tilde{\mu}^2 r^2 \phi = 0
\end{equation}
where $\tilde{\mu} = m\sqrt{\Lambda}/\hbar$ and $\Box = \nabla_{a} \nabla^{a}$ is the Laplace-Beltrami operator. Notice that in our equation of motion we have explicit reference to the particular inertial reference frame in which we are analyzing the behavior of the field as opposed to the Klein-Gordon equation which has no explicit reference to any inertial system. This seems to be consistent with the idea that the proper time is {\it not} the invariant quantity associated with our theory of inertial centers, and therefore our choice of proper time reflects the choice of {\it scale} in which we must work to analyze the progression of our field within this inertial system. One can also apply this substitution in an analogous manner to other equations of motion/Lagrangians, yet in the following we will only address the simple case of a free real massive scalar field.

Then, as outlined in Chapter 4.2 of \cite{Wald_QFT} and briefly reviewed in Appendix D, we must ``slice" our manifold $M$ into space-like hypersurfaces each indexed by $t$ ($\Sigma_t$). For our radial Rindler chart, the future-directed unit normal to each $\Sigma_t$ is given by
\begin{equation}
n^{a} = \frac{1}{\sqrt{\Lambda}|r|}\bigg(\frac{\partial}{\partial t} \bigg)^a
\end{equation}
where the absolute value is necessary to keep $n^{a}$ future-directed for all values of $r: 0 < r^2 < \infty$, allowing for positive and negative values. We will interpret the physical significance of this relaxation on the domain restrictions for our radial coordinate later in our analysis. We see that our hypersurface can be decomposed into the union of two surfaces for each of the Rindler wedges ($r<0$ and $r>0$), and thus the inner product of our Klein-Gordon extension is given by
\begin{eqnarray}
(\phi_1, \phi_2) = -i\Omega([\bar{\phi}_1,\bar{\pi}_1], [\phi_2, \pi_2]) \nonumber
\\ = -i \int_{\Sigma_{{\rm I}} \cup \Sigma_{{\rm II}}} d^3x \sqrt{|h|} [\phi_2 n^{a} \nabla_{a} \bar{\phi}_1 - \bar{\phi}_1 n^{a} \nabla_{a} \phi_2 ]
\end{eqnarray}
where the bar symbol indicates complex conjugation (i.e. $\bar{\phi}_i$ is the complex conjugate of $\phi_i$), $\Sigma_0 = \Sigma_{{\rm I}} \cup \Sigma_{{\rm II}}$ is the union of these two radial Rindler wedge space-like hypersurfaces, $n^{a}$ is the unit normal to our space-like hypersurface $\Sigma_0$, $h_{ab}$ is the induced Riemannian metric on $\Sigma_0$ ($h = \mathrm{det}(h_{\nu \beta})$; $(h_{\nu \beta})$ denotes the matrix associated with these Riemannian metric components), and $\Omega$ refers to the symplectic structure for our extension of the Klein-Gordon equation.

We should be rather concerned considering the discontinuous nature of the time-orientation of $n^{a}$ (the absolute value is not a smooth function) as well as the undefined behavior of our unit normal for $r=0$, the location of our inertial center. However, given the solutions we find below, it seems to be an important question whether or not we are forced to treat each Rindler wedge separately as its own globally hyperbolic space-time or the combination of these wedges as the entire space-time over which we must analyze solutions to our extension of the Klein-Gordon equation. The difference between these two formulations will be that in the former we must define separate creation and annihilation operators for each wedge as in the analysis of \cite{Wald}. Whereas in the latter, we have one set of creation and annihilation operators for all values of $r$ over the range: $0 < r^2 < \infty$, where $r$ can take both positive and negative values. It also seems likely that a greater understanding of our inertial centers and their physical significance (i.e. how are these inertial centers established?) will provide far more insight into the proper way to treat this situation. In this paper, however, we assume the latter approach requiring that we use all values of $r$ (positive and negative) to cover our inertial reference frame and naively ignore the issues with $r=0$ mentioned above. This approach seems to be far more consistent with the idea implicit in our theory of inertial centers that the radial Rindler chart covers the entire flat space-time manifold for the inertial system in question, except of course for the location of each of our inertial centers (i.e. $r=0$). We find that our inner product is given by
\begin{eqnarray}
(\phi_1, \phi_2) = -i \int_{-\infty}^{\infty}dr \int_{0}^{\pi} d\theta \int_{0}^{2\pi} d\phi \bigg[ r^2 \cosh^2(\sqrt{\Lambda}t) \sin{\theta} \bigg(\frac{\phi_2}{\sqrt{\Lambda}|r|} \partial_{t} \bar{\phi}_1 - \frac{\bar{\phi}_1}{\sqrt{\Lambda}|r|} \partial_{t} \phi_2 \bigg) \bigg]\bigg|_{t=0}  \nonumber
\\ = -\frac{i}{\sqrt{\Lambda}} \cosh^2(\sqrt{\Lambda}t) \bigg[ \int_{0}^{\infty} r dr \int_{0}^{\pi} \sin{\theta} d\theta \int_{0}^{2 \pi} d\phi \bigg(\phi_2 \partial_t \bar{\phi}_1 - \bar{\phi}_1 \partial_t \phi_2 \bigg) \nonumber
\\ - \int_{- \infty}^{0} r dr \int_{0}^{\pi} \sin{\theta} d\theta \int_{0}^{2 \pi} d\phi \bigg(\phi_2 \partial_t \bar{\phi}_1 - \bar{\phi}_1 \partial_t \phi_2 \bigg) \bigg] \bigg|_{t = 0}
\end{eqnarray}
Our remaining task reduces to solving for solutions ($\phi_i$) to our extension of the Klein-Gordon equation (\ref{eq:KG extend}). From Appendix F which utilizes \cite{Arfken}, \cite{Abramowitz}, \cite{Watson}, \cite{Titchmarsh}, and \cite{Passian}, we find
\begin{equation} \label{eq:wavefcn}
\phi_{\alpha, l, m} =  \sqrt{\frac{\tilde{\mu} \alpha}{2\pi \cosh(\pi \alpha)}} \cdot  \sqrt{1- \eta^2} \cdot P_{l}^{-2i \alpha}(\eta) \cdot \frac{K_{i \alpha}(\frac{\rho^2}{2})}{\rho} \cdot Y_l^m(\theta, \phi)
\end{equation}
where $\eta = \tanh(\sqrt{\Lambda}t)$ and $\rho = \sqrt{\tilde{\mu}} r$. $Y_l^m$ is the spherical harmonic of degree $l$ and order $m$. We maintain convention and use $m$ to denote the order of $Y_l^m$. However, this $m$ is a quantum number very different from the mass of our scalar field. The mass term is contained solely in our expression for $\tilde{\mu}$. $K_{i \alpha}$ is the Macdonald function (modified Bessel function) of imaginary order $\alpha$. $P_l^{-2i\alpha}$ is the Legendre function of degree $l$ and imaginary order $-2\alpha$.

\begin{figure}[!ht]
\begin{center}
\includegraphics[width = 8.3cm]{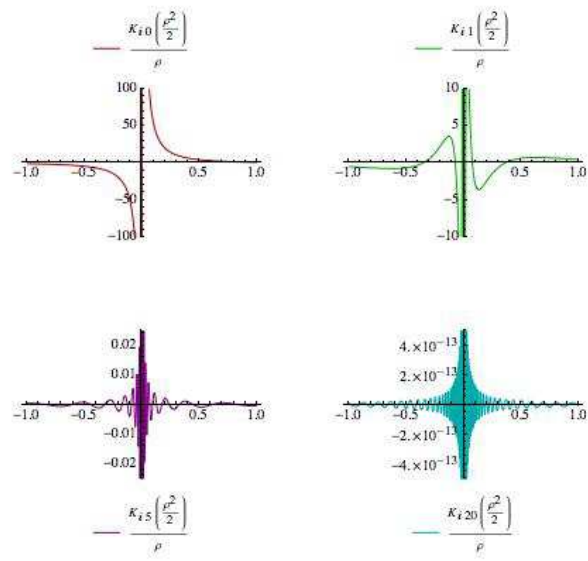}
\end{center}
\caption{{\bf Plots of $h_{\alpha}(\rho)$ for $\alpha = 0,1,5,$ and $20$}}
\label{fig3}
\end{figure}

Notice, we allow $K_{i \alpha}(\frac{\rho^2}{2})/\rho$ to have domain: $0 < \rho^2 < \infty$ where $\rho$ can take both positive and negative values. Physically, this interpretation requires the existence of the field in {\it both} the negative and positive $r$ regions of the inertial system which brings us back to the discussion earlier in this section of our concern with $n^a$. From \cite{Dunster}, the limiting behavior of $K_{i \alpha}$ expressed as
\begin{displaymath}
\lim_{y \rightarrow 0^+} K_{i \alpha}(y) = -\bigg(\frac{\pi}{\alpha \sinh(\alpha \pi)} \bigg)^{1/2} \bigg[\sin(\alpha \ln(y/2)-\phi_{\alpha,0}) + \mathcal{O}(y^2)\bigg]
\end{displaymath}
where $\phi_{\alpha, 0} = \arg\{\Gamma(1+i \alpha)\}$ and $\Gamma(z)$ is the gamma function along with
\begin{displaymath}
\lim_{y \rightarrow \infty} K_{i \alpha}(y) = \bigg(\frac{\pi}{2y}\bigg)^{1/2} e^{-y}\bigg[1+ \mathcal{O}\bigg(\frac{1}{y}\bigg) \bigg]
\end{displaymath}
shows that $K_{i \alpha}(\frac{\rho^2}{2})/\rho$ oscillates for small $|\rho|$ when $\alpha \neq 0$ and exponentially decays for large $|\rho|$. In addition, from Figure \ref{fig3}, we see that our radial `wave function' spreads out away from $\rho = 0$ for larger `momentum' values of $\alpha$, allowing for oscillatory behavior at larger values of $|\rho|$ and thus an increased likelihood of observing quanta farther away from the inertial center of the reference frame in question.

Our Heisenberg field operator can be expanded in the following manner (see Chapters 3.1 and 3.2 of \cite{Wald_QFT}):
\begin{eqnarray} \label{eq:field_operator}
\hat{\Phi}(t,r,\theta,\phi) = \int_{0}^{\infty} d\alpha \sum_{l=0}^{\infty} \sum_{m = -l}^{l} [\phi_{\alpha, l, m} \hat{a}(\bar{\phi}_{\alpha,l,m}) + \bar{\phi}_{\alpha,l,m} \hat{a}^{\dagger}(\phi_{\alpha,l,m}) ] \nonumber
\\ = \int_{0}^{\infty} d\alpha \sum_{l=0}^{\infty} \sum_{m = -l}^{l} \sqrt{\frac{\tilde{\mu} \alpha}{2\pi \cosh(\pi \alpha)}} \cdot \frac{K_{i \alpha}(\frac{\rho^2}{2})}{\rho} \cdot \sqrt{1- \eta^2} \cdot \bigg[ \hat{a}(\bar{\phi}_{\alpha, l, m}) P^{-2i\alpha}_{l}(\eta) Y_l^m(\theta, \phi) \nonumber
\\ + \hat{a}^{\dagger}(\phi_{\alpha, l, m}) P^{2i\alpha}_{l}(\eta) \bar{Y}_l^m(\theta, \phi)\bigg]
\end{eqnarray}
where the annihilation and creation operators in terms of our inner product are
\begin{eqnarray}
\hat{a}(\bar{\phi}_{\alpha, l, m}) = (\phi_{\alpha, l, m}, \hat{\Phi}) \nonumber
\\ \hat{a}^{\dagger}(\phi_{\alpha, l, m}) = - (\bar{\phi}_{\alpha, l, m}, \hat{\Phi}) \nonumber
\end{eqnarray}
and $\{ \phi_{\alpha, l, m}\}$ comprise an orthonormal basis of the ``positive frequency" solutions to the extended version of the Klein-Gordon equation for the theory of inertial centers. For a real scalar field, these annihilation and creation operators satisfy the commutation relations (bosonic statistics):
\begin{equation}
[\hat{a}(\bar{\phi}_{\alpha,l,m}), \hat{a}^{\dagger} (\phi_{\alpha', l', m'})] = (\phi_{\alpha,l,m}, \phi_{\alpha', l', m'}) = \delta(\alpha - \alpha') \delta_{l l'} \delta_{m m'} \nonumber
\end{equation}

A very important point for the reader to take away from our analysis in this section is that our field operator as defined in (\ref{eq:field_operator}) exists in {\it both} the $r>0$ and $r<0$ portions of space-time. In other words, we take space-time to be comprised of both the $r>0$ and $r<0$ regions of the inertial system, and thus the Minkowski chart is {\it not} able to cover all of space-time in our theory. It then appears that a potential explanation for the matter/antimatter asymmetry in our observable universe within the framework of our theory of inertial centers would be that there exists a parallel region of each inertial system embodied mathematically above by the existence of our field operator in the hypothetical $r<0$ region of space-time. Logically, if we exist in our region of space-time with an imbalance toward matter, one would then assume that in this parallel region there exists an imbalance in favor of antimatter as the total charge of the field throughout {\it all} of space-time should be conserved. We are, of course, operating under the assumption that the solutions to our equation of motion extend in a similar manner as in special relativity when one allows for complex fields of non-zero spin (e.g. solutions to a Dirac equation \cite{Dirac} extension are also solutions to our Klein-Gordon extension) since we should not worry about antiparticles with a real scalar field. Therefore, we must extend our work on the theory of inertial centers to incorporate spin in order to see the full significance of this possible explanation for the matter/antimatter asymmetry in our observable universe.

To conclude our discussion, we assume throughout the rest of this section that $\sqrt{\Lambda}$ is a universal constant for all inertial systems, taken to be the Hubble constant as proposed in our introduction, and imagine that there exists an observer very near to an inertial center point such that his/her motion in this particular reference frame is approximately stationary (i.e. spatial `four-velocities' are very much outweighed by `velocity' in time, $dt/d\chi$). Then from our classical analysis of geodesic paths, our observer experiences a radial acceleration according to (\ref{eq:stationary_force}) of
\begin{displaymath}
\frac{d^2r}{dt^2} = - \Lambda r
\end{displaymath}
where $t$ coincides with the proper time $\tau$ for our nearly stationary observer in this system. However, say we wish to understand our observer's motion {\it not} in terms of his/her proper time in this particular inertial frame but instead in terms of his/her proper time in an external inertial frame of reference where these two different systems do not share a common inertial center point. We know that our invariant interval is given by
\begin{displaymath}
-d\chi^2 = -\Lambda r_l^2 d\tau_l^2 = -\Lambda r_e^2 d\tau_e^2
\end{displaymath}
where the $e$ ($l$) subscript refers to quantities in the external (local) inertial reference frame. Assuming our observer is nearly stationary in {\it both} inertial systems (i.e. coordinate times for each system coincide with proper times within each reference frame respectively), his/her clock in the local frame progresses by
\begin{equation} \label{eq:eff_time}
\frac{dt_l}{dt_e} = \frac{c_e}{c_l}
\end{equation}
Thus,
\begin{displaymath}
\frac{d}{dt_l} = \frac{c_l}{c_e} \cdot \frac{d}{dt_e}
\end{displaymath}
where the $c$'s refer to the Minkowski constants for each particular reference frame (\ref{eq:speedoflight}). Plugging in above,
\begin{equation}
\frac{d^2r_l}{dt_e^2} = - \Lambda_{{\rm eff}} \cdot r_l
\end{equation}
where $\sqrt{\Lambda_{{\rm eff}}} = \sqrt{\Lambda} \cdot c_e/c_l$.

According to Newtonian mechanics which is a good approximation here since we assume our observer is nearly stationary in the local inertial system, one would attribute this radial acceleration to a `force' (even though we know that there really is no force here), and associated with this `force' is a potential ($\vec{F} = - \vec{\nabla}V$; see Chapters 1 and 2 of \cite{Goldstein}). So for the acceleration above, one would assume while working in Newtonian mechanics that there exists a potential causing this movement of the form:
\begin{equation}
V = \frac{1}{2} m \Lambda_{{\rm eff}} \cdot r_l^2
\end{equation}
Then our Hamiltonian ($H = T + V$; see Chapter 8 of \cite{Goldstein}) for this system is given by
\begin{equation}
H = \frac{p^2}{2m} + \frac{1}{2}m \Lambda_{{\rm eff}} \cdot r_l^2
\end{equation}
where $m$ is the mass of our observer and $T = p^2/2m$ is the kinetic energy associated with his/her motion as observed in the external frame. If our observer is on the order of $10^{-15}$~m\cite{De_Jager} away from his/her local inertial center and $c_e$ is found in the external frame to be $\approx 3.0 \times 10^8$~m/s\cite{Mohr}, we find $\sqrt{\Lambda_{{\rm eff}}}\sim 10^{23}$~s$^{-1}$. We remark for the reader less acquainted with nuclear theory that the Hamiltonian above is referred to as the isotropic harmonic oscillator and was used as a starting point for nuclear shell models due to its ability to reproduce the ``magic numbers" associated with stable configurations of nucleons within the nucleus (see Chapter 4 of \cite{Mayer} and Chapter 3.7 of \cite{Sakurai}). In addition, the energy scale associated with the Hamiltonian above (i.e. $\hbar \cdot \sqrt{\Lambda_{{\rm eff}}} \sim 10^8$~eV) is of a similar order as the scale inputted into these isotropic harmonic oscillator models for the magnitude of the nuclear `force' \cite{Brown}. Thus, our ability to replicate the same features as those of the simplest nuclear shell model compels us to ask the following question with regard to the theory of inertial centers: {\it Is there an inertial center point at the center of the nucleus of every atom?}

\section*{Limitations of the study, open questions, and future work}

There is a plethora of data for us to critically investigate the validity of this theory of inertial centers. Nevertheless, we have chosen to leave these detailed investigations for future work as the purpose of this paper is to lay out the theoretical foundations to illicit these types of rigorous comparisons with experiment for all aspects of our model. As we have mentioned briefly at certain points within our discussion, there are many open questions that must be addressed. The most pressing of these appears to be how to explain the cosmic microwave background (CMB) within our theory of inertial centers. One may be tempted to immediately point to the Fulling-Davies-Unruh effect\cite{Unruh} as the source of this cosmic radiation since the Unruh effect predicts that an ``accelerating" observer in Minkowski vacuum, who can be described by orbits of constant spatial coordinate in the classic Rindler chart, detects black-body radiation that appears to be nearly homogeneous and isotropic with predicted anisotropies due to the orientation of this observer throughout his/her ``accelerated" path \cite{Hinton}. However, we must keep in mind that the scale associated with the temperature of Unruh radiation \cite{Unruh}\cite{Wald}
\begin{displaymath}
T =\frac{\hbar a}{2\pi k_B c}
\end{displaymath}
requires $a\sim 10^{20}$~{m/s$^2$} to produce a temperature on the order of the CMB, $T \approx 2.7$~K \cite{Fixsen}, where $k_B \approx 1.38 \times 10^{-23}$~J K$^{-1}$ is Boltzmann's constant, $\hbar \approx 1.05 \times 10^{-34}$~J s\cite{Mohr}, and $a$ is the proper acceleration of the observer. If we approximate the original analysis of \cite{Unruh} by working in 1+1 space-time (i.e. 1 time and 1 spatial dimension), the acceleration would be proportional to the inverse of $r = r_0$ for observers moving along orbits of constant $r$ \cite{Wald}. This then requires $r_0 = c^2/a \sim 10^{-3}$~m for the CMB temperature scale, which clearly makes no sense since we would be millimeters away from the center of our observable universe. Nevertheless, the analysis used to derive the Unruh effect implicitly operates under the assumption of the validity of special relativity in flat space-time and therefore takes $d\chi^2 = c^2 d\tau^2$. Yet, as we have emphasized repeatedly above, in our theory of inertial centers, the invariant interval associated with the metric is given in terms of proper time by $d\chi^2 = \Lambda r^2 d\tau^2$. Therefore, we must extend these ideas to apply to our model where we are observers existing within multiple inertial systems (universe $\rightarrow \ldots \rightarrow$ Local Group $\rightarrow$ Milky Way). In addition, for our situation, this radiation would not be interpreted physically as due to the ``acceleration" of the observer as in the case of \cite{Unruh}, but instead one would have to think of this effect as simply the result of the restriction of the Minkowski vacuum to each of the radial Rindler wedges (see Chapters 4.5 and 5.1 of \cite{Wald_QFT}). We are still encouraged that this course of action may result in a plausible interpretation as experimental evidence of large-scale temperature anomalies appears to suggest a significant orientation to the CMB \cite{Planck}.

At this point in our discussion, we offer a brief review of the literature concerning both the Pioneer anomaly as well as the other known astrometric anomalies within our own solar system. First, however, we mention other theories which contrast with our own study but are relevant for the discussion below. The authors of \cite{Kopeikin} and \cite{Iorio_localcos} investigate the potential effects of an expanding universe which could be induced on objects within our solar system. Furthermore, \cite{Iorio_solarsystemcosm} attempts to model the consequences of an extra radial acceleration on the orbital motion of a planet within our solar system. As well, \cite{Grumiller} provides an alternative model for gravitation resulting in an additional ``Rindler-like" term at large distances which the author claims can potentially model the plateauing nature of observed orbital velocity curves. We must stress that the model proposed in \cite{Grumiller} is in fact very different from the model that we have proposed above as our theory of inertial centers does {\it not} attempt to reformulate gravity. As we emphasized earlier, our model is an attempt to reformulate the motion of objects when no net external forces are acting upon said objects in empty flat space-time. Nevertheless, \cite{Iorio_solar_modgrav}, \cite{Iorio_oort}, \cite{Grumiller_solar}, \cite{Grumiller_large_dist}, and \cite{Culetu_time} use these ideas of an additional ``Rindler-like" term in gravitation to examine the possible observable effects of the aforementioned extension to general relativity. For a background reference concerning phenomenology in the context of general relativity, we refer the reader to \cite{Iorio_phenom} as preparation for our presentation of the known anomalies exhibited within our own solar system.

Besides the Pioneer anomaly, there are experimental claims of possible anomalies alluding to inconsistencies with our current model for the Solar System. These include:

1. An anomalous secular increase in the eccentricity of the orbit of the Moon

2. The ``flyby" anomaly

3. An anomalous correction to the precession of the orbit of Saturn

4. A secular variation of the gravitational parameter $GM_{\odot}$ where $M_{\odot}$ is the mass of the Sun

5. A secular variation of the astronomical unit (AU)

The anomalous secular increase in the eccentricity of the orbit of the Moon was originally found in the experimental analysis of the Lunar Laser Ranging (LLR) data in \cite{Williams_lunarrot} and expanded upon in \cite{Williams_lunargeophys}, \cite{Williams_lunarcore}, \cite{Iorio_moon}, and \cite{Iorio_empMoon}. The ``flyby" anomaly refers to an anomalous shift in the Doppler residuals received from spacecrafts when comparing signals before and after these spacecrafts undergo gravitational assists about planets within the Solar System \cite{Anderson_Nieto_flyby}\cite{Gravassist}\cite{Iorio_effectgenrelhyperbolic}. The anomalous perihelion precession of Saturn appears to be a more controversial claim as the work of \cite{Fienga_gravtests} and \cite{Pitjeva_epm2008} seems to suggest the validity of this observation with further investigation in \cite{Iorio_perihelionSaturn} and \cite{Iorio_perihelionSaturnPlanetX}. However, work such as \cite{Fienga_INPOP10a}, \cite{Pitjeva_epm2010}, and \cite{Pitjev_Pitjeva_constraintsdarkmatter} seems to show that this reported anomaly is an experimental artifact. Finally, the last two anomalies of a secular variation in the product of the mass of our Sun and the gravitational constant $G$ as well as the astronomical unit are more difficult claims to understand in the context of our model as there are many complex mechanisms which could affect our measurements of these quantities (e.g. rate of mass accretion of the Sun from infalling objects versus depletion through expelled radiation resulting from nuclear fusion) in addition to the fact that our measurement of the AU is implicitly linked to our measurement of $GM_{\odot}$\cite{Pitjeva_Pitjev_EstimationSun}. Nevertheless, \cite{Pitjeva_Pitjev_EstimationSun} and \cite{Iorio_effect_sun} are useful references for these anomalies. Additionally, \cite{Nieto} provides a detailed summary of the majority of the anomalies listed above.

Returning to the Pioneer anomaly, the reader may have concerns with our earlier analysis as recent simulations such as \cite{ThermalPioneer} suggest that this anomaly should be taken as a thermal effect from the spacecraft itself instead of evidence linked to ``new physics". For a selection of work concerning the possible thermal explanation of the Pioneer anomaly, see \cite{ThermalPioneer}, \cite{Rievers_ThermalDiss}, \cite{Rievers_Newpowerful}, \cite{Rievers_ModelThermal}, \cite{Bertolami_Radiative}, \cite{Bertolami_Thermal}, and \cite{Francisco}. Nevertheless, this analysis still does not address the asymmetric nature of the ``flyby" anomaly \cite{Anderson_Nieto_flyby}\cite{Gravassist} as well as the other significant astrometric Solar System anomalies summarized in \cite{Nieto}. By ``asymmetric nature", we are referring to the fact that the magnitude of the ``flyby" anomaly appears to depend upon the direction of approach of the space probe toward Earth as well as the angle of deflection away after ``flyby". Furthermore, as mentioned in \cite{TuryshevThermal}, the ``onset" of the Pioneer anomaly after Pioneer 11's encounter with Saturn is still of concern when explaining these observables as the result of systemic thermal effects. While \cite{ThermalPioneer} briefly addresses this ``onset" in their conclusion, future analysis of the early data points for Pioneer 11 near its gravitational assist about Saturn appears to be of the utmost importance, especially considering before its encounter with Saturn this spacecraft moved nearly tangentially to the direction of Sagittarius A*, whereas after it traveled nearly toward the Milky Way center. Thus, in the context of our own model, this ``onset" has the potential to be interpreted as the consequence of the spacecraft's {\it change in direction relative to the inertial center associated with the center of the Milky Way}, similar to ideas we will have to explore for the asymmetric nature of the Earth ``flyby" anomalies (for potential connections between the Pioneer and ``flyby" anomalies, see \cite{Anderson_Nieto_flyby}). Therefore, we choose not to rule out the possibility that the Pioneer anomaly may be support for our theory of inertial centers as this effect as modeled in our earlier analysis in fact must be observed in order for our theory to have physical relevance. As mentioned earlier, we will have to address in far more rigorous detail in future work the dual nature of the Pioneer residuals in order to possibly explain the blueshifts from both Pioneer 10 and Pioneer 11 data.

In addition, others such as \cite{Grumiller}, \cite{Iorio_solar_modgrav}, and \cite{Iorio_oort} have used a ``Rindler-like force" emanating from the center of a gravitational source to supplement general relativistic gravity as a model that can potentially explain orbital velocity curves as well as the Pioneer anomaly \cite{Grumiller_solar}\cite{Grumiller_large_dist}. For a review of how this and other gravitational supplements would impact current expectations for the orbits of other major bodies in the Solar System, see \cite{Iorio_Giudice}, \cite{Fienga_gravtests}, \cite{Standish_testing}, \cite{Standish}, \cite{Iorio_orbeffects}, \cite{Iorio_lensethirringPioneer}, \cite{Iorio_NeptunePioneer}, \cite{Iorio_velocitydepend}, \cite{Iorio_Pioneergravorigin}, \cite{Iorio_JupiterSaturn}, \cite{Iorio_oort}, \cite{Page_HowWell}, \cite{Varieschi}, \cite{Page_MinorPlanets}, \cite{Page_WeakLimit}, \cite{Wallin_TestinggravOutersolar}, and \cite{Tangen}. However, these supplements all require spherical symmetry about the center of the gravitational source in question and are {\it very different} from our reformulation of flat space-time where in our theory we do {\it not} assume that there exists a gravitational source at the center of galaxies, groups, clusters, etc. Recall that we are concerned with reformulating inertial motion and inertial reference frames in flat space-time (i.e. our description of the way in which objects move in {\it flat} space-time when subjected to no net external forces). Additionally, we maintain that locally within confined regions of the inertial system of our theory of inertial centers Einstein's version of gravitation seen as the consequence of space-time curvature induced by the energy-momentum of a massive object in his theory of general relativity still applies in the same manner. In other words, in our theory of inertial centers, this observed deviation from assumed special relativistic flat space-time geodesics arises from our redefinition of the inertial system itself instead of some modification to gravitation. Consequently, when attempting to explain these astrometric Solar System anomalies in the context of our theory, we focus on the difference in geodesics in the galactic inertial reference frame when compared to assumed special relativistic geodesics for flat space-time and assume that {\it all} of the objects in our Solar System including the Sun orbit about the inertial center point associated with the center of the Milky Way (again, we assume that there is no gravitational source at the center of our galaxy). Meaning, the Pioneer anomaly is not taken to be a phenomenon due to gravity in the theory of inertial centers. Instead the Pioneer anomaly and possibly the other astrometric Solar System anomalies which we have listed above are taken to be the result of our redefinition of inertial systems as well as the change in our expectations for what constitutes inertial motion. Consequently, the relative acceleration between massive objects in our solar system is nearly unchanged from what one would expect from general relativity as all objects within our solar system orbit about the center of the Milky Way along relatively similar paths. Therefore, we are not modifying our expectations for the interactions between objects within the Solar System. We are modifying our expectations for the paths of all objects in the Solar System through the Milky Way. While internally within our solar system the planets remain nearly unchanged in their paths as they move slowly in the ``Newtonian limit" (i.e. their speeds are much less than that of light), light propagating between these massive objects in our theory won't behave as one would expect from general relativity as at these speeds one must take into account the properties of the larger inertial system associated with our galaxy. 

One must bear in mind that these anomalies are linked to the propagation of electromagnetic radiation throughout our solar system as our experimental apparatuses use light for precision measurements. While the work of \cite{Kopeikin} attributes the Pioneer anomaly to the local effects of light signal propagation in an expanding universe as expressed by a ``post-Friedmannian" metric decomposition, these claims would not be able to explain the asymmetric nature of the wavelength shift residuals in the ``flyby" anomaly as the FLRW metric requires homogeneous and isotropic expansion of space in all directions \cite{Friedmann}. However, there is {\it no} expansion in our theory of inertial centers and our inertial reference frames {\it do} have an orientation. Therefore, we must take into consideration, when comparing with our own model in future work, two important ideas: in this theory of inertial centers, the speed of light is not constant in flat space-time and objects follow inertial paths described by geodesics about inertial centers in the radial Rindler chart, where we assume that the inertial center associated with the Milky Way is in the direction of Sagittarius A*. Thus, in our model, the observables associated with the astrometric Solar System anomalies listed above do not necessarily reflect the existence of an additional acceleration in the Solar System since our theory's radial acceleration would be imposed on {\it all} objects within the Solar System {\it including the Sun} and in the same direction toward the center of the Milky Way (\ref{eq:stationary_force}) with seemingly negligible difference in magnitude depending upon the position of the massive object in question (i.e. changes in position within our solar system are negligible relative to the distance of our solar system from the center of the Milky Way when considering the motion of massive satellites, planets, etc.). In other words, in sharp contrast with the analysis in papers such as \cite{Grumiller}, \cite{Tangen}, \cite{Standish} and \cite{Iorio_Giudice}, we assume that there is no additional acceleration associated with the Sun's gravitational pull on other objects within the Solar System, and thus the relative acceleration of a satellite, planet, etc. with respect to the center of the Sun remains nearly unaffected in our model when we compare with general relativity. Instead, it appears that in the theory of inertial centers these anomalies should more likely be interpreted as a consequence of the non-constant nature of the speed of light within our galactic inertial system as well as of the expected shifts in wavelength when light propagates between differing distances from an inertial center point. Future experiments within the vicinity of our solar system to test the validity of the theory of inertial centers could include sending a spacecraft to the outer edges of our solar system along a closed orbit about the Sun or using identical spacecrafts along open orbits in different directions with respect to the galactic center (e.g. one travels tangentially to the direction of the center of the Milky Way while another moves directly toward/away from the center; for a hyperbolic orbit proposal, see \cite{Dittus}). To test the positional dependence aspects for electromagnetic radiation in this theory, these hypothetical missions should measure the potential variations in wavelength shift and time delay for light signals sent and received at different positions along these orbits with respect to the center of the Milky Way. As well, future theoretical work will require us to explicitly detail observational effects on our astrometric measurements of the planetary ephemerides that are unique to the theory of inertial centers. One could then potentially find these predicted deviations from current models when comparing with the experimental work of \cite{Fienga_INPOP10a} and \cite{Pitjev_Pitjeva_constraintsdarkmatter}.

Using the measured value for the speed of light on Earth ($c_{\rm Earth} \approx 3.0 \times 10^8$~m/s) and the value for the time-scale given from the ``time acceleration" in \cite{Anderson_Nieto_flyby}, we find that our distance to the center of the Milky Way is approximately $r_0|_{{\rm MW}} \approx 1.03 \times 10^{23}$~km. We see that the value obtained for our galactic radial distance is far larger than the predicted value from models requiring a supermassive black hole at the center of the Milky Way (intimidatingly, nearly six orders of magnitude \cite{Eisenhauer}). It is imperative then that we reconcile this calculated value with observational data. Not only will this maintain consistency with experiment but it will also provide accurate distance scales within our galaxy. This will allow us to further understand the large observed wavelength shifts near Sagittarius A* within the framework of our theory of inertial centers and potentially explain the paradox of youth \cite{Ghez} through concrete analysis of star formation near the Milky Way center.

Addressing our classical inertial motion analysis, one can immediately tell from the theoretical approach in our discussion that this paper is limited by the lack of necessary quantitative comparison with orbital velocity curves, redshift surveys, and lensing observations. Future work will require modeling using computer simulations of our equations of motion not only to produce orbital velocity curves that will facilitate comparison with data but to also give us a far more thorough understanding of classical inertial motion outside of the limiting behavior examined in this paper. To implement, it appears that we should use a finite difference method with the component form of our geodesic equation parametrized in terms of the proper time of the object in question within a particular inertial system as expanded upon at the end of Appendix B. Furthermore, we will have to apply this same finite difference method to our normalization condition for the `four-velocity' but parametrized in terms of the proper time in this inertial frame. We also have to attend to a pressing issue with regard to the ``Hubble behavior" associated with wavelength shifts within our inertial system. As outlined earlier, this theory requires that we observe {\it both} significant redshifts and blueshifts, yet on scales larger than the Local Group, blueshifted emitters are reportedly scarce. Thus, if our theory is to be considered seriously, we must provide an explanation for why there is such an imbalance towards reported redshifted emitters at the largest observable scales. Nevertheless, one apparent resolution lies in the possible alternative ``blueshift interpretation" of spectroscopic profiles as mentioned and subsequently applied in \cite{Basu_1998}, \cite{Basu_Copilah}, and \cite{Basu_2011} with possible support for the re-examination of spectroscopic profiles in the blueshifted emission lines found in other work such as \cite{Yaqoob}.

Proceeding to our quantum concerns, our seemingly shocking proposal that at the center of the nucleus of every atom there could potentially exist an inertial center point raises many more questions for our theory of inertial centers. Of course, this type of claim requires thorough and rigorous justification in both future theoretical work and even more importantly in comparison with experiment. For example, a simple comparison with experiment would be to determine how accurate of a fit our ``n-particle amplitudes" (reviewed in Appendix D) with individual solutions for quantum numbers $(\alpha, l, m)$ given by (\ref{eq:wavefcn}) are with current experimental knowledge of the nucleus. Nevertheless, we have chosen to mention these ideas in this paper in order to highlight to the reader how much of a potential impact this redefinition of inertial motion and inertial reference frames could possibly have on our understanding of structure formation for all scales from the largest to the smallest. As for questions: for one, can we reconcile these claims with our current knowledge of the electronic and nuclear structure of the atom when we factor in charge, spin, and electromagnetism? Additionally, how much of our current model for the nucleus is affected by these ideas? It also becomes ever more important to answer the following: What establishes one of these inertial centers as well as the orientation of one of our inertial systems?

\section*{Conclusions}

All of our assumptions within this work in one way or another are built upon the idea that objects do {\it not} move in a straight line at a constant speed when no external forces are acting upon them in empty flat space-time. In other words, we assume that Newton's first law does not give the correct characterization of inertial motion. Therefore, we essentially ``start from scratch" and concentrate on how to incorporate all of the following observed features into a revised understanding of inertial motion: accelerated redshifts and the Hubble relation, plateauing orbital velocity curves at large distances from a central point about which objects move, consistent velocity ``flow" on the largest of scales directed toward a central point, and an orientation associated with each of these central points. We take an inertial frame of reference to be the system within which objects follow these revised inertial trajectories and begin our reformulation with the knowledge that our theory of globally flat space-time must reduce to special relativity within confined regions of our newly defined inertial systems. Consequently, it appears natural to approach this reformulation from the notion that we should have a metric theory of flat space-time, and within this metric theory objects still follow along geodesic trajectories when no external forces are acting upon them as in special and general relativity. However, in order to distinguish our metric theory of flat space-time from special relativity, we must require that our affine parameter {\it not} be proper time globally throughout these reference frames. In addition, we find that we are able to reproduce the previously listed features with the radial Rindler chart as the coordinate parametrization of our flat space-time manifold, thereby assuming the physical significance of special central points which we deem ``inertial center points" situated throughout all of space-time. As one would expect from their given name, these inertial center points describe the centers of each of our inertial systems, and our inertial trajectories are then assumed to be the orbits of objects about these inertial centers. Meaning, inertial motion must be thought of relative to both the center point and the orientation (i.e. location of the poles) of each of these inertial reference frames. Consequently, it is assumed that the observed motion of objects about central points on the largest of scales (e.g. stars orbiting the center of a galaxy, galaxies orbiting the center of a group/cluster, etc.) is {\it not} due to gravitational effects but is instead a manifestation of inertial motion within our theory of flat space-time, which we term our ``Theory of Inertial Centers".

This redefinition of inertial motion then allows us to no longer assume the existence of `dark energy', `dark matter', and `dark flow'. Furthermore, as we have the ability to model the Hubble relation within our theory, we do not require the occurrence of a `Big-Bang' event, and therefore we also do not require `inflation' nor an expanding universe (i.e. we do not operate under the assumptions of $\Lambda$CDM). The cornerstone of our theory is embodied in the statement that within our inertial systems, time and space are fundamentally intertwined such that time- and spatial-translational invariance are {\it not} inherent symmetries of flat space-time. Meaning, our invariant interval associated with the metric incorporates {\it both} time and spatial distance. Therefore, observable clock rates depend upon not only the relative velocity of observers within these inertial systems but also on the difference in distance of each observer from an inertial center, expressed mathematically by relation (\ref{eq:invariant}). Given this relation, we find that our theory of globally flat space-time in fact reduces to special relativity for observers which we can consider as nearly stationary with respect to the inertial center point about which they orbit (i.e. the local stationary limit). As well, our ideas then require that the local speed of light which we measure within a confined region of these newly defined inertial systems is linearly dependent upon our distance away from the inertial center about which we orbit (\ref{eq:speedoflight}). Thus, the speed of light throughout each of these redefined inertial systems in flat space-time is not constant.

With these theoretical foundations presented, we proceeded by examining the local consequences of our theory for a gravitational system located within one of these inertial systems as an observer should be able to measure with a detector of the necessary sensitivity the deviation of an object's (specifically light's) inertial path in flat space-time away from special relativistic geodesics and into the geodesics of our theory as outlined in the local stationary limit. Thus, within the framework of the theory of inertial centers, we interpret the Pioneer anomaly as an observable consequence of our revised ideas on inertial motion. However, as mentioned later in our paper, there are many open questions that must be answered with regard to the propagation of light signals within our solar system in the context of our theory. Specifically, can our revision of inertial motion and inertial reference frames explain the other known astrometric Solar System anomalies (i.e. ``flyby" anomaly, the anomalous increase in the eccentricity of the Moon, and the variation in the AU)? And, can we explain the blueshifted nature of both Pioneer 10 and Pioneer 11 Doppler data once we factor in the two-way nature of these residuals as well as the change in clock rates for observers located at different distances from the center of the Milky Way in our model?

Furthermore, after quantizing for a real massive scalar field, we came upon a potential explanation for the asymmetry between matter and antimatter in our observable universe within the context of our theory of inertial centers. If we allow for the possibility that our field exists in both radial Rindler wedges (i.e. $r>0$ and $r<0$), it appears that a logical explanation for the observable imbalance toward matter would be that our antimatter counterparts are located in the ``other" radial Rindler wedge for each of our inertial systems, as the charge of each field in these systems should be conserved (e.g. abundance of electrons in one wedge should imply an abundance of positrons in the ``other" wedge). Nevertheless, this logic relies on the consistency of our extension for a real scalar field to complex fields with spin. Thus, in future work, we will have to address the validity of this interpretation when we extend our analysis (e.g. Dirac spinors). In addition, we concluded our discussion by examining the nearly stationary limit for particles close to an inertial center point. Using expression (\ref{eq:stationary_force}), we chose to work naively under Newton's assumptions and take this acceleration on our observer to be the result of a Newtonian force derived from a conservative potential. Then, the stationary Hamiltonian associated with this Newtonian approximation would take the form of the isotropic harmonic oscillator. Taking the perspective of an observer exterior to the inertial system in question (i.e. the external observer orbits a {\it different} inertial center), we found the observed oscillator energy scale using relation (\ref{eq:invariant}) while operating under the assumption that the time-scale for each inertial system is a universal constant and therefore the same for each. A simple potential explanation for the ability of the isotropic harmonic oscillator to explain the ``magic numbers" associated with stable arrangements of nucleons within the nucleus of an atom then arose in the context of our model. Since both the form of our stationary Hamiltonian as well as the determined energy scale match that of the starting point for our nuclear shell models, it appears that we must seriously consider the possibility that there exists an inertial center point at the center of the nucleus of every atom when working under the assumptions of our theory of inertial centers as, in our stationary limit, the acceleration of each particle within the inertial system mimics what one would find if he/she naively assumed a Newtonian Hamiltonian of the form of the isotropic harmonic oscillator. In other words, within the context of our theory, the ability of the isotropic harmonic oscillator to model the simplest nuclear configurations would be interpreted as a consequence of the physical existence of an inertial center located at the center of the nucleus of every atom, where these simple configurations of nucleons arise from the stationary limit for objects very near to an inertial center. Although these claims are radical in nature, we are still compelled to question whether or not the nuclear `force' is even really a force within the framework of our model. Future theoretical and experimental work will be required in order to fully understand the nature of these ideas.

% Do NOT remove this, even if you are not including acknowledgments
\section*{Acknowledgments}

I would like to thank several anonymous reviewers as well as both editors for useful comments and critiques which very much helped to improve the clarity of this work.

\appendix
\section {Affine connection terms, Ricci and Riemann tensors} \label{appendixa}

Following \cite{Wald_GR} for our general expressions below, we work in the metric
\begin{displaymath}
d s^2 = -\Lambda r^2 d t^2 + d r^2 + r^2 \cosh^2(\sqrt{\Lambda}t)[d\theta^2 + d\phi^2\sin^2{\theta}]
\end{displaymath}
Our affine connection tensor for this choice of coordinates is given by the expression:
\begin{displaymath}
\Gamma^{c}_{a b} = \frac{1}{2} g^{c d} [\partial_{a} g_{d b} + \partial_{b} g_{d a} - \partial_{d} g_{ab} ]
\end{displaymath}
where in component form $\partial_{\mu} = \partial/\partial x^{\mu}$ is our ordinary partial derivative and in radial Rindler coordinates our metric components gathered in matrix form are given by
\begin{eqnarray}
(g_{\mu \nu}) =
\begin{pmatrix}
-\Lambda r^2 & 0 & 0 & 0
\\ 0 & 1 & 0 & 0
\\ 0 & 0 & r^2 \cosh^2(\sqrt{\Lambda}t) & 0
\\ 0 & 0 & 0& r^2 \cosh^2(\sqrt{\Lambda}t) \sin^2{\theta}
\end{pmatrix} \label{eq:metricmatrix}
\\ (g^{\mu \nu}) =
\begin{pmatrix}
-\frac{1}{\Lambda r^2} & 0 & 0 & 0
\\ 0 & 1 & 0 & 0
\\ 0 & 0 & \frac{1}{r^2 \cosh^2(\sqrt{\Lambda}t)} & 0
\\ 0 & 0 & 0& \frac{1}{r^2 \cosh^2(\sqrt{\Lambda}t) \sin^2{\theta}}
\end{pmatrix}\label{eq:metricinverse}
\end{eqnarray}
with $(g^{\mu \nu})$ corresponding to the inverse of the matrix associated with $(g_{\mu \nu})$ ($g_{tt} = -\Lambda r^2, g_{rr} = 1, g_{\theta \theta} = r^2\cosh^2(\sqrt{\Lambda}t), g_{\phi \phi} = r^2 \cosh^2(\sqrt{\Lambda}t) \sin^2{\theta}$). For the reader who may be unfamiliar with abstract index notation, we look for each affine connection term associated with the tensor above by examining this expression in component form:
\begin{displaymath}
\Gamma^{\lambda}_{\mu \nu} = \frac{1}{2} \sum_{\rho} g^{\lambda \rho} [\partial_{\mu} g_{\rho \nu} + \partial_{\nu} g_{\rho \mu} - \partial_{\rho} g_{\mu \nu}]
\end{displaymath}
where we use greek indices (e.g. $\lambda, \mu, \nu$) for components and latin indices (e.g. $a, b, c$) for the tensor itself. Just as an example, we look for the component $\Gamma^{t}_{t r}$:
\begin{displaymath}
\Gamma^{t}_{t r} = \frac{1}{2} \sum_{\beta} g^{t \beta} [\partial_{t} g_{\beta r} + \partial_{r} g_{\beta t} - \partial_{\beta} g_{tr}]
\end{displaymath}
where we sum over like indices for $\beta = t$, $r$, $\theta$, $\phi$. Then,
\begin{eqnarray}
\Gamma^{t}_{t r} = \frac{1}{2}\bigg\{ g^{t t} [\partial_{t} g_{t r} + \partial_{r} g_{t t} - \partial_{t} g_{tr}] + g^{t r} [\partial_{t} g_{r r} + \partial_{r} g_{r t} - \partial_{r} g_{tr}] + g^{t \theta} [\partial_{t} g_{\theta r} + \partial_{r} g_{\theta t} - \partial_{\theta} g_{tr}] \nonumber
\\ + g^{t \phi} [\partial_{t} g_{\phi r} + \partial_{r} g_{\phi t} - \partial_{\phi} g_{tr}] \bigg\} \nonumber
\end{eqnarray}
However, from (\ref{eq:metricinverse}), we know that $g^{tr} = g^{t\theta} = g^{t \phi} = 0$. Therefore, our expression reduces to
\begin{displaymath}
\Gamma^{t}_{t r} = \frac{1}{2}g^{t t} [\partial_{t} g_{t r} + \partial_{r} g_{t t} - \partial_{t} g_{tr}]
\end{displaymath}
Yet from (\ref{eq:metricmatrix}), we see that $g_{tr} = 0$. This leaves us with
\begin{displaymath}
\Gamma^{t}_{t r} = \frac{1}{2}g^{t t} \partial_{r} g_{t t} = \frac{1}{2}\bigg(- \frac{1}{\Lambda r^2}\bigg) \cdot \partial_{r} (-\Lambda r^2) = \frac{1}{r}
\end{displaymath}
For the reader who wishes to derive the rest of these affine connection components, we notice from (\ref{eq:metricmatrix}) that $g_{\mu \nu} = 0$ for $\mu \neq \nu$ (our metric is diagonal in radial Rindler coordinates). Then the expression for our affine connection terms reduces to
\begin{displaymath}
\Gamma^{\lambda}_{\mu \nu} = \frac{1}{2} g^{\lambda \lambda}[\partial_{\mu} g_{\lambda \nu} + \partial_{\nu} g_{\lambda \mu} - \partial_{\lambda}g_{\mu \nu}]
\end{displaymath}
Yet, because $\Gamma^{c}_{a b}$ is symmetric in $a \Leftrightarrow b$ (i.e. if we swap $a$ and $b$ indices, our tensor remains the same as one can see above since $g_{a b}$ is also symmetric under the same exchange by definition), our possibilities for the affine connection terms are limited to three cases: $\lambda = \nu \neq \mu$; $\lambda = \nu = \mu$; $\mu = \nu \neq \lambda$. For $\lambda = \nu \neq \mu$,
\begin{displaymath}
\Gamma^{\lambda}_{\mu \lambda} = \frac{1}{2}g^{\lambda \lambda}[\partial_{\mu}g_{\lambda \lambda} + \partial_{\lambda}g_{\lambda \mu} - \partial_{\lambda}g_{\mu \lambda}] = \frac{1}{2}g^{\lambda \lambda} \partial_{\mu} g_{\lambda \lambda} \indent (\lambda \neq \mu)
\end{displaymath}
where we used the diagonal property of our metric parametrization in the last equality. Applying similar logic to our other two cases, we obtain
\begin{displaymath}
\Gamma^{\lambda}_{\lambda \lambda} = \frac{1}{2}g^{\lambda \lambda} \partial_{\lambda} g_{\lambda \lambda} \indent {\rm and} \indent \Gamma^{\lambda}_{\mu \mu} = - \frac{1}{2}g^{\lambda \lambda} \partial_{\lambda} g_{\mu \mu} \indent (\lambda \neq \mu)
\end{displaymath}
Using these identities, one finds that our non-zero affine connection terms are:
\begin {eqnarray}
\Gamma^{t}_{tr} = \frac{1}{r} \indent \Gamma^{t}_{\theta \theta} = \frac{1}{\sqrt{\Lambda}}\cosh(\sqrt{\Lambda}t)\sinh(\sqrt{\Lambda}t) \indent \Gamma^{t}_{\phi \phi} = \frac{1}{\sqrt{\Lambda}}\cosh(\sqrt{\Lambda}t)\sinh(\sqrt{\Lambda}t)\sin^2{\theta} \nonumber
\\ \Gamma^{r}_{tt} = \Lambda r \indent \Gamma^{r}_{\theta \theta} = -r\cosh^2(\sqrt{\Lambda}t) \indent \Gamma^{r}_{\phi \phi} = -r\cosh^2(\sqrt{\Lambda}t)\sin^2{\theta} \nonumber
\\ \Gamma^{\theta}_{\theta t} = \Gamma^{\phi}_{\phi t} = \sqrt{\Lambda}\tanh(\sqrt{\Lambda}t) \indent \Gamma^{\theta}_{\theta r} = \Gamma^{\phi}_{\phi r} = \frac{1}{r} \indent \Gamma^{\theta}_{\phi \phi} = -\sin{\theta}\cos{\theta} \indent \Gamma^{\phi}_{\phi \theta} = \cot{\theta} \nonumber
\end{eqnarray}

We define the curvature tensor by the action of the linear map $(\nabla_a \nabla_b - \nabla_b \nabla_a)$ on a dual vector field $\omega_c$ (for more information on vector fields, see Chapter 2 of \cite{Wald_GR}):
\begin{displaymath}
\nabla_a \nabla_b \omega_c - \nabla_b \nabla_a \omega_c = {R_{a b c}}^d \omega_d
\end{displaymath}
where $\nabla_{a}$ is the derivative operator compatible with our metric (or covariant derivative; $\nabla_a g_{bc} = 0$) and we refer to ${R_{a b c}}^d$ as the Riemann curvature tensor. Our Riemann curvature tensor can be expressed in terms of the affine connection associated with a particular choice of coordinate chart:
\begin{displaymath}
{R_{a b c}}^d = \partial_{b} \Gamma^{d}_{a c} - \partial_{a} \Gamma^{d}_{b c} + \Gamma^{e}_{c a} \Gamma^{d}_{b e} - \Gamma^{e}_{c b} \Gamma^{d}_{a e}
\end{displaymath}
And in component form,
\begin{displaymath}
{R_{\mu \nu \rho}}^{\sigma} = \partial_{\nu} \Gamma^{\sigma}_{\mu \rho} - \partial_{\mu} \Gamma^{\sigma}_{\nu \rho} + \sum_{\lambda} \bigg[ \Gamma^{\lambda}_{\rho \mu} \Gamma^{\sigma}_{\nu \lambda} - \Gamma^{\lambda}_{\rho \nu} \Gamma^{\sigma}_{\mu \lambda}\bigg]
\end{displaymath}
Using our affine connection terms, we discover ${R_{\mu \nu \rho}}^{\sigma} = 0$ $\forall \mu, \nu,\rho, \sigma$ as we should expect since the radial Rindler chart is just a coordinate transformation away from the Minkowski chart (the geometric properties of the manifold are independent of coordinate parametrization). Therefore, $R_{\mu \beta} = \sum_{\lambda}{R_{\mu \lambda \beta}}^{\lambda} = 0$ and $\sum_{\lambda, \mu, \nu, \beta} R^{\lambda \mu \nu \beta}R_{\lambda \mu \nu \beta} = 0$. The metric satisfies the Einstein field equations in vacuum without a cosmological constant \cite{Einstein_genrel} and represents flat space-time.

\section{Equations of motion}\label{appendixb}

\begin{equation} \label{eq:geodesic}
0 = U^{a}\nabla_{a}U^{b}
\end{equation}
where $\nabla_{a}$ is the derivative operator compatible with our metric (or covariant derivative; $\nabla_a g_{bc} = 0$). For the reader who may be unfamiliar with concepts in differential geometry, the action of this derivative operator on an arbitrary vector field (a vector field is an assignment of a vector at each point on the manifold) can be expressed in terms of our more familiar partial derivatives through the affine connection tensor associated with a particular coordinate system. When our derivative operator acts on an arbitrary vector field $v^{a}$, we have
\begin{displaymath}
\nabla_{a} v^{b} = \partial_{a} v^{b} + \Gamma^{b}_{a c} v^{c}
\end{displaymath}
and for this same derivative operator acting upon a dual vector field,
\begin{displaymath}
\nabla_{a} v_{b} = \partial_{a} v_{b} - \Gamma^{c}_{a b} v_{c}
\end{displaymath}
Without going into further detail with regard to vector spaces, the reader may feel more informed to know that we can relate a vector with its dual space counterpart through the metric:
\begin{displaymath}
v_{a} = g_{a b} v^{b}
\end{displaymath}
In addition, a vector $v^a$ given at each point on a curve $C$ is said to be parallelly transported as one moves along this curve if
\begin{displaymath}
t^a \nabla_a v^{b} = 0
\end{displaymath}
where $t^a$ refers to the tangent vector to the curve. We then define a geodesic to be a curve whose tangent denoted $U^a$ satisfies (\ref{eq:geodesic}) (for more on parallel transport, see Chapter 3.3 of \cite{Wald_GR}) and assume that our particles travel along these curves when subjected to no net external forces. Additionally, a parametrization of a curve which yields (\ref{eq:geodesic}) is called an affine parametrization, and thus by definition a geodesic is required to be affinely parametrized.

For a geodesic along which one of our particles moves denoted $x^{\mu}(\sigma)$ in our particular coordinate system and parametrized in terms of the affine parameter $\sigma$, our tangent vector to this curve in component form is given by $U^{\mu} = dx^{\mu}/d\sigma$ (where $\sigma = \chi$ for massive particles) and is said to be the `proper velocity' or `four-velocity' of this particle. We also define
\begin{displaymath}
\chi = \int (- g_{ab} T^a T^b)^{1/2} dt
\end{displaymath}
where $T^a$ is the tangent vector to any particular time-like (i.e. $g_{ab}T^a T^b < 0$) curve and $t$ is an arbitrary parametrization of this curve. Thus, along a time-like geodesic affinely parameterized by $\chi$, we have
\begin{displaymath}
g_{ab}U^a U^b = -1
\end{displaymath}

Applying all of the above concepts to expand our equation of motion (\ref{eq:geodesic}) in a particular coordinate system,
\begin{displaymath}
0 = U^{a} \bigg[\partial_{a}U^{b} + \Gamma^{b}_{a c} U^{c}\bigg]
\end{displaymath}
Using our expression for the `proper velocity' in component form, we come upon the geodesic equation of motion for particles in terms of our affine connection terms:
\begin{eqnarray}
0 = \sum_{\alpha} \frac{dx^{\alpha}}{d\sigma} \cdot \frac{\partial}{\partial x^{\alpha}}(\frac{dx^{\nu}}{d\sigma}) + \sum_{\mu, \rho} \Gamma^{\nu}_{\mu \rho} \frac{dx^{\mu}}{d\sigma} \frac{dx^{\rho}}{d\sigma} \nonumber
\\ = \frac{d^2 x^{\nu}}{d\sigma^2} + \sum_{\mu, \rho} \Gamma^{\nu}_{\mu \rho} \frac{dx^{\mu}}{d\sigma} \frac{dx^{\rho}}{d\sigma} \nonumber
\end{eqnarray}
Therefore, for our radial Rindler metric, we can plug in the affine connection terms found in Appendix \ref{appendixa} where in addition $x^{\mu}(\sigma) \rightarrow \langle t(\sigma), r(\sigma), \theta(\sigma), \phi(\sigma)  \rangle$. This notation for our vector components signifies
\begin{displaymath}
x^a = t(\sigma) \bigg(\frac{\partial}{\partial t} \bigg)^a + r(\sigma) \bigg(\frac{\partial}{\partial r} \bigg)^a + \theta(\sigma) \bigg(\frac{\partial}{\partial \theta} \bigg)^a + \phi(\sigma) \bigg(\frac{\partial}{\partial \phi} \bigg)^a
\end{displaymath}
where $(\partial/\partial t)^a$, $(\partial/\partial r)^a$, $(\partial/\partial \theta)^a$, and $(\partial/\partial \phi)^a$ are linearly independent tangent vectors which span the tangent spaces at each point on the manifold. For example, we take our equation of motion for $t(\sigma)$:
\begin{eqnarray}
0 = \frac{d^2 t}{d\sigma^2} + \sum_{\mu, \rho} \Gamma^{t}_{\mu \rho} \frac{dx^{\mu}}{d\sigma} \frac{dx^{\rho}}{d\sigma} \nonumber
\\ = \frac{d^2 t}{d\sigma^2} + \Gamma^{t}_{t r} \frac{dt}{d\sigma} \frac{dr}{d\sigma} + \Gamma^{t}_{r t} \frac{dr}{d\sigma} \frac{dt}{d\sigma} + \Gamma^{t}_{\theta \theta} \frac{d\theta}{d\sigma} \frac{d\theta}{d\sigma} + \Gamma^{t}_{\phi \phi} \frac{d\phi}{d\sigma} \frac{d\phi}{d\sigma} \nonumber
\end{eqnarray}
However, we know from our work in Appendix \ref{appendixa} that $\Gamma^{t}_{t r} = \Gamma^{t}_{r t}$. Consequently, we find after plugging in for each affine connection term
\begin{displaymath}
0 = \frac{d^2 t}{d\sigma^2} + \frac{2}{r} \frac{dt}{d\sigma} \frac{dr}{d\sigma} + \frac{1}{\sqrt{\Lambda}} \cosh(\sqrt{\Lambda}t) \sinh(\sqrt{\Lambda}t) \bigg[ \bigg(\frac{d\theta}{d\sigma}\bigg)^2 + \sin^2{\theta} \bigg(\frac{d\phi}{d\sigma}\bigg)^2 \bigg]
\end{displaymath}
Applying similar logic for $\nu = r$, $\theta$, and $\phi$:
\begin{eqnarray}
0 = \frac{d^2r}{d \sigma^2} + \Lambda r \bigg(\frac{d t}{d \sigma}\bigg)^2 - r\cosh^2(\sqrt{\Lambda}t)\bigg[ \bigg(\frac{d\theta}{d \sigma}\bigg)^2 + \sin^2{\theta}\bigg(\frac{d\phi}{d \sigma}\bigg)^2 \bigg] \nonumber
\\ 0 = \frac{d^2\theta}{d \sigma^2} + 2\frac{d \theta}{d \sigma}\bigg[\sqrt{\Lambda}\tanh(\sqrt{\Lambda}t)\frac{d t}{d \sigma} + \frac{1}{r}\frac{d r}{d \sigma}\bigg] - \sin{\theta}\cos{\theta}\bigg(\frac{d \phi}{d \sigma}\bigg)^2 \nonumber
\\ 0 = \frac{d^2\phi}{d \sigma^2} + 2\frac{d \phi}{d \sigma}\bigg[\sqrt{\Lambda}\tanh(\sqrt{\Lambda}t)\frac{d t}{d \sigma} + \frac{1}{r}\frac{d r}{d \sigma} + \cot{\theta}\frac{d\theta}{d \sigma}\bigg] \nonumber
\end{eqnarray}
As briefly mentioned above, if one evaluates the norm of the `proper velocity', he/she will find:
\begin{displaymath}
U^{a}U_{a} = \sum_{\mu, \nu} g_{\mu \nu} U^{\mu} U^{\nu} = \sum_{\mu, \nu} g_{\mu \nu}\frac{dx^{\mu}}{d\sigma}\frac{dx^{\nu}}{d \sigma} = \left\{\begin{array}{l l} 0 & \quad \textrm{null geodesics} \\ -1 & \quad \textrm{time-like geodesics} \\ \end{array}\right.
\end{displaymath}
In a relatively simple way, one can see this from our line element where $-d\chi^2 = \sum_{\mu, \nu} g_{\mu \nu} dx^{\mu} dx^{\nu}$. Massless particles travel along null geodesics (i.e. our norm vanishes) whereas massive particles travel along time-like geodesics. For comparison with special relativity and general relativity, we express the component form of our time-like geodesics where $\sigma = \chi$ in terms of the physically observable elapsed time as measured by a clock carried along the given curve in a particular inertial system, $\tau$:
\begin{displaymath}
0 = \frac{d^2\tau}{d\chi^2} \frac{dx^{\nu}}{d\tau} + \bigg(\frac{d\tau}{d\chi}\bigg)^2 \cdot \bigg[\frac{d^2x^{\nu}}{d\tau^2} + \sum_{\mu, \rho} \Gamma^{\nu}_{\mu \rho}  \frac{dx^{\mu}}{d\tau} \frac{dx^{\rho}}{d\tau}\bigg]
\end{displaymath}
One immediately notices that the term in brackets represents the component form of the geodesic equation for special and general relativity and would be set equal to zero in both of these theories. However, since in the theory of inertial centers $d^2\tau/{d\chi^2}\neq 0$ as $d\chi/d\tau = \sqrt{\Lambda} \cdot r(\tau)$, the term in brackets is {\it not} necessarily zero for our theory, and thus the observed inertial motion of massive objects in our model characterized by the equation above is in fact very different from inertial motion as seen in special and general relativity.

\section{Killing vector fields}\label{appendixc}

As in our previous appendices, we provide a summary of \cite{Wald_GR} with regard to the more general statements below (see Appendix C and Chapter 2 of \cite{Wald_GR}). In order to understand the relevance of Killing vector fields with respect to inherent symmetries associated with our manifold, we must begin with a brief introduction to isometries and Lie derivatives. For two manifolds $M$ and $N$, let $\phi$ be a smooth map from $M$ to $N$ ($\phi: M \rightarrow N$) and $f$ be a function from $N$ to the reals ($f: N \rightarrow \mathbb{R}$). Then the composition of $f$ with $\phi$, $f\circ \phi$, produces a function from $M \rightarrow \mathbb{R}$ and $\phi$ is said to ``pull back" $f$. In addition $\phi$ ``carries along" tangent vectors at a particular point $p \in M$ to tangent vectors at $\phi(p) \in N$, and therefore defines a map $\phi^{\star}: V_p \rightarrow V_{\phi(p)}$ in the following manner:
\begin{displaymath}
(\phi^{\star} v)(f) = v(f \circ \phi)
\end{displaymath}
where $v \in V_p$, $\phi^{\star} v \in V_{\phi(p)}$, and $V_p$ denotes the tangent vector space at $p$. One can also use $\phi$ to ``pull back" dual vectors at $\phi(p)$ to dual vectors at $p$ by defining a map $\phi_{\star}: V^{\star}_{\phi(p)} \rightarrow V^{\star}_p$ requiring for all $v^a \in V_p$
\begin{displaymath}
(\phi_{\star} \mu)_a v^a = \mu_a (\phi^{\star} v)^a
\end{displaymath}
where $V^{\star}_p$ denotes the dual vector space at $p$. If $\phi: M \rightarrow N$ is a diffeomorphism (i.e. a smooth function that is one-to-one, onto, and its inverse is also smooth), then for an arbitrary tensor $T^{b_1 \ldots b_k} {}_{a_1 \ldots a_l}$ of type $(k,l)$ at $p$ (type $(k,l)$ refers to the number of dual vector ``slots" and vector ``slots", respectively), the tensor $(\phi^{\star} T)^{b_1 \ldots b_k} {}_{a_1 \ldots a_l}$ at $\phi(p)$ is defined by
\begin{displaymath}
(\phi^{\star}T)^{b_1 \ldots b_k}{}_{a_1 \ldots a_l} (\mu_1)_{b_1} \cdots (\mu_k)_{b_k}(t_1)^{a_1} \cdots (t_l)^{a_l} = T^{b_1 \ldots b_k}{}_{a_1 \ldots a_l} (\phi_{\star} \mu_1)_{b_1} \cdots ([\phi^{-1}]^{\star} t_l)^{a_l}
\end{displaymath}
as $(\phi^{-1})^{\star}: V_{\phi(p)} \rightarrow V_p$. If $\phi: M \rightarrow M$ is a diffeomorphism and $T$ is a tensor field on $M$, then we refer to $\phi$ as a symmetry transformation for the tensor field $T$ if $\phi^{\star}T = T$. In addition, if
\begin{displaymath}
(\phi^{\star} g)_{a b} = g_{a b}
\end{displaymath}
we refer to $\phi$ as an isometry.

To introduce the notion of Lie derivatives, we come back to diffeomorphisms and define a one-parameter group of diffeomorphisms $\phi_t$ as a smooth map from $\mathbb{R} \times M \rightarrow M$ such that for fixed $t \in \mathbb{R}$, $\phi_t : M \rightarrow M$ is a diffeomorphism. As well, for all $t,s \in \mathbb{R}$, $\phi_t \circ \phi_s = \phi_{t+s}$. In particular, this requires $\phi_{t=0}$ to be the identity map. A vector field $v^a$ can be thought of as the infinitessimal generator of a one-parameter group of finite transformations of $M$ in the following manner. For fixed $p \in M$, we refer to the curve $\phi_t (p): \mathbb{R} \rightarrow M$ as an orbit of $\phi_t$ which passes through $p$ at $t=0$. $v|_p$ is defined to be the tangent to this curve at $t=0$. We also define the Lie derivative with respect to $v^a$ by
\begin{displaymath}
\mathfrak{L}_{v} T^{b_1 \ldots b_k}{}_{a_1 \ldots a_l} = \lim_{t \rightarrow 0} \bigg \{\frac{\phi^{\star}_{-t}T^{b_1 \ldots b_k}{}_{a_1 \ldots a_l} - T^{b_1 \ldots b_k}{}_{a_1 \ldots a_l} }{t} \bigg \}
\end{displaymath}
where all tensors above are evaluated at a point $p$. $\mathfrak{L}_{v}$ is then a linear map from smooth tensor fields of type $(k,l)$ to smooth tensor fields of type $(k,l)$ and satisfies the Leibniz rule on outer products of tensors. Since $v^a$ is tangent to the integral curves of $\phi_t$, for functions $f: M \rightarrow \mathbb{R}$
\begin{displaymath}
\mathfrak{L}_v (f) = v(f)
\end{displaymath}
In addition, if $\phi_t$ is a symmetry transformation for $T$, we have $\mathfrak{L}_v T^{b_1 \ldots b_k}{}_{a_1 \ldots a_l} = 0$. Furthermore, it is found that the Lie derivative with respect to $v^a$ of a vector field $w^a$ is given by the commutator:
\begin{displaymath}
\mathfrak{L}_v w^a = [v,w]^a
\end{displaymath}
where
\begin{displaymath}
[v,w]^a = v^b \nabla_b w^a - w^b \nabla_b v^a
\end{displaymath}
and for a dual vector,
\begin{displaymath}
\mathfrak{L}_v \mu_a = v^b \nabla_b \mu_a + \mu_b \nabla_a v^b
\end{displaymath}
The more general action of a Lie derivative with respect to $v^a$ on a general tensor field $T^{b_1 \ldots b_k}{}_{a_1 \ldots a_l}$ is given by
\begin{displaymath}
\mathfrak{L}_v T^{b_1 \ldots b_k}{}_{a_1 \ldots a_l} = v^c \nabla_c T^{b_1 \ldots b_k}{}_{a_1 \ldots a_l} - \sum_{i=1}^{k} T^{b_1 \ldots c \ldots b_k}{}_{a_1 \ldots a_l}\nabla_c v^{b_i} + \sum_{j=1}^{l} T^{b_1 \ldots b_k}{}_{a_1 \ldots c \ldots a_l} \nabla_{a_j}v^c
\end{displaymath}
where $\nabla_a$ is our derivative operator compatible with the metric $g_{ab}$ (i.e. $\nabla_c g_{ab} =0$). Then a Killing vector field $\xi^a$ is defined to be the vector field which generates a one-parameter group of isometries $\phi_t: M \rightarrow M$ of the metric, $(\phi^{\star}_t g)_{a b} = g_{a b}$. As remarked earlier, the necessary condition for $\phi_t$ to be a group of isometries is $\mathfrak{L}_{\xi} g_{a b} = 0$. Using the expression above for the action of a Lie derivative on a tensor field,
\begin{eqnarray}
\mathfrak{L}_{\xi} g_{a b} = \xi^c \nabla_c g_{ab} + g_{cb}\nabla_a \xi^c + g_{ac} \nabla_b \xi^c \nonumber
\\ = \nabla_a \xi_b + \nabla_b \xi_a \nonumber
\end{eqnarray}
Thus, we come upon Killing's equation:
\begin{displaymath}
\nabla_{a}\xi_{b} + \nabla_{a}\xi_{b} = 0
\end{displaymath}
For any particular Killing vector field $\xi^a$, along a geodesic $\gamma$ with tangent vector $U^a$ one finds
\begin{eqnarray}
U^b \nabla_b (\xi_a U^a) = U^b U^a \nabla_b \xi_a + \xi^a U^b \nabla_b U^a \nonumber
\\ = \frac{1}{2}U^a U^b [\nabla_b \xi_a + \nabla_a \xi_b] + \xi^a U^b \nabla_b U^a = 0 \nonumber
\end{eqnarray}
where the first term vanishes by Killing's equation and the second by the geodesic equation (\ref{eq:geodesic}). Meaning, along $\gamma$, $\xi^a U_a$ is constant (Noether's theorem).

Using our affine connection component terms, Killing's equation takes the form
\begin{displaymath}
\partial_{\mu} \xi_{\nu} + \partial_{\nu} \xi_{\mu} = 2\sum_{\rho} \Gamma^{\rho}_{\mu \nu} \xi_{\rho}
\end{displaymath}
which gives for each pair ($\mu, \nu$),
\begin{eqnarray}
(t,t): \partial_{t}\xi_{t} = \Lambda r \xi_{r} \indent (t,r): \partial_{t}\xi_{r} + \partial_{r}\xi_{t} = \frac{2}{r} \xi_{t} \indent (t,\theta): \partial_{t}\xi_{\theta} + \partial_{\theta}\xi_{t} = 2\sqrt{\Lambda}\tanh(\sqrt{\Lambda}t) \xi_{\theta} \nonumber
\\ (t,\phi): \partial_{t}\xi_{\phi} + \partial_{\phi}\xi_{t} = 2\sqrt{\Lambda}\tanh(\sqrt{\Lambda}t) \xi_{\phi} \indent (r,r): \partial_{r}\xi_{r} = 0 \indent (r, \theta): \partial_{\theta}\xi_{r} + \partial_{r}\xi_{\theta} = \frac{2}{r} \xi_{\theta} \nonumber
\\ (r, \phi): \partial_{\phi}\xi_{r} + \partial_{r}\xi_{\phi} = \frac{2}{r}\xi_{\phi} \indent (\theta, \phi): \partial_{\theta}\xi_{\phi} + \partial_{\phi}\xi_{\theta} = 2\cot{\theta}\xi_{\phi} \nonumber
\\ (\theta, \theta): \partial_{\theta}\xi_{\theta} = \frac{1}{\sqrt{\Lambda}}\cosh(\sqrt{\Lambda}t)\sinh(\sqrt{\Lambda}t)\xi_{t} - r\cosh^2(\sqrt{\Lambda}t)\xi_{r} \nonumber
\\ (\phi, \phi): \partial_{\phi}\xi_{\phi} = \sin^2{\theta}\bigg[\frac{1}{\sqrt{\Lambda}}\cosh(\sqrt{\Lambda}t)\sinh(\sqrt{\Lambda}t)\xi_{t} - r\cosh^2(\sqrt{\Lambda}t)\xi_{r}\bigg]  -\sin{\theta}\cos{\theta}\xi_{\theta} \nonumber
\end{eqnarray}
Immediately we notice from the $(t,t)$ equation that $\xi_{t}=0 \Longrightarrow \xi_{r} = 0$ and from the $(\theta, \theta)$ equation that $\xi_{r} = \xi_{\theta} = 0 \Longrightarrow \xi_{t} = 0$. For $\xi_{t} = \xi_{r} = 0$, we find the three rotational Killing vector fields:
\begin{eqnarray}
\Omega^{\mu}_{1} \rightarrow \langle 0,0,\cos{\phi},-\cot{\theta}\sin{\phi} \rangle \nonumber
\\ \Omega^{\mu}_{2} \rightarrow \langle 0,0,\sin{\phi},\cot{\theta}\cos{\phi} \rangle \nonumber
\\ \psi^{\mu} \rightarrow \langle 0,0,0,1 \rangle \nonumber
\end{eqnarray}
For $\xi_{\theta}=\xi_{\phi} = 0$, we have a time and radial Killing vector field:
\begin{displaymath}
\rho^{\mu} \rightarrow \langle \frac{1}{\sqrt{\Lambda}r}\cosh(\sqrt{\Lambda}t), -\sinh(\sqrt{\Lambda}t), 0, 0 \rangle
\end{displaymath}
For $\xi_{r} = \xi_{\phi} = 0$, a time and $\theta$ Killing vector field:
\begin{displaymath}
\Theta^{\mu} \rightarrow \langle \frac{1}{\sqrt{\Lambda}}\cos{\theta}, 0, -\sin{\theta}\tanh(\sqrt{\Lambda}t),  0 \rangle
\end{displaymath}
For $\xi_{\phi}=0$, a time, radial, and $\theta$ Killing vector field:
\begin{displaymath}
 \xi^{\mu}_{(t,r,\theta)} = \langle -\frac{\sinh(\sqrt{\Lambda}t)\cos{\theta}}{\sqrt{\Lambda}r} , \cosh(\sqrt{\Lambda}t)\cos{\theta}, -\frac{\sin{\theta}}{r\cosh(\sqrt{\Lambda}t)}, 0 \rangle
\end{displaymath}
In addition, for only $\xi_r = 0$, we find two Killing vector fields:
\begin{eqnarray}
\xi^{\mu}_{(t,\theta, \phi), 1} \rightarrow \langle \frac{1}{\Lambda}\sin{\theta}\sin{\phi}, 0, \frac{1}{\sqrt{\Lambda}}\tanh(\sqrt{\Lambda}t)\cos{\theta}\sin{\phi}, \frac{1}{\sqrt{\Lambda}\sin{\theta}}\tanh(\sqrt{\Lambda}t)\cos{\phi} \rangle \nonumber
\\ \xi^{\mu}_{(t,\theta, \phi), 2} \rightarrow  \langle -\frac{1}{\Lambda}\sin{\theta}\cos{\phi}, 0, -\frac{1}{\sqrt{\Lambda}}\tanh(\sqrt{\Lambda}t)\cos{\theta}\cos{\phi}, \frac{1}{\sqrt{\Lambda}\sin{\theta}}\tanh(\sqrt{\Lambda}t)\sin{\phi} \rangle \nonumber
\end{eqnarray}
Finally, taking all components to be non-zero, we have the last two Killing vector fields:
\begin{eqnarray}
\xi^{\mu}_{(t,r,\theta, \phi), 1} \rightarrow \langle -\frac{1}{\sqrt{\Lambda}r}\sinh(\sqrt{\Lambda}t)\sin{\theta}\sin{\phi}, \cosh(\sqrt{\Lambda}t)\sin{\theta}\sin{\phi}, \frac{1}{r\cosh(\sqrt{\Lambda}t)}\cos{\theta}\sin{\phi},
 \nonumber \\ \frac{1}{r\cosh(\sqrt{\Lambda}t)\sin{\theta}}\cos{\phi} \rangle \nonumber
\\ \xi^{\mu}_{(t,r,\theta, \phi), 2} \rightarrow \langle -\frac{1}{\sqrt{\Lambda}r}\sinh(\sqrt{\Lambda}t)\sin{\theta}\cos{\phi},
\cosh(\sqrt{\Lambda}t)\sin{\theta}\cos{\phi}, \frac{1}{r\cosh(\sqrt{\Lambda}t)}\cos{\theta}\cos{\phi},
\nonumber \\ -\frac{1}{r\cosh(\sqrt{\Lambda}t)\sin{\theta}}\sin{\phi} \rangle \nonumber
\end{eqnarray}
Summarizing, we have ten linearly independent Killing vector fields for this metric.

\section{Symplectic structure} \label{appendixd}

Within this appendix, we'll briefly address and apply the concepts presented in \cite{Wald_QFT} for the formulation of a quantum field theory of a real scalar field with a general background metric, where we shall not concern ourselves with the interaction between matter and space-time at the quantum level and instead treat the metric as non-dynamic (hence the term ``background"). However, we strongly encourage the reader to review \cite{Wald_QFT} in order to fully understand all the material presented below.

The information associated with the dynamical evolution of a physical system can be conveyed within the symplectic structure $\Omega$ for a particular action $S$ given by
\begin{displaymath}
\Omega ( [\phi_1, \pi_1], [\phi_2, \pi_2] ) = \int_{\Sigma_0} d^3x \bigg(\pi_{1} \phi_{2} - \pi_{2} \phi_{1} \bigg)
\end{displaymath}
where $\Omega$ is a non-degenerate antisymmetric bilinear map from the solutions of the equation of motion associated with our action to the real numbers. In addition, a point in phase-space (Hamiltonian formalism) corresponds to the specification of our field solution $\phi$ and its conjugate momentum $\pi = \partial \mathcal{L}/\partial \dot{\phi}$ on a space-like hypersurface $\Sigma_0$ associated with our ``initial value" configuration ($\mathcal{L}$ is the Lagrangian density associated with the action $S$ which we'll give below). The fundamental Poisson brackets in classical theory can then be expressed as
\begin{displaymath}
\{\Omega([\phi_1, \pi_1], \cdot), \Omega([\phi_2, \pi_2], \cdot)\} = -\Omega([\phi_1, \pi_1],[\phi_2, \pi_2])
\end{displaymath}
where $\Omega(y, \cdot)$ is a linear function assuming our choice of $y$ does not vary (our input argument is only `$\cdot$'). If we arbitrarily choose $[\phi_1, \pi_1] = [0, f_1]$ and $[\phi_2, \pi_2] = [f_2, 0]$, our classical Poisson brackets reduce to
\begin{displaymath}
\bigg \{\int d^3x f_1 (x) \phi(x), \int d^3y f_2(y) \pi(y) \bigg \} = \int d^3x f_1(x) f_2(x)
\end{displaymath}
which we can think of as the more familiar canonical relations
\begin{displaymath}
\{\phi(x), \pi(y) \} = \delta(x - y)
\end{displaymath}

Then to construct our quantum theory of a scalar field, we extend the functions $\Omega([\phi, \pi], \cdot)$ to operators $\hat{\Omega}([\phi, \pi], \cdot)$ satisfying the commutation relations
\begin{displaymath}
[\hat{\Omega}([\phi_1, \pi_1], \cdot), \hat{\Omega}([\phi_2, \pi_2], \cdot) ] = -i \hbar \Omega([\phi_1, \pi_1], [\phi_2, \pi_2]) \hat{\mathrm{I}}
\end{displaymath}
($\hat{\mathrm{I}}$ denotes the identity operator) and introduce the inner product associated with this system
\begin{displaymath}
(\psi^+, \chi^+) = -i \Omega(\bar{\psi}^+, \chi^+)
\end{displaymath}
where $\bar{\psi}^{+}$ represents the complex conjugate of $\psi^{+}$ and we have decomposed our full solutions $\psi, \chi \in \mathcal{S}$ of the equation of motion for our action into
\begin{displaymath}
\psi = \psi^{+} + \psi^{-}
\end{displaymath}
such that our inner product with respect to these ``positive frequency" solutions $\psi^{+}, \chi^{+}$ is positive-definite. We denote the solution space spanned by these ``positive frequency" parts as $\mathcal{S}^{\mathbb{C}+}$. In addition, we have expressed our inner product only in terms of solutions to our equation of motion $\psi \in \mathcal{S}$ as for each solution there corresponds a point in phase space $[\psi, \pi_{\psi}]$. One proceeds to ``Cauchy-complete" in the norm defined by this inner product to obtain our complex Hilbert space $\mathcal{H}$ (see Chapter 3.2 and Appendix A.1 of \cite{Wald_QFT}). Thus, we represent our classical observables $\Omega(\psi, \cdot)$ for each solution $\psi$ by the operator
\begin{displaymath}
\hat{\Omega} (\psi, \cdot) = i \hat{a}(\bar{\mathrm{K} \psi}) - i \hat{a}^{\dagger} (\mathrm{K} \psi) 
\end{displaymath}
where $\mathrm{K}: \mathcal{S} \rightarrow \mathcal{H}$ is a map from the full solutions to our complex Hilbert space. As well, $\hat{a}(\cdot)$ and $\hat{a}^{\dagger}(\cdot)$ denote the annihilation and creation operators, respectively, which act on a general state $\Psi$ in the symmetric Fock space $\mathcal{F}_s(\mathcal{H})$ in the following manner. For a general state $\Psi = \langle\psi, \psi^{a_1}, \psi^{a_1 a_2}, \ldots, \psi^{a_1 \ldots a_n}, \dots\rangle$ representing our ``n-particle amplitudes" where for scalar theory $\psi^{a_1\ldots a_n} = \psi^{(a_1 \ldots a_n)}$ $\forall n$ (round parantheses denote symmetrization when dealing with abstract indices here and below) and $\xi^a \in \mathcal{H}$, $\bar{\xi}_a \in \bar{\mathcal{H}}$, we have
\begin{eqnarray}
\hat{a}(\bar{\xi}) \Psi = \langle \bar{\xi}_a \psi^a, \sqrt{2} \bar{\xi}_a \psi^{a a_1}, \sqrt{3}  \bar{\xi}_a \psi^{a a_1, a_2}, \ldots \rangle \nonumber
\\ \hat{a}^{\dagger}(\xi) \Psi = \langle 0, \psi \xi^{a_1}, \sqrt{2} \xi^{(a_1} \psi^{a_2)}, \sqrt{3} \xi^{(a_1} \psi^{a_2 a_3)}, \ldots \rangle \nonumber
\end{eqnarray}
Indices on $\bar{\xi}_a$ and $\xi^a$ are dropped in our expressions on the left-hand side of these equations for notational convenience. In this paper, we operate under the assumption that the norms of these two expressions are finite. In addition, the inner product of two vectors $\xi, \eta \in \mathcal{H}$ is denoted by
\begin{displaymath}
(\xi, \eta) = \bar{\xi}_a \eta^a
\end{displaymath}
In this notation, $\psi \in \otimes^n \mathcal{H}$ is denoted $\psi^{a_1 \ldots a_n}$ and $\bar{\psi} \in \otimes^n \bar{\mathcal{H}}$ as $\bar{\psi}_{a_1 \ldots a_n}$ where $\otimes^n \mathcal{H} = \mathcal{H}_1 \otimes \ldots \otimes \mathcal{H}_n$ for $\mathcal{H}_1 = \ldots = \mathcal{H}_n = \mathcal{H}$ (n-fold tensor product space). As well, $\mathcal{F}_s(\mathcal{H}) = \oplus_{n=0}^{\infty} (\otimes^{n}_s \mathcal{H})$ where $\otimes^{n}_s \mathcal{H}$ is the symmetric n-fold tensor product space and $\otimes^0 \mathcal{H}$ is defined to be the complex numbers $\mathbb{C}$ (see Appendix A of \cite{Wald_QFT}).

To clarify further with regard to Fock space notation, we relate back to Dirac ``bra-ket" notation:
\begin{displaymath}
|0\rangle_{\hat{a}} \equiv \langle 1,0,0,\ldots \rangle
\end{displaymath}
where
\begin{displaymath}
\hat{a}(\bar{\xi}) |0\rangle_{\hat{a}} = 0
\end{displaymath}
for some general $\xi^a$ given $|0\rangle_{\hat{a}}$ denotes the vacuum state associated with the creation and annihilation operators on our Fock space (i.e. the $\hat{a}$'s). Given our general state $\Psi$, the probability of finding only a single `$\hat{a}$ particle' in state $\beta \in \mathcal{H}$ is taken to be $|\bar{\beta}_a \psi^a|^2$ (i.e. $\psi^a$ is the ``one-particle amplitude", $\psi^{a_1 a_2}$ is the ``two-particle amplitude", etc.). Here, the term `particle' really refers to an excitation of the particular field associated with our $\hat{a}, \hat{a}^{\dagger}$ operators (quanta). Therefore, one can think of $\hat{a}(\bar{\xi})$ as an operator annihilating a quantum of state $\xi^a$ from each of the ``n-particle" states in the general state $\Psi$, and analogously $\hat{a}^{\dagger}(\xi)$ as creating a quantum in each. Our annihilation and creation operators also satisfy the commutation relation:
\begin{displaymath}
[\hat{a}(\bar{\xi}), \hat{a}^{\dagger}(\eta))] = \bar{\xi}_a \eta^{a} \hat{\mathrm{I}}
\end{displaymath}
Note that our use of abstract index notation in this paragraph does not refer to the metric. In other words, when working with Hilbert space vectors, we always assume contraction occurs over the inner product of the respective Hilbert space as defined earlier in this appendix and not with regard to the metric.

Then for our theory of inertial centers, the action associated with the equation of motion for our Klein-Gordon extension in a particular inertial system (\ref{eq:KG extend}) takes the form
\begin{equation} \label{eq:KG extend action}
S = -\frac{1}{2}\int d^4 x \sqrt{|g|} \bigg(\nabla^a \phi \nabla_a \phi + {\tilde{\mu}}^2 r^2 \phi^2 \bigg)
\end{equation}
where one can verify this by extremizing the action ($\delta S = 0$) to obtain our equation of motion. We proceed with our formulation by ``slicing" our manifold $M$ into space-like hypersurfaces each indexed by a time parameter $t$ ($\Sigma_t$). Then, we introduce a vector field on $M$ associated with our time evolution and defined by $t^a \nabla_a t = 1$, which we can decompose in the following manner:
\begin{displaymath}
t^a = N n^a + N^a
\end{displaymath}
(in contrast with our previous paragraph, abstract index notation here employs the metric). $n^a$ is the future-directed unit normal vector field to our space-like hypersurfaces $\Sigma_t$ (future-directed in the sense that $n^a$ lies in the same direction as $t^a$), and $N^a$ represents the remaining tangential portion of $t^a$ to $\Sigma_t$. In addition, we introduce coordinates $t, x^1, x^2, x^3$ such that $t^a \nabla_a x^i = 0$ for $i = 1,2,3$ which allows $t^a = (\partial/\partial t)^a$. Our action in (\ref{eq:KG extend action}) can then be rewritten in terms of the integral of a Lagrangian density $\mathcal{L}$ over our time parameter $t$ and our space-like hypersurface $\Sigma_t$:
\begin{displaymath}
S = \int dt \int_{\Sigma_t} d^3x \mathcal{L}
\end{displaymath}
with
\begin{displaymath}
\mathcal{L} = \frac{1}{2} N \sqrt{|h|} \bigg( (n^a \nabla_a \phi)^2 - h^{ab} \nabla_a \phi \nabla_b \phi - {\tilde{\mu}}^2 r^2 \phi^2 \bigg)
\end{displaymath}
where $h_{ab}$ is the induced Riemannian metric on $\Sigma_t$ and $h = \mathrm{det}(h_{\beta \nu})$. Yet, since
\begin{displaymath}
n^a \nabla_a \phi = \frac{1}{N} (t^a - N^a)\nabla_a \phi = \frac{1}{N} \dot{\phi} - \frac{1}{N} N^a \nabla_a \phi
\end{displaymath}
where $\dot{\phi} = t^a \nabla_a \phi$, we find that our conjugate momentum density on $\Sigma_t$ takes the form
\begin{displaymath}
\pi = \frac{\partial \mathcal{L}}{\partial \dot{\phi}} = (n^a \nabla_a \phi)\sqrt{|h|}
\end{displaymath}
as it does with our original Klein-Gordon action. Consequently, our symplectic structure for a free scalar field in the theory of inertial centers is given by
\begin{eqnarray}
\Omega([\phi_1, \pi_1],[\phi_2, \pi_2]) = \int_{\Sigma_0} d^3x (\pi_1 \phi_2 - \pi_2 \phi_1) \nonumber
\\ = \int_{\Sigma_0} d^3x \sqrt{|h|} [\phi_2 n^a \nabla_a \phi_1 - \phi_1 n^a \nabla_a \phi_2] \nonumber
\end{eqnarray}
where $\Sigma_0$ is the space-like hypersurface associated with our ``initial value" configuration at $t=0$.

\section{Divergence of a vector field} \label{appendixe}

Following Chapter 3.4 of \cite{Wald_GR}, the divergence of a vector field $v^a$ is given by
\begin{displaymath}
\nabla_a v^a = \partial_a v^a + \Gamma^{a}_{a b}v^b
\end{displaymath}
where we have used our knowledge from Appendix \ref{appendixb} to expand this expression. However, in component form
\begin{displaymath}
\Gamma^a_{a \nu} = \sum_{\mu} \Gamma^{\mu}_{\mu \nu} = \frac{1}{2} \sum_{\mu, \rho} g^{\mu \rho} [\partial_{\mu} g_{\rho \nu} + \partial_{\nu} g_{\rho \mu} - \partial_{\rho} g_{\mu \nu} ]
\end{displaymath}
Yet the first and last of these terms cancel as we are summing over both $\mu$ and $\rho$ and $g^{ab} = g^{ba}$. This leaves us with
\begin{displaymath}
\Gamma^{a}_{a \nu} = \frac{1}{2} \sum_{\mu, \rho} g^{\mu \rho} \partial_{\nu} g_{\mu \rho}
\end{displaymath}
but if we think in terms of the matrix form of our components, $(g_{\mu \nu})$, we have
\begin{displaymath}
\sum_{\mu, \rho} g^{\mu \rho} \partial_{\nu} g_{\mu \rho} = \frac{\partial_{\nu} g}{g}
\end{displaymath}
where $g = \mathrm{det}(g_{\mu \nu})$. Therefore,
\begin{displaymath}
\Gamma^{a}_{a \nu} = \frac{1}{2} \frac{\partial_{\nu} g}{g} = \partial_{\nu} \ln{\sqrt{|g|}}
\end{displaymath}
Plugging in above for our divergence term,
\begin{displaymath}
\nabla_a v^a = \sum_{\mu} \bigg[\partial_{\mu} v^{\mu} + v^{\mu} \partial_{\mu} \ln{\sqrt{|g|}} \bigg] = \sum_{\mu} \frac{1}{\sqrt{|g|}} \partial_{\mu} (\sqrt{|g|} v^{\mu})
\end{displaymath}
Then for a scalar field $f$,
\begin{displaymath}
\nabla_a \nabla^a f = \sum_{\mu} \frac{1}{\sqrt{|g|}} \partial_{\mu} (\sqrt{|g|} g^{\mu \nu} \partial_{\nu}f)
\end{displaymath}
as $\nabla_a f = \partial_a f$.

\section{Scalar field solutions} \label{appendixf}

We look for solutions ($\phi_i$) to our extension of the Klein-Gordon equation:
\begin{displaymath}
(\nabla_a \nabla^a - \tilde{\mu}^2 r^2) \phi_i = 0
\end{displaymath}
As shown in Appendix \ref{appendixe}, for a real scalar field
\begin{displaymath}
\nabla_{a} \nabla^{a} \phi_i = \sum_{\nu, \beta} \frac{1}{\sqrt{|g|}}\partial_{\nu}(\sqrt{|g|} g^{\nu \beta}\partial_{\beta}\phi_i)
\end{displaymath}
where for our purposes $g_{\nu \beta}$ refers to the radial Rindler metric components and $\sqrt{|g|} = \sqrt{\Lambda} r^3 \cosh^2(\sqrt{\Lambda} t) \sin{\theta}$. Expanding (\ref{eq:KG extend}),
\begin{eqnarray}
0 = -\tilde{\mu}^2 r^2 \phi_i - \frac{1}{\Lambda r^2 \cosh^2(\sqrt{\Lambda}t)} \partial_t (\cosh^2(\sqrt{\Lambda}t)\partial_t \phi_i) + \frac{1}{r^3}\partial_r(r^3 \partial_r \phi_i)
\nonumber \\ + \frac{1}{r^2 \cosh^2(\sqrt{\Lambda}t)} \bigg[\frac{1}{\sin{\theta}}\partial_{\theta}(\sin{\theta} \partial_{\theta}\phi_i) + \frac{1}{\sin^2{\theta}}\partial^2_{\phi}\phi_i \bigg]
\end{eqnarray}
We look for separable solutions of the form, $\phi_i = Z_i \cdot g(t) \cdot h(r) \cdot Y_l^m(\theta, \phi)$, where $Z_i$ is a normalization constant and the $Y_l^m$ are spherical harmonics satisfying
\begin{equation}
\frac{1}{\sin{\theta}}\partial_{\theta}(\sin{\theta} \partial_{\theta}Y_l^m) + \frac{1}{\sin^2{\theta}}\partial^2_{\phi}Y_l^m = -l(l+1) Y_l^m
\end{equation}
where
\begin{equation}
Y_l^m (\theta, \phi) = \sqrt{\frac{(2l+1)}{4\pi}\frac{(l-m)!}{(l+m)!}} P_l^m(\cos{\theta}) \cdot e^{im \phi}
\end{equation}
with $l$ as a non-negative integer, $|m| \leq l$, and $m$ also as an integer (see Chapter 3.6 of \cite{Sakurai} or Chapter 15.5 of \cite{Arfken}). $P_l^m$ is an associated Legendre function which satisfies the differential equation
\begin{displaymath}
\bigg[(1-x^2)\partial^2_x - 2x\partial_x + \bigg(l[l+1] - \frac{m^2}{1-x^2} \bigg)\bigg] P^m_l(x) = 0
\end{displaymath}
and can be expressed in terms of Rodrigues' formula \cite{Abramowitz}:
\begin{displaymath}
P_l^m(x) = \frac{(-1)^m}{2^l l!} (1-x^2)^{m/2} \frac{d^{l+m}}{dx^{l+m}} (x^ 2 - 1)^l
\end{displaymath}
with
\begin{displaymath}
P_l^{-m} = (-1)^m \frac{(l-m)!}{(l+m)!}P_l^m
\end{displaymath}
The spherical harmonics $Y_l^m$ as expressed above obey the orthogonality relation:
\begin{displaymath}
\int_{0}^{\pi} \sin{\theta} d\theta \int_{0}^{2 \pi} d\phi Y_l^m \bar{Y}_{l'}^{m'} = \delta_{l l'} \delta_{m m'}
\end{displaymath}
Plugging in and dividing by $\phi_i$,
\begin{displaymath}
0 = -\tilde{\mu}^2 r^2 - \frac{1}{g \Lambda r^2 \cosh^2(\sqrt{\Lambda}t)} \partial_t (\cosh^2(\sqrt{\Lambda}t)\partial_t g) + \frac{1}{h r^3 }\partial_r(r^3 \partial_r h) - \frac{l(l+1)}{r^2 \cosh^2(\sqrt{\Lambda}t)}
\end{displaymath}
Multiplying through by $r^2$ and grouping functions of $t$ and $r$:
\begin{displaymath}
\frac{1}{h r}\partial_r(r^3 \partial_r h) - \tilde{\mu}^2 r^4 = \frac{1}{g \Lambda \cosh^2(\sqrt{\Lambda}t)}\partial_t(\cosh^2(\sqrt{\Lambda}t)\partial_t g) + \frac{l(l+1)}{\cosh^2(\sqrt{\Lambda}t)} = - (4 \alpha^2 + 1)
\end{displaymath}
where $\alpha$ is a constant. Thus, we have two differential equations:
\begin{eqnarray}
0 = r^2 \partial^2_r h + 3r\partial_r h + \bigg[(4 \alpha^2 + 1) - \tilde{\mu}^2 r^4 \bigg] h
\\ 0 = \frac{1}{\Lambda \cosh^2(\sqrt{\Lambda}t)}\partial_t(\cosh^2(\sqrt{\Lambda}t)\partial_t g) + \bigg[\frac{l(l+1)}{\cosh^2(\sqrt{\Lambda}t)} + (4 \alpha^2 + 1)\bigg] g
\end{eqnarray}
Focusing on our radial equation first, we set $\rho = \sqrt{\tilde{\mu}} r$:
\begin{equation}
0 = \rho^2 \partial^2_{\rho} h + 3\rho \partial_{\rho} h + \bigg[(4 \alpha^2 + 1)- \rho^4 \bigg]h
\end{equation}
Letting $h(\rho) = z(\rho)/\rho$
\begin{displaymath}
0 = \rho^2 \partial^2_{\rho}z + \rho \partial_{\rho} z + \bigg[ 4 \alpha^2 -\rho^4 \bigg] z
\end{displaymath}
and setting $y = \rho^2/2$, we find
\begin{displaymath}
0 = y^2 \partial^2_y z + y \partial_y z + [\alpha^2 - y^2]z
\end{displaymath}
But this is just the modified Bessel equation of pure imaginary order (see Chapter 3 of \cite{Watson}). Choosing the physically realistic solution (we expect $h$ to decay for large $r$ since in our classical analysis massive objects are ``confined" to motion about their inertial centers), our full expression takes the form
\begin{equation} \label{eq:h_expr}
h_{\alpha}(\rho) = \frac{K_{i \alpha}(\frac{\rho^2}{2})}{\rho}
\end{equation}
where $K_{i \alpha}$ is the Macdonald function (modified Bessel function) of imaginary order $\alpha$ given in integral form (for $y>0$):
\begin{displaymath}
K_{i \alpha}(y) = \int_{0}^{\infty} d\eta \cos(\alpha \eta) e^{- y \cosh{\eta}}
\end{displaymath}
and $\alpha$ is restricted to the range: $0 \leq \alpha < \infty$ (see Chapter 4.15 of \cite{Titchmarsh} and \cite{Fulling} for its application to quantization in the classic Rindler case with the Klein-Gordon equation).

The Macdonald function of imaginary order obeys an orthogonality relation which will be useful to us for determining part of our normalization constant. From \cite{Passian},
\begin{displaymath}
\int_{0}^{\infty} dy \frac{K_{i\alpha}(y) K_{i\alpha '}(y)}{y} = \frac{\pi^2} {2 \alpha \sinh(\pi \alpha)} \delta(\alpha - \alpha')
\end{displaymath}
Later we'll need:
\begin{eqnarray}
\int_{0}^{\infty} d\rho \frac{K_{i\alpha}(\frac{\rho^2}{2}) K_{i\alpha '}(\frac{\rho^2}{2})}{\rho} = \int_{0}^{\infty} d\eta \int_{0}^{\infty} d\eta' \cos(\alpha \eta) \cos(\alpha' \eta') \int_{0}^{\infty} \frac{d\rho}{\rho} e^{-(\rho^2/2)(\cosh{\eta} + \cosh{\eta'})}
\nonumber \\ = \int_{0}^{\infty} d\eta \int_{0}^{\infty} d\eta' \cos(\alpha \eta) \cos(\alpha' \eta') \int_{0}^{\infty} \frac{dy}{2y} e^{-y(\cosh{\eta} + \cosh{\eta'})}
\nonumber \\ = \frac{1}{2} \int_{0}^{\infty} dy \frac{K_{i\alpha}(y) K_{i\alpha '}(y)}{y}
= \frac{\pi^2} {4 \alpha \sinh(\pi \alpha)} \delta(\alpha - \alpha') \nonumber
\end{eqnarray}

Examining our second differential equation, we let $\eta = \tanh(\sqrt{\Lambda}t)$:
\begin{equation}\label{eq:time_behavior}
0 = (1-\eta^2)^2 \partial^2_{\eta} g + [l(l+1)(1-\eta^2) + (4\alpha^2 + 1)]g
\end{equation}
where $\eta^2 < 1$. Before proceeding any further, we must make one remark that will be crucial for evaluating our inner product. Taking the complex conjugate of (\ref{eq:time_behavior})
\begin{displaymath}
0 = (1-\eta^2)^2 \partial^2_{\eta} \bar{g} + [l(l+1)(1-\eta^2) + (4\alpha^2 + 1)]\bar{g}
\end{displaymath}
Multiplying (\ref{eq:time_behavior}) by $\bar{g}$ and subtracting by $g$ times the complex conjugate of (\ref{eq:time_behavior}), we find
\begin{displaymath}
\bar{g}\partial^2_{\eta}g - g \partial^2_{\eta}\bar{g} = 0
\end{displaymath}
or
\begin{equation} \label{eq:constant_g}
\bar{g}\partial_{\eta}g - g \partial_{\eta}\bar{g} = {\rm constant}
\end{equation}
Returning to (\ref{eq:time_behavior}), we divide through by $1-\eta^2$ and make the substitution $g(\eta) = \sqrt{1- \eta^2} \cdot p(\eta)$:
\begin{displaymath}
0 = (1-\eta^2)\sqrt{1-\eta^2}\bigg[\partial^2_{\eta}p - \frac{2 \eta}{1-\eta^2}\partial_{\eta}p - \frac{p}{(1-\eta^2)^2} \bigg] + \bigg[l(l+1) + \frac{4\alpha^2}{1-\eta^2} + \frac{1}{1-\eta^2}\bigg]\sqrt{1-\eta^2} \cdot p
\end{displaymath}
which reduces to
\begin{displaymath}
0 = (1-\eta^2) \partial^2_{\eta}p - 2\eta \partial_{\eta}p+ \bigg[l(l+1) + \frac{4\alpha^2}{1-\eta^2}\bigg]p
\end{displaymath}
But the solution to this differential equation is the Legendre function of the first kind \cite{Abramowitz} (since our domain is restricted to $\eta^2 < 1$) which can be expressed in the following manner:
\begin{equation}\label{eq:genLegendre}
P^{\pm 2i \alpha}_{l}(\eta) = \frac{1}{\Gamma(1 \mp 2i\alpha)} \bigg[\frac{1+\eta}{1-\eta} \bigg]^{\pm i \alpha} \,_2F_1 (-l,l+1; 1 \mp 2i\alpha, \frac{1-\eta}{2})
\end{equation}
where $\,_2F_1$ is the hypergeometric function which for our parameters can take the form
\begin{equation}
\,_2F_1(-l,l+1; 1 \mp 2i\alpha, \frac{1-\eta}{2}) = \frac{\Gamma(1 \mp 2i\alpha)}{\Gamma(-l)\Gamma(l+1)} \sum_{k=0}^{\infty} \frac{\Gamma(k-l) \Gamma(k+l+1)}{k! \cdot \Gamma(k+1 \mp 2i\alpha)} \bigg(\frac{1-\eta}{2} \bigg)^k
\end{equation}
and $\Gamma(z)$ is the gamma function which can be written in integral form for $\Re(z)>0$ where $z$ is a complex variable as
\begin{displaymath}
\Gamma(z) = \int_{0}^{\infty} t^{z-1} e^{-t} dt
\end{displaymath}
Note that this Legendre function, $P_{\nu}^{\mu}$, is in fact a generalization of the Legendre function used earlier for our angular dependence where the parameters $\mu$, $\nu$ here are allowed to be complex numbers instead of solely real integers. Our full expression for $g$ is then
\begin{equation} \label{eq:g_expr}
g(\eta) = \sqrt{1- \eta^2} \cdot P_{l}^{-2i \alpha}(\eta)
\end{equation}
where as we'll see below, our choice of $-2i \alpha$ is necessary in order to ensure that our inner product is positive-definite with respect to our solutions for $\phi_i$ so that we may properly construct our field operator (i.e. we take the ``positive frequency" solutions; see Chapter 3.2 of \cite{Wald_QFT}). 

To find our normalization constant, we'll need to evaluate (\ref{eq:constant_g}). For the rest of our analysis in this appendix, we use Chapter 8 of \cite{Abramowitz} as a reference for our general expressions. Beginning with $\partial_{\eta}g$:
\begin{displaymath}
\partial_{\eta}g = - \frac{\eta}{\sqrt{1-\eta^2}} P_{l}^{-2i \alpha} + \sqrt{1- \eta^2} \partial_{\eta} P_{l}^{-2i \alpha}
\end{displaymath}
But
\begin{displaymath}
\partial_{\eta} P_{\nu}^{\mu} = - \frac{\nu \eta}{1- \eta^2} P_{\nu}^{\mu} + \frac{\mu + \nu}{1-\eta^2} P_{\nu-1}^{\mu}
\end{displaymath}
Plugging in above
\begin{eqnarray}
\partial_{\eta}g = - \frac{\eta}{\sqrt{1-\eta^2}} P_{l}^{-2i \alpha} + \sqrt{1- \eta^2}\bigg(- \frac{l \eta}{1- \eta^2} P_l^{-2i \alpha} + \frac{l-2i \alpha}{1- \eta^2} P_{l-1}^{-2i \alpha}  \bigg) \nonumber
\\ = -\frac{\eta (1+l)}{\sqrt{1- \eta^2}} P_l^{-2i \alpha} + \frac{l-2i \alpha}{\sqrt{1- \eta^2}}P_{l-1}^{-2i \alpha} \nonumber
\end{eqnarray}
Then
\begin{eqnarray}
\bar{g}\partial_{\eta}g - g \partial_{\eta}\bar{g} = \sqrt{1- \eta^2} \bigg[\bar{P}_{l}^{-2i \alpha} \bigg(-\frac{\eta (1+l)}{\sqrt{1- \eta^2}} P_l^{-2i \alpha} + \frac{l-2i \alpha}{\sqrt{1- \eta^2}}P_{l-1}^{-2i \alpha} \bigg) \nonumber
\\ - P_{l}^{-2i \alpha}\bigg(-\frac{\eta (1+l)}{\sqrt{1- \eta^2}} \bar{P}_l^{-2i \alpha} + \frac{l+2i \alpha}{\sqrt{1- \eta^2}}\bar{P}_{l-1}^{-2i \alpha}  \bigg) \bigg] \nonumber
\\ = (l-2i \alpha) \bar{P}_{l}^{-2i \alpha} P_{l-1}^{-2i \alpha} - (l+2i \alpha)P_{l}^{-2i \alpha} \bar{P}_{l-1}^{-2i\alpha} \nonumber
\end{eqnarray}
However, as one can tell from (\ref{eq:genLegendre}), $\bar{P}_{l}^{-2i \alpha} = P_l^{2i \alpha}$. Therefore,
\begin{displaymath}
\bar{g}\partial_{\eta}g - g \partial_{\eta}\bar{g} = (l-2i\alpha)P_{l}^{2i \alpha} P_{l-1}^{-2i\alpha} - (l+2i \alpha)P_{l}^{-2i \alpha} P_{l-1}^{2i \alpha} = {\rm constant}
\end{displaymath}
Yet since this expression must be constant, we can evaluate for any particular value of $\eta$. Because we have expressions for the Legendre functions at $\eta=0$, we'll make this convenient choice where
\begin{displaymath}
P_{\nu}^{\mu}(0) = 2^{\mu} \pi^{-1/2} \cos\bigg[\frac{\pi}{2}(\nu + \mu) \bigg] \frac{\Gamma(\frac{1}{2} + \frac{1}{2} \nu + \frac{1}{2} \mu)}{\Gamma(1 + \frac{1}{2}\nu - \frac{1}{2} \mu)}
\end{displaymath}
For $\eta = 0$:
\begin{eqnarray}
P_{l-1}^{-2i \alpha}(0) \cdot P_{l}^{2i \alpha}(0) = 2^{-2i \alpha} \pi^{-1/2} \cos\bigg[\frac{\pi}{2}(l - 1 - 2i \alpha) \bigg] \frac{\Gamma(\frac{1}{2} + \frac{l-1}{2} - i \alpha)}{\Gamma(1 + \frac{l-1}{2} + i \alpha)} \nonumber 
\\ \cdot 2^{2i \alpha} \pi^{-1/2} \cos\bigg[\frac{\pi}{2}(l + 2i \alpha) \bigg] \frac{\Gamma(\frac{1}{2} + \frac{l}{2} + i \alpha)}{\Gamma(1 + \frac{l}{2} - i \alpha)} \nonumber
\\ = \frac{1}{\pi} \cos \bigg[\frac{\pi}{2}(l-1-2i \alpha) \bigg] \cos \bigg[\frac{\pi}{2}(l + 2i \alpha) \bigg] \frac{\Gamma(\frac{l}{2} - i \alpha)}{\Gamma(1 + \frac{l}{2} - i \alpha)} \nonumber
\end{eqnarray}
But from properties of the gamma function (see Chapter 6 of \cite{Abramowitz}),
\begin{displaymath}
\Gamma(1+z) = z \Gamma(z)
\end{displaymath}
Using this property above,
\begin{eqnarray}
P_{l-1}^{-2i \alpha}(0) \cdot P_{l}^{2i \alpha}(0) = \frac{1}{\pi(\frac{l}{2} - i \alpha)} \cos \bigg[\frac{\pi}{2}(l-1-2i \alpha) \bigg] \cos \bigg[\frac{\pi}{2}(l + 2i \alpha) \bigg] \nonumber
\end{eqnarray}
And with the expressions
\begin{eqnarray}
\cos\bigg[\frac{\pi}{2}(l - 1 - 2i \alpha) \bigg] = \frac{1}{2} \bigg[e^{i(\pi/2)(l - 1 - 2i\alpha)} + e^{-i(\pi/2)(l-1-2i \alpha)} \bigg] \nonumber
\\ \cos\bigg[\frac{\pi}{2}(l + 2i \alpha) \bigg] = \frac{1}{2} \bigg[e^{i(\pi/2)(l +2i\alpha)} + e^{-i(\pi/2)(l+2i \alpha)} \bigg] \nonumber
\end{eqnarray}
we have
\begin{displaymath}
\cos\bigg[\frac{\pi}{2}(l -1- 2i \alpha) \bigg]\cos\bigg[\frac{\pi}{2}(l + 2i \alpha) \bigg] = \frac{1}{4} \bigg[ e^{i\pi(l - 1/2)} + e^{-i \pi(l-1/2)} + e^{i\pi(2i \alpha + 1/2)} + e^{-i \pi(2i\alpha + 1/2)} \bigg]
\end{displaymath}
Thus,
\begin{displaymath}
(l-2i\alpha) P_{l-1}^{-2i \alpha} P_{l}^{2i \alpha}|_{\eta = 0} = \frac{1}{2\pi} \bigg[ e^{i\pi(l - 1/2)} + e^{-i \pi(l-1/2)} + e^{i\pi(2i \alpha + 1/2)} + e^{-i \pi(2i\alpha + 1/2)} \bigg]
\end{displaymath}
Applying similar logic to the second term in our expression above, we find
\begin{displaymath}
(l+2i\alpha)P_{l}^{-2i\alpha}P_{l-1}^{2i\alpha}|_{\eta = 0} = \frac{1}{2\pi}\bigg[ e^{i\pi(l - 1/2)} + e^{-i \pi(l-1/2)} + e^{i\pi( 2i \alpha - 1/2)} + e^{-i \pi(2i\alpha - 1/2)} \bigg]
\end{displaymath}
Then
\begin{eqnarray}
\bar{g}\partial_{\eta}g - g \partial_{\eta}\bar{g} = \frac{1}{2\pi}\bigg[e^{i\pi/2 - 2\pi \alpha} + e^{-i\pi/2 + 2\pi\alpha} - e^{i\pi/2 + 2\pi \alpha} - e^{-i\pi/2 - 2\pi \alpha} \bigg]
\nonumber \\ = \frac{1}{2\pi}\bigg[e^{i\pi/2}\bigg(e^{-2\pi \alpha} - e^{2\pi \alpha} \bigg) + e^{-i\pi/2}\bigg(e^{2\pi\alpha} - e^{-2\pi\alpha} \bigg) \bigg] \nonumber
\\ = - \frac{(e^{2\pi \alpha} - e^{-2\pi \alpha})(e^{i\pi/2} - e^{-i\pi/2})}{2\pi} \nonumber
\end{eqnarray}
But
\begin{displaymath}
e^{i\pi/2} - e^{-i\pi/2} = 2i \sin(\pi/2) = 2i
\end{displaymath}
and
\begin{displaymath}
e^{2\pi \alpha} - e^{-2\pi \alpha} = 2\sinh(2\pi \alpha) = 4\sinh(\pi \alpha) \cosh(\pi \alpha)
\end{displaymath}
Therefore, plugging in above
\begin{displaymath}
\bar{g}\partial_{\eta}g - g \partial_{\eta}\bar{g} = - \frac{4i}{\pi}\sinh(\pi \alpha) \cosh(\pi \alpha)
\end{displaymath}
Addressing our inner product where our solutions are of the form $\phi_i = Z_i \cdot g \cdot h \cdot Y_l^m$
\begin{eqnarray}
(\phi_1, \phi_2) = -\frac{i}{\sqrt{\Lambda}} Z_1 Z_2 \cosh^2(\sqrt{\Lambda}t)[g_2 \partial_{t} \bar{g}_1 - \bar{g}_1 \partial_{t} g_{2}]|_{t = 0} \int_{0}^{\pi} \sin{\theta} d\theta \int_{0}^{2 \pi} d\phi \bar{Y}_{l_1}^{m_1} Y_{l_2}^{m_2} \nonumber
\\ \cdot \bigg[ \int_{0}^{\infty} r \bar{h}_{\alpha_1}h_{\alpha_2} dr - \int_{-\infty}^{0} r \bar{h}_{\alpha_1}h_{\alpha_2} dr \bigg]
\nonumber \\ = i Z_1 Z_2  \delta_{l_1 l_2}\delta_{m_1 m_2}  [\bar{g}_1 \partial_{\eta} g_{2} - g_2 \partial_{\eta} \bar{g}_1]|_{\eta = 0} \cdot \bigg[ \int_{0}^{\infty} dr \frac{K_{i \alpha_1}(\frac{\tilde{ \mu} r^2}{2})K_{i \alpha_2}(\frac{\tilde {\mu} r^2}{2})}{\tilde{\mu} r}
\nonumber \\ - \int_{-\infty}^{0} dr \frac{K_{i \alpha_1}(\frac{\tilde {\mu} r^2}{2})K_{i \alpha_2}(\frac{\tilde {\mu} r^2}{2})}{\tilde{\mu} r} \bigg]
\nonumber \\ = i Z_1 Z_2 \delta_{l_1 l_2}\delta_{m_1 m_2} [\bar{g}_1 \partial_{\eta} g_{2} - g_2 \partial_{\eta} \bar{g}_1]|_{\eta = 0} \cdot \frac{2}{\tilde{\mu}} \int_{0}^{\infty} d\rho \frac{K_{i \alpha_1}(\frac{\rho^2}{2})K_{i \alpha_2}(\frac{\rho^2}{2})}{\rho}
\nonumber \\ = i Z_1 Z_2 \delta_{l_1 l_2}\delta_{m_1 m_2}\delta(\alpha_1 - \alpha_2) \cdot \frac{\pi^2} {2  \tilde{\mu} \alpha_1 \sinh(\pi \alpha_1)} \cdot [\bar{g}_1 \partial_{\eta} g_{2} - g_2 \partial_{\eta} \bar{g}_1]|_{\eta = 0}
\nonumber \\ = i Z_1^2 \delta_{l_1 l_2}\delta_{m_1 m_2}\delta(\alpha_1-\alpha_2) \cdot \frac{\pi^2} {2  \tilde{\mu} \alpha_1 \sinh(\pi \alpha_1)} \cdot [\bar{g}_1 \partial_{\eta} g_{1} - g_1 \partial_{\eta} \bar{g}_1]|_{\eta = 0} \nonumber
\\ = i Z_1^2 \delta_{l_1 l_2}\delta_{m_1 m_2}\delta(\alpha_1-\alpha_2) \cdot \frac{\pi^2} {2  \tilde{\mu} \alpha_1 \sinh(\pi \alpha_1)} \bigg[-\frac{4i \sinh(\pi \alpha_1) \cosh(\pi \alpha_1)}{\pi} \bigg] \nonumber
\\ = Z_1^2 \delta_{l_1 l_2}\delta_{m_1 m_2}\delta(\alpha_1-\alpha_2) \cdot \frac{2 \pi \cosh(\pi \alpha_1)}{\tilde{\mu} \alpha_1} \nonumber
\end{eqnarray}
where we keep in mind the fact that
\begin{displaymath}
\frac{\partial}{\partial t} = \frac{\partial \eta}{\partial t} \frac{\partial}{\partial \eta} = \frac{\sqrt{\Lambda}}{\cosh^2(\sqrt{\Lambda}t)} \cdot \frac{\partial}{\partial \eta}
\end{displaymath}
Thus, our normalization constant is found to be
\begin{equation} \label{eq:norm_constant}
Z_{\alpha} = \sqrt{\frac{\tilde{\mu} \alpha}{2\pi \cosh(\pi \alpha)}}
\end{equation}

\bibliography{dark_R2.bib}

\section*{Figure Legends}

\section*{Tables}

\begin{table}[!ht]
\caption{
\bf{Redshift from objects within the Local Group}}
\begin{tabular}{|c|c|c|c|c|}
\hline
Object & RA (J2000.0) & Dec (J2000.0) & Redshift & Distance Mod (mag) \\
\hline
Andromeda V & 01h10m17.10s & +47d37m41.0s & -0.001344 & 24.52 \\
\hline
Andromeda I & 00h45m39.80s & +38d02m28.0s & -0.001228 & 24.46 \\
\hline
Andromeda VI & 23h51m46.30s & +24d34m57.0s & -0.001181 & 24.58\\
\hline
Andromeda III & 00h35m33.78s & +36d29m51.9s & -0.001171 & 24.38\\
\hline
IC 0010 & 00h20m17.34s & +59d18m13.6s & -0.001161 & 24.57\\
\hline
Andromeda VII & 23h26m31.74s & +50d40m32.6s & -0.001024 & 24.7\\
\hline
MESSIER 031 & 00h42m44.35s & +41d16m08.6s & -0.001001 & 24.46\\
\hline
Draco Dwarf & 17h20m12.39s & +57d54m55.3s & -0.000974 & 19.61\\ 
\hline
Pisces I & 01h03m55.00s & +21d53m06.0s & -0.000956 & 24.5\\
\hline
UMi Dwarf & 15h09m08.49s & +67d13m21.4s & -0.000824 & 19.3 \\
\hline
MESSIER 110 & 00h40m22.08s & +41d41m07.1s & -0.000804 & 24.5\\
\hline
IC 1613 & 01h04m47.79s & +02d07m04.0s & -0.000781 & 24.33\\
\hline
NGC 0185 & 00h38m57.97s & +48d20m14.6s & -0.000674 & 24.13\\
\hline
MESSIER 032 & 00h42m41.83s & +40d51m55.0s & -0.000667 & 24.42\\
\hline
NGC 0147 & 00h33m12.12s & +48d30m31.5s & -0.000644 & 24.3\\
\hline
Andromeda II & 01h16m29.78s & +33d25m08.8s & -0.000627 & 24.03\\
\hline
Pegasus Dwarf & 23h28m36.25s & +14d44m34.5s & -0.000612 & 26.34\\
\hline
MESSIER 033 & 01h33m50.89s & +30d39m36.8s & -0.000597 & 24.69\\
\hline
Aquarius dIrr & 20h46m51.81s & -12d50m52.5s & -0.00047 & 26\\
\hline
WLM & 00h01m58.16s & -15d27m39.3s & -0.000407 & 25.09\\
\hline
Cetus Dwarf Spheroidal & 00h26m11.03s & -11d02m39.6s & -0.00029 & 24.51\\
\hline
SagDIG & 19h29m59.58s & -17d40m51.3s & -0.000264 & 25.03\\
\hline
NGC 6822 & 19h44m57.74s & -14d48m12.4s & -0.00019 & 23.41\\
\hline
Leo A & 09h59m26.46s & +30d44m47.0s & 0.000067 & \\
\hline
Fornax Dwarf Spheroidal & 02h39m59.33s & -34d26m57.1s & 0.000178 & 20.7\\
\hline
Phoenix Dwarf & 01h51m06.34s & -44d26m40.9s & 0.000187 & 23.08\\
\hline
Leo B & 11h13m28.80s & +22d09m06.0s & 0.000264 & 21.67\\
\hline
Sculptor Dwarf Elliptical & 01h00m09.36s & -33d42m32.5s & 0.000367 & 19.67 \\
\hline
Sagittarius Dwarf Spheroidal & 18h55m19.50s & -30d32m43.0s & 0.000467 & 17.17\\
\hline
Small Magellanic Cloud & 00h52m44.78s & -72d49m43.0s & 0.000527 & 18.95\\
\hline 
Tucana Dwarf & 22h41m49.60s & -64d25m10.0s & 0.000647 & 24.74 \\
\hline
Sextans Dwarf Spheroidal & 10h13m02.96s & -01d36m52.6s & 0.000747 & 19.73\\ 
\hline
Carina Dwarf & 06h41m36.69s & -50d57m58.3s & 0.000764 & 20.02 \\
\hline
Large Magellanic Cloud & 05h23m34.53s & -69d45m22.1s & 0.000927 & 18.46 \\
\hline
Leo I & 10h08m28.10s & +12d18m23.0s & 0.000951 & 21.91\\
\hline
\end{tabular}
\begin{flushleft} Data retrieved from NASA/IPAC Extragalactic Database (NED): http://ned.ipac.caltech.edu
\end{flushleft}
\label{tab:gen_redshift_LG}
\end{table}

\begin{table}[!ht]
\caption{
\bf{Redshift from galaxies within $\approx \pm 2$~h of 0~h in RA within Local Group}}
\begin{tabular}{|c|c|c|c|c|}
\hline
Object & RA (J2000.0) & Dec (J2000.0) & Redshift & Distance Mod (mag) \\
\hline
Andromeda V & 01h10m17.10s & +47d37m41.0s & -0.001344 & 24.52 \\
\hline
Andromeda I & 00h45m39.80s & +38d02m28.0s & -0.001228 & 24.46 \\
\hline
Andromeda VI & 23h51m46.30s & +24d34m57.0s & -0.001181 & 24.58\\
\hline
Andromeda III & 00h35m33.78s & +36d29m51.9s & -0.001171 & 24.38\\
\hline
IC 0010 & 00h20m17.34s & +59d18m13.6s & -0.001161 & 24.57\\
\hline
Andromeda VII & 23h26m31.74s & +50d40m32.6s & -0.001024 & 24.7\\
\hline
MESSIER 031 & 00h42m44.35s & +41d16m08.6s & -0.001001 & 24.46\\
\hline
Pisces I & 01h03m55.00s & +21d53m06.0s & -0.000956 & 24.5\\
\hline
MESSIER 110 & 00h40m22.08s & +41d41m07.1s & -0.000804 & 24.5\\
\hline
IC 1613 & 01h04m47.79s & +02d07m04.0s & -0.000781 & 24.33\\
\hline
NGC 0185 & 00h38m57.97s & +48d20m14.6s & -0.000674 & 24.13\\
\hline
MESSIER 032 & 00h42m41.83s & +40d51m55.0s & -0.000667 & 24.42\\
\hline
NGC 0147 & 00h33m12.12s & +48d30m31.5s & -0.000644 & 24.3\\
\hline
Andromeda II & 01h16m29.78s & +33d25m08.8s & -0.000627 & 24.03\\
\hline
Pegasus Dwarf & 23h28m36.25s & +14d44m34.5s & -0.000612 & 26.34\\
\hline
MESSIER 033 & 01h33m50.89s & +30d39m36.8s & -0.000597 & 24.69\\
\hline
WLM & 00h01m58.16s & -15d27m39.3s & -0.000407 & 25.09\\
\hline
Cetus Dwarf Spheroidal & 00h26m11.03s & -11d02m39.6s & -0.00029 & 24.51\\
\hline
Fornax Dwarf Spheroidal & 02h39m59.33s & -34d26m57.1s & 0.000178 & 20.7\\
\hline
Phoenix Dwarf & 01h51m06.34s & -44d26m40.9s & 0.000187 & 23.08\\
\hline
Sculptor Dwarf Elliptical & 01h00m09.36s & -33d42m32.5s & 0.000367 & 19.67 \\
\hline
Small Magellanic Cloud & 00h52m44.78s & -72d49m43.0s & 0.000527 & 18.95\\
\hline
\end{tabular}
\begin{flushleft} Data retrieved from NASA/IPAC Extragalactic Database (NED): http://ned.ipac.caltech.edu
\end{flushleft}
\label{tab:red_shift_Dec_pattern}
\end{table}

\end{document}

\begin{flushleft}
{\Large
\textbf{Recent further research regarding the theory of inertial centers (unpublished)}
}
% Insert Author names, affiliations and corresponding author email.
\\
Michael R. Feldman$^{1,\ast}$
\\
\bf{1} Michael R. Feldman Private researcher, New York, NY, United States of America
\\
$\ast$ E-mail: mrf44@cornell.edu
\end{flushleft}

\section*{Geodesic solutions}
Our geodesic equation is given by (see (72) of \cite{Feldman})
\begin{displaymath}
0 = U^a \nabla_a U^b
\end{displaymath}
To find solutions, we start by expressing this equation in terms of Minkowski coordinates $(T, X, Y, Z)$, where the spatial coordinate origin of this system is located at the position of our inertial center (i.e. $r=0 \rightarrow (X,Y,Z)=(0,0,0)$). Since our metric in terms of this Minkowski coordinate parametrization takes the form
\begin{displaymath}
-d\chi^2 = -c^2 dT^2 + \sum_{i} dX_i^2
\end{displaymath}
where $i =1,2,3$ and $X^1 = X, X^2 = Y, X^3 = Z$, our metric components in Minkowski coordinates $g_{\mu \nu}$ are all constants. Thus, our affine connection tensor for Minkowski coordinates is zero: $\Gamma^{c}_{a b}|_{\rm Minkowski} =0$. Our equations of motion in Minkowski coordinates are then found to be
\begin{equation}
\frac{d^2 T}{d\sigma^2} = 0 \indent \indent {\rm and} \indent \indent \frac{d^2 X^i}{d \sigma^2} = 0
\end{equation}
where $\sigma \rightarrow \chi = \sqrt{\Lambda} \cdot \int r(\tau) d\tau$ for massive geodesics in the theory of inertial centers (see the introduction or equation (57) of \cite{Feldman}). Examining only our spatial components, we see that
\begin{eqnarray}
0 = \frac{d^2X^i}{d\sigma^2} = \frac{d}{d\sigma}\bigg(\frac{dT}{d\sigma}\frac{dX^i}{dT}\bigg) \nonumber \\
= \frac{d^2 T}{d \sigma^2}\frac{dX^i}{dT} + \bigg(\frac{dT}{d\sigma}\bigg)^2 \frac{d^2X^i}{dT^2} \nonumber
\end{eqnarray}
which leads us to
\begin{equation} \label{diffMinkgeo}
\frac{d^2 X^i}{dT^2} = 0
\end{equation}
where we have used the equation of motion for our Minkowski time coordinate $T$.

The solutions to (\ref{diffMinkgeo}) are of the form
\begin{equation} \label{Minkgeo}
X^i (T) = v^i T + X^i_0
\end{equation}
where $X^i_0$ is our spatial Minkowski coordinate location when $T=0$. Nevertheless, we know from the form of our affine parameter $\chi$ that we wish to express this coordinate parametrization of our geodesics in terms of the more physically applicable radial Rindler coordinates, where the coordinate transformations from the Minkowski coordinate parameterization employed above are
\begin{eqnarray}
T = \frac{r}{c}\sinh(\sqrt{\Lambda}t) \nonumber \\
X = r \cosh(\sqrt{\Lambda}t) \sin{\theta} \cos{\phi} \nonumber \\
Y = r \cosh(\sqrt{\Lambda}t) \sin{\theta} \sin{\phi} \nonumber \\
Z = r \cosh(\sqrt{\Lambda}t) \cos{\theta} \nonumber
\end{eqnarray}
Using these transformations in (\ref{Minkgeo}), we find
\begin{eqnarray}
r \cosh(\sqrt{\Lambda}t) \sin{\theta} \cos{\phi} = \bigg(\frac{v^X}{c}\bigg) r\sinh(\sqrt{\Lambda}t) + X_0 \nonumber \\
r \cosh(\sqrt{\Lambda}t) \sin{\theta} \sin{\phi} = \bigg(\frac{v^Y}{c}\bigg) r\sinh(\sqrt{\Lambda}t) + Y_0 \nonumber \\
r \cosh(\sqrt{\Lambda}t) \cos{\theta} = \bigg(\frac{v^Z}{c}\bigg) r\sinh(\sqrt{\Lambda}t) + Z_0 \nonumber
\end{eqnarray}
Solving for $r$ by squaring each one of these equations and adding the results together
\begin{displaymath}
r^2 \cosh^2(\sqrt{\Lambda}t) = \sum_{i} \bigg(\frac{v^i}{c} r\sinh(\sqrt{\Lambda}t) + X^i_0 \bigg)^2
\end{displaymath}
where we choose our time coordinates $T$ and $t$ to coincide for our initial conditions (i.e. $t=0 \iff T=0$ and $[dt = dT]|_{t=T=0}$). This requirement forces our Minkowski constant $c$ to be equal to $\sqrt{\Lambda}r_0$ where $r_0$ here is the initial radial position of our object along the geodesic and $\sqrt{\Lambda}$ is fixed by the value of the Hubble constant. Remember that our inertial time coordinate $t$ progresses at the same rate as the clock of a stationary observer located the same distance away from our inertial center as our object. This will hold true when the object is located at any particular position along its path, since at each radial location, our line element for a stationary observer reduces to $-\Lambda r^2 d\tau_S^2 = - \Lambda r^2 dt^2$ (see (57) and (1) of \cite{Feldman}). Thus, coordinate time $t$ is physically equivalent to the elapsed time measured by an observer carried with the object in question but in the limit where the speed of this `carried' observer relative to the inertial center in question is zero at all points along the traversed path (i.e. the `carried' observer remains stationary at each point that the object reaches).

Expanding the expression above,
\begin{displaymath}
r^2 \cosh^2(\sqrt{\Lambda}t) = \sum_{i} \bigg[\bigg(\frac{v^i}{c}\bigg)^2 r^2 \sinh^2(\sqrt{\Lambda}t) + 2\frac{v^i}{c}X_0^i r\sinh(\sqrt{\Lambda}t) +(X^i_0)^2 \bigg]
\end{displaymath}
But
\begin{displaymath}
\sum_i (X^i_0)^2 = \sum_i (X^i(T))^2|_{T=0} = \sum_i (X^i(t))^2|_{t=0} = r_0^2
\end{displaymath}
Using this relation,
\begin{eqnarray}
r_0^2 + 2r\sinh(\sqrt{\Lambda}t) \sum_i \bigg(\frac{v^i}{c} \bigg) X_0^i = r^2\bigg[\cosh^2(\sqrt{\Lambda}t) - \sinh^2(\sqrt{\Lambda}t) \sum_{i} \bigg(\frac{v^i}{c}\bigg)^2 \bigg] \nonumber \\
= r^2\bigg[\cosh^2(\sqrt{\Lambda}t) - \sinh^2(\sqrt{\Lambda}t) \bigg(1 - 1 + \sum_{i} \bigg(\frac{v^i}{c}\bigg)^2 \bigg)\bigg] \nonumber \\
= r^2\bigg[\cosh^2(\sqrt{\Lambda}t) - \sinh^2(\sqrt{\Lambda}t) + \sinh^2(\sqrt{\Lambda}t) \bigg(1 - \sum_{i} \bigg(\frac{v^i}{c}\bigg)^2 \bigg)\bigg] \nonumber \\
= r^2\bigg[1 + \sinh^2(\sqrt{\Lambda}t) \bigg(1 - \sum_{i} \bigg(\frac{v^i}{c}\bigg)^2 \bigg)\bigg] \nonumber
\end{eqnarray}
However, for $R^2 = \sum_i (X^i)^2$, we see that $v^R(T) = dR/dT$ is given by
\begin{displaymath}
Rv^{R} = \sum_i X^i \frac{dX^i}{dT} = \sum_i X^i v^i
\end{displaymath}
Yet, since $dX^i/dT$ is a constant for all $T$, we have
\begin{equation}
\sum_i X_0^i v^i = Rv^R|_{T=0} = Rv^R |_{t=0} = r_0 v_0^r
\end{equation}
where $v_0^r = dr/dt|_{t=0}$ is the initial radial velocity of the object in the inertial system. Plugging in above, we find
\begin{displaymath}
0 = r^2\bigg[1 + \sinh^2(\sqrt{\Lambda}t) \bigg(1 - \sum_{i} \bigg(\frac{v^i}{c}\bigg)^2 \bigg)\bigg] - 2r\sinh(\sqrt{\Lambda}t) \cdot r_0 \frac{v^r_0}{c} - r_0^2
\end{displaymath}
Solving for $r(t)$,
\begin{displaymath}
r(t) = r_0 \bigg[\frac{\frac{v_0^r}{c}\sinh(\sqrt{\Lambda}t) \pm \sqrt{(\frac{v_0^r}{c})^2 \sinh^2(\sqrt{\Lambda}t) + [1 + \sinh^2(\sqrt{\Lambda}t) (1 - \sum_{i} (\frac{v^i}{c})^2 )] }}{1 + \sinh^2(\sqrt{\Lambda}t) (1 - \sum_{i} (\frac{v^i}{c})^2 )} \bigg]
\end{displaymath}
When $t=0$, we want our expression to reduce to $r_0$ and not $-r_0$. Therefore, we take only the positive root for our solution.

Additionally, we see that $\sum_{i} (v^i)^2$ can be expressed in terms of the induced Riemannian metric components $h_{ij}$ for a space-like hypersurface in our original Minkowski coordinates in the following manner:
\begin{displaymath}
\sum_{i} (v^i)^2 = \sum_{i,j} h_{ij} v^i v^j = v_0^2
\end{displaymath}
However, this is simply the squared magnitude of the $v^i$'s in Minkowski coordinates along this hypersurface. Furthermore, since our time coordinates coincide at $t=T=0$, $v_0$ can be interpreted physically as the initial speed of our inertially traveling object measured in terms of the clock of a stationary observer located at $r=r_0$. In other words, since $v^i = dX^i/dT = dX^i/dT|_{T=0}$,
\begin{eqnarray}
v_0^2 = \sum_{i,j} h_{ij} \frac{dX^i}{dT} \frac{dX^j}{dT} = \sum_{i,j} h_{ij} \frac{dX^i}{dT} \frac{dX^j}{dT} \bigg|_{T=0} \nonumber \\
= \sum_{i,j} \sum_{k,l} h_{ij} \frac{\partial X^i}{\partial \tilde{x}^k}\frac{d \tilde{x}^k}{dT} \frac{\partial X^j}{\partial \tilde{x}^l} \frac{d\tilde{x}^l}{dT}\bigg|_{T=0} \nonumber \\
= \sum_{k,l} \bigg(\sum_{i,j}h_{ij}\frac{\partial X^i}{\partial \tilde{x}^k} \frac{\partial X^j}{\partial \tilde{x}^l} \bigg)\frac{d\tilde{x}^k}{dt}\frac{d\tilde{x}^l}{dt}\bigg|_{t=0} \nonumber \\
= \sum_{k,l} \tilde{h}_{kl} \frac{d\tilde{x}^k}{dt} \frac{d\tilde{x}^l}{dt}\bigg|_{t=0}
\end{eqnarray}
where $\tilde{h}_{kl}$ are induced Riemannian spatial metric components along the $t=T=0$ hypersurface and $\tilde{x}^k$ are spatial coordinates. Both tilded expressions are in radial Rindler coordinates. Plugging back into our expression for $r(t)$ where we remember to take only the positive root, we find for the radial position of our object
\begin{equation} \label{radialmotion}
r(t) = r_0 \bigg[\frac{\frac{v_0^r}{c}\sinh(\sqrt{\Lambda}t) + \sqrt{1 + \sinh^2(\sqrt{\Lambda}t) (1 + (\frac{v_0^r}{c})^2 - (\frac{v_0}{c})^2) }}{1 + \sinh^2(\sqrt{\Lambda}t) (1 - (\frac{v_0}{c})^2 )} \bigg]
\end{equation}
where $c = \sqrt{\Lambda}r_0$ and
\begin{eqnarray}
(v_0^r)^2 - v_0^2 = (v_0^r)^2 - \sum_{k,l}\tilde{h}_{kl}\frac{d\tilde{x}^k}{dt}\frac{d\tilde{x}^l}{dt}\bigg|_{t=0} \nonumber \\
= (v_0^r)^2 - \bigg[(v_0^r)^2 + r_0^2\bigg((v_0^{\theta})^2 + \sin^2{\theta_0}(v_0^{\phi})^2 \bigg)\bigg] \nonumber \\
= - r_0^2\bigg((v_0^{\theta})^2 + \sin^2{\theta_0}(v_0^{\phi})^2 \bigg)
\end{eqnarray}

Furthermore, our normalization condition for the `four-velocity' along these geodesics is
\begin{displaymath}
- k = g_{ab}U^a U^b
\end{displaymath}
where $U^{\mu} = dx^{\mu}/d\sigma$ and
\begin{displaymath}
k = \left\{\begin{array}{l l} 0 & \quad \textrm{null geodesics} \\ 1 & \quad \textrm{time-like geodesics} \\ \end{array}\right.
\end{displaymath}
We can express this normalization condition in our original Minkowski coordinates in the following manner:
\begin{displaymath}
- k = -c^2 \bigg(\frac{dT}{d\sigma} \bigg)^2 + \sum_i \bigg(\frac{dX^i}{d\sigma} \bigg)^2 = -\bigg(\frac{dT}{d\sigma}\bigg)^2 \bigg[c^2 - \sum_i (v^i )^2 \bigg] = -\bigg(\frac{dT}{d\sigma}\bigg)^2 \bigg[c^2 - v_0^2 \bigg]
\end{displaymath}
which provides us with the constraints that one would expect
\begin{eqnarray}
v_0 = c \indent \textrm{massless objects} \\
v_0 < c \indent \textrm{massive objects}
\end{eqnarray}
Returning to our coordinate transformations, the angular positions of our object as a function of $t$ are found to be
\begin{eqnarray}
\phi(t) = \tan^{-1}\bigg[\frac{\frac{v^y_0}{c} r(t) \sinh(\sqrt{\Lambda}t) + Y_0}{\frac{v^x_0}{c} r(t) \sinh(\sqrt{\Lambda}t) + X_0} \bigg] \nonumber \\
= \phi_0 + \tan^{-1}\bigg[\frac{r\sinh(\sqrt{\Lambda}t) \frac{v_0^{\phi}}{c} r_0 \sin{\theta_0}}{r_0 \sin{\theta_0} + r\sinh(\sqrt{\Lambda}t)[\frac{v_0^r}{c} \sin{\theta_0} + \frac{v_0^{\theta}}{c} r_0 \cos{\theta_0}]} \bigg]
\end{eqnarray}
and
\begin{eqnarray}
\theta (t) = \cos^{-1}\bigg[\frac{v^z_0}{c}\tanh(\sqrt{\Lambda}t) + \frac{Z_0}{r(t)\cosh(\sqrt{\Lambda}t)} \bigg] \nonumber \\
= \cos^{-1}\bigg[\tanh(\sqrt{\Lambda}t) \bigg(\frac{v_0^r}{c}\cos{\theta_0} - \frac{v_0^{\theta}}{c}r_0 \sin{\theta_0} \bigg) + \frac{r_0 \cos{\theta_0}}{r \cosh(\sqrt{\Lambda}t)} \bigg]
\end{eqnarray}
where
\begin{eqnarray}
X(t)|_{t=0} = X_0 = r_0 \sin{\theta_0}\cos{\phi_0} \\
Y(t)|_{t=0} = Y_0 = r_0 \sin{\theta_0}\sin{\phi_0} \\
Z(t)|_{t=0} = Z_0 = r_0 \cos{\theta_0}
\end{eqnarray}
and
\begin{eqnarray}
v^X = v_0^x = v_0^r \sin{\theta_0} \cos{\phi_0} + v_0^{\theta} r_0 \cos{\theta_0}\cos{\phi_0} - v^{\phi}_0 r_0 \sin{\theta_0} \sin{\phi_0}\label{v1} \\
v^Y = v_0^y = v_0^r \sin{\theta_0} \sin{\phi_0} + v_0^{\theta} r_0 \cos{\theta_0}\sin{\phi_0} + v^{\phi}_0 r_0 \sin{\theta_0} \cos{\phi_0}\label{v2} \\
v^Z = v_0^z = v_0^r \cos{\theta_0} - v_0^{\theta} r_0 \sin{\theta_0}\label{v3}
\end{eqnarray}
One can obtain initial velocity expressions (\ref{v1})-(\ref{v3}) by plugging our spatial coordinate transformations into each of our $dX^i/dT|_{T=0}$ terms.

We have yet to address the relationship between our time coordinate $t$ and the proper time of our object $\tau$ in this particular inertial reference frame. For clarification, we remember that $\tau$ represents the physical elapsed time as measured by a clock {\it moving} with our object along the traversed geodesic. From our normalization condition for massive objects ($\sigma = \chi$),
\begin{displaymath}
\frac{dT}{d\chi} = \frac{1}{c\sqrt{1 - (\frac{v_0}{c})^2}}
\end{displaymath}
Choosing to examine the proper time of our clock in the particular inertial frame of reference in which the object orbits about our inertial center, we have $d\chi/d\tau = \sqrt{\Lambda}r$ and so
\begin{displaymath}
\frac{dT}{d\tau} = \frac{\sqrt{\Lambda}r}{c} \frac{1}{\sqrt{1-(\frac{v_0}{c})^2}}
\end{displaymath}
Expressing $T$ in radial Rindler coordinates,
\begin{eqnarray}
\frac{dT}{d\tau} = \frac{d}{d\tau}\bigg(\frac{r}{c}\sinh(\sqrt{\Lambda}t)\bigg) \nonumber \\
= \frac{1}{c}\frac{dt}{d\tau}\bigg(\frac{dr}{dt} \sinh(\sqrt{\Lambda}t) + \sqrt{\Lambda}r \cosh(\sqrt{\Lambda}t) \bigg) \nonumber
\end{eqnarray}
and plugging in above
\begin{equation} \label{clockrateswithr}
\frac{dt}{d\tau} = \frac{1}{\sqrt{1- (\frac{v_0}{c})^2}} \frac{\sqrt{\Lambda}r(t)}{\frac{dr}{dt}\sinh(\sqrt{\Lambda}t) + \sqrt{\Lambda}r(t) \cosh(\sqrt{\Lambda}t)}
\end{equation}
Notice, when $t=0$ we have
\begin{equation}
\frac{dt}{d\tau}\bigg|_{t = 0} = \frac{1}{\sqrt{1- (\frac{v_0}{c})^2}}
\end{equation}
which is simply the familiar special relativistic relation.

Using (\ref{radialmotion}) in (\ref{clockrateswithr}), we find
\begin{equation} \label{clockdiff}
\frac{dt}{d\tau} = \frac{\gamma_0}{\cosh(\sqrt{\Lambda}t)} \bigg[\frac{2}{1+ \frac{\sinh^2(\sqrt{\Lambda}t)}{\gamma_0^2}} - \frac{1}{\sqrt{}|_{+v} \cdot [(\frac{v_0^r}{c})\sinh(\sqrt{\Lambda}t) + \sqrt{}|_{+v}]} \bigg]^{-1}
\end{equation}
where
\begin{displaymath}
\gamma_0 = \frac{1}{\sqrt{1-(\frac{v_0}{c})^2}} \indent \indent \textrm{and} \indent \indent \sqrt{}|_{+v} = \sqrt{1+\sinh^2(\sqrt{\Lambda}t)\bigg[\bigg(\frac{v_0^r}{c}\bigg)^2 + \frac{1}{\gamma_0^2}\bigg]}
\end{displaymath}
In the limit as $t\rightarrow \pm \infty$, this expression reduces to
\begin{displaymath}
\lim_{t\rightarrow \pm \infty} \frac{dt}{d\tau} = \gamma_0 \frac{\sinh^2(\sqrt{\Lambda}t)}{\cosh(\sqrt{\Lambda}t)} \bigg[2\gamma_0^2 - \frac{1}{\sqrt{(\frac{v_0^r}{c})^2 + \frac{1}{\gamma_0^2}}}\frac{1}{(\frac{v_0^r}{c}) + \sqrt{(\frac{v_0^r}{c})^2 + \frac{1}{\gamma_0^2}}} \bigg]^{-1} \bigg|_{t\rightarrow \pm \infty} \rightarrow \infty
\end{displaymath}
However, examining the radial velocity of our object as $t\rightarrow \pm \infty$,
\begin{displaymath}
\lim_{t\rightarrow \pm \infty} \frac{dr}{dt} = - \sqrt{\Lambda} r_0 \frac{\cosh(\sqrt{\Lambda}t)}{\sinh^2(\sqrt{\Lambda}t)}\gamma_0^2 \bigg[\bigg(\frac{v_0^r}{c} \bigg) + \sqrt{\bigg(\frac{v_0^r}{c} \bigg)^2+ \frac{1}{\gamma_0^2}} \bigg] \bigg|_{t\rightarrow \pm \infty}
\end{displaymath}
where
\begin{displaymath}
\lim_{t\rightarrow \pm \infty} r(t) = r_0 \frac{\gamma_0^2}{\sinh^2(\sqrt{\Lambda}t)}\bigg[\bigg(\frac{v_0^r}{c} \bigg) + \sqrt{\bigg(\frac{v_0^r}{c} \bigg)^2+ \frac{1}{\gamma_0^2}} \bigg] \bigg|_{t\rightarrow \pm \infty}
\end{displaymath}
And so we find that
\begin{displaymath}
\lim_{t\rightarrow \pm \infty} \frac{dr}{dt} = \mp \sqrt{\Lambda} r|_{t \rightarrow \pm \infty}
\end{displaymath}
Therefore, since the radial velocity of our object approaches that of light in the limit as $t\rightarrow \pm \infty$ (see (30) and (37) of \cite{Feldman} as well), we should expect the above result for our difference in clock rates (\ref{clockdiff}) which says that the clock of a stationary observer located at the same radial position as our object in this limit will see the clock of our now nearly light-speed object not progressing in time.

Furthermore, after integrating (\ref{clockdiff}), we find that the elapsed proper time in terms of the elapsed time measured by our `carried' stationary observer is
\begin{equation}
\tau (t) = \frac{2}{\sqrt{\Lambda}}\bigg\{\tan^{-1}\bigg[\frac{\sinh(\sqrt{\Lambda}t)}{\gamma_0} \bigg] + \tan^{-1}\bigg[\frac{1}{\gamma_0}\bigg(\frac{1- \beta_0 \sinh(\sqrt{\Lambda}t) - \sqrt{}|_{+v}}{\beta_0 - \frac{v_0^r}{c} + (\beta_0 + \frac{v_0^r}{c})[\beta_0 \sinh(\sqrt{\Lambda}t) + \sqrt{}|_{+v}]}\bigg) \bigg] \bigg\}
\end{equation}
where
\begin{displaymath}
\beta_0 = \sqrt{\bigg(\frac{v_0^r}{c}\bigg)^2+\frac{1}{\gamma_0^2}}
\end{displaymath}
and we take $\tau(t=0) = 0$ so that these clocks coincide for our initial conditions. One could also express our geodesics in terms of the clock of an observer located at a different spatial position than that of our object by comparing these two clocks using the form of our invariant interval, similar to how we obtained (66) at the end of ``Quantization of a real scalar field" in \cite{Feldman}.

\section*{Massless solutions to Klein-Gordon extension}
We begin from the proposed extension of the Klein-Gordon equation for a real scalar field in the theory of inertial centers (see (60) of \cite{Feldman})
\begin{displaymath}
(\Box - \tilde{\mu}^2 r^2) \phi = 0
\end{displaymath}
where $\Box = \nabla_a \nabla^a$ is the Laplace-Beltrami operator and $\nabla_a$ is the derivative operator compatible with the metric $g_{ab}$ (i.e. $\nabla_a g_{bc} = 0$). Additionally, $\tilde{\mu} = m\sqrt{\Lambda}/\hbar$ where $m$ is the mass of our scalar field, $\hbar$ is the reduced Planck constant, and $\sqrt{\Lambda}$ is taken to be the Hubble constant. As we are examining solutions ($\phi_i$) to this equation for a massless real scalar field, we set $m=0$ and find
\begin{equation} \label{eq:masslessKG}
\Box \phi_i = 0
\end{equation}
Expanding (\ref{eq:masslessKG}) in radial Rindler coordinates:
\begin{displaymath}
0 = -\frac{1}{\Lambda r^2 \cosh^2(\sqrt{\Lambda}t)} \partial_{t}(\cosh^2(\sqrt{\Lambda}t) \partial_{t}\phi_i) + \frac{1}{r^3} \partial_{r} (r^3 \partial_{r}\phi_i) + \frac{1}{r^2 \cosh^2(\sqrt{\Lambda}t)}\bigg[\frac{1}{\sin{\theta}}\partial_{\theta}(\sin{\theta}\partial_{\theta} \phi_i) + \frac{1}{\sin^2{\theta}} \partial^2_{\phi}\phi_i \bigg]
\end{displaymath}
Then for $\phi_i = Z_i \cdot g(t) \cdot h(r) \cdot Y_{l}^m (\theta,\phi)$ where $Z_i$ is a normalization constant, we find
\begin{displaymath}
0 = -\frac{1}{g \cdot \Lambda r^2 \cosh^2(\sqrt{\Lambda}t)} \partial_{t}(\cosh^2(\sqrt{\Lambda}t) \partial_{t}g) + \frac{1}{h \cdot r^3} \partial_{r} (r^3 \partial_{r}h) - \frac{l(l+1)}{r^2 \cosh^2(\sqrt{\Lambda}t)}
\end{displaymath}
Grouping functions of $t$ and $r$:
\begin{displaymath}
\frac{1}{h\cdot r}\partial_{r}(r^3\partial_{r}h) = \frac{1}{g\cdot \Lambda \cosh^2(\sqrt{\Lambda}t)}\partial_{t}(\cosh^2(\sqrt{\Lambda} t)\partial_{t} g) + \frac{l(l+1)}{\cosh^2(\sqrt{\Lambda}t)} = - (4\alpha^2 + 1)
\end{displaymath}
where $\alpha$ is a constant. This gives us two differential equations
\begin{eqnarray}
0 = r^2 \partial^2_r h + 3r\partial_r h + (4\alpha^2 + 1)h \label{eq:r behavior}
\\ 0 = \frac{1}{\Lambda \cosh^2(\sqrt{\Lambda}t)} \partial_t (\cosh^2(\sqrt{\Lambda}t)\partial_t g) + \bigg[\frac{l(l+1)}{\cosh^2(\sqrt{\Lambda}t)} + (4\alpha^2 +1) \bigg] \label{eq:t behavior}
\end{eqnarray}
The ``positive frequency" solution to (\ref{eq:t behavior}) was found in Appendix F of \cite{Feldman} to be
\begin{displaymath}
g(\eta) = \sqrt{1- \eta^2} \cdot P_{l}^{-2i \alpha}(\eta)
\end{displaymath}
where $\eta = \tanh(\sqrt{\Lambda}t)$ and $P_{l}^{-2i\alpha}$ is the Legendre function of degrees $l$ and imaginary order $-2\alpha$. Thus, the only difference between our solutions for the massive and massless cases occurs in the radial function $h$ as one can see from (\ref{eq:r behavior}). Yet (\ref{eq:r behavior}) is the Cauchy-Euler equation of second order given in general form by (see Chapter 61 of \cite{Zwillinger})
\begin{displaymath}
x^2 y'' + ax y' + by = 0
\end{displaymath}
with solution $y(x) = c_+ x^{\lambda_+} + c_{-} x^{\lambda_{-}}$ for distinct roots $\lambda_+$, $\lambda_-$ or $y(x) = c_1 x^{\lambda} + c_2 \ln(x)x^{\lambda}$ for double roots to the characteristic polynomial equation
\begin{displaymath}
0 = \lambda^2 + (a-1)\lambda + b
\end{displaymath}
For our case, we have $a = 3$ and $b = 4\alpha^2 +1$, which gives
\begin{displaymath}
\lambda_{\pm} = -1 \pm 2i \alpha 
\end{displaymath}
And so we have for possible solutions $h_{\alpha}$ in the massless case
\begin{equation}
h^{\pm}_{\alpha}(r) \propto \frac{1}{r} \cdot (r^2)^{\pm i \alpha}
\end{equation}

The inner product of two positive frequency solutions is given by
\begin{displaymath}
(\phi_1, \phi_2) = -i \int_{-\infty}^{\infty}dr \int_{0}^{\pi} d\theta \int_{0}^{2\pi} d\phi \bigg[ r^2 \cosh^2(\sqrt{\Lambda}t) \sin{\theta} \bigg(\frac{\phi_2}{\sqrt{\Lambda}|r|} \partial_{t} \bar{\phi}_1 - \frac{\bar{\phi}_1}{\sqrt{\Lambda}|r|} \partial_{t} \phi_2 \bigg) \bigg]\bigg|_{t=0}
\end{displaymath}
where $\bar{\phi_i}$ is the complex conjugate of $\phi_i$. Plugging in for our massless solutions,
\begin{displaymath}
(\phi^{\pm}_1, \phi^{\pm}_2) = - i Z_1 Z_2 \delta_{l_1 l_2} \delta_{m_1 m_2} [g_2 \partial_{\eta} \bar{g}_1 - \bar{g}_1 \partial_{\eta}g_2]\|_{\eta = 0} \cdot \bigg[\int_0^{\infty} \frac{dr}{r} \cdot (r^2)^{\pm i (\alpha_2 - \alpha_1)} - \int_{- \infty}^{0} \frac{dr}{r} (r^2)^{\pm i(\alpha_2 - \alpha_1)} \bigg]
\end{displaymath}
But letting $\tilde{r} = -r$ in the second integral, we see that
\begin{eqnarray}
- \int_{- \infty}^{0} \frac{dr}{r} (r^2)^{\pm i(\alpha_2 - \alpha_1)} = - \int_{\infty}^{0} \frac{d\tilde{r}}{\tilde{r}} (\tilde{r}^2)^{\pm i(\alpha_2 - \alpha_1)} \nonumber
\\ = \int^{\infty}_{0} \frac{d\tilde{r}}{\tilde{r}} (\tilde{r}^2)^{\pm i(\alpha_2 - \alpha_1)} \nonumber
\end{eqnarray}
Our inner product then reduces to
\begin{displaymath}
(\phi^{\pm}_1, \phi^{\pm}_2) = - 2i Z_1 Z_2 \delta_{l_1 l_2} \delta_{m_1 m_2} [g_2 \partial_{\eta} \bar{g}_1 - \bar{g}_1 \partial_{\eta}g_2]\|_{\eta = 0} \cdot \int_0^{\infty} \frac{dr}{r} \cdot (r^2)^{\pm i (\alpha_2 - \alpha_1)}
\end{displaymath}
However, if we take $x = \ln{r}$ such that $dx = dr/r$, our integral becomes
\begin{displaymath}
\int_0^{\infty} \frac{dr}{r} \cdot (r^2)^{\pm i (\alpha_2 - \alpha_1)} = \int_{-\infty}^{\infty} dx(e^{2x})^{\pm i (\alpha_2 - \alpha_1)}
\end{displaymath}
Notice that we do not have to worry about negative values of $r$ with this transformation to $x$ as our limits of integration for the integral on the left-hand side of the above equation are only over positive values of $r$. Furthermore, letting $y = 2x$, our inner product becomes
\begin{displaymath}
(\phi^{\pm}_1, \phi^{\pm}_2) = - i Z_1 Z_2 \delta_{l_1 l_2} \delta_{m_1 m_2} [g_2 \partial_{\eta} \bar{g}_1 - \bar{g}_1 \partial_{\eta}g_2]\|_{\eta = 0} \cdot \int_{-\infty}^{\infty} e^{\pm iy (\alpha_2 - \alpha_1)} dy
\end{displaymath}
But
\begin{displaymath}
\int_{-\infty}^{\infty} e^{\pm iy (\alpha_2 - \alpha_1)} dy = 2\pi \delta (\alpha_2 - \alpha_1)
\end{displaymath}
where the $\pm$ sign has no impact on the form of the delta function on the right-hand side of this equation. Then,
\begin{displaymath}
(\phi^{\pm}_1, \phi^{\pm}_2) = - 2\pi i Z_1^2 \delta_{l_1 l_2} \delta_{m_1 m_2} \delta(\alpha_2 - \alpha_1) [g_1 \partial_{\eta} \bar{g}_1 - \bar{g}_1 \partial_{\eta}g_1]\|_{\eta = 0}
\end{displaymath}
In Appendix F of \cite{Feldman}, we found
\begin{displaymath}
\bar{g}\partial_{\eta}g - g \partial_{\eta}\bar{g} = - \frac{4i}{\pi}\sinh(\pi \alpha) \cosh(\pi \alpha)
\end{displaymath}
So
\begin{displaymath}
(\phi^{\pm}_1, \phi^{\pm}_2) = 8 \sinh(\pi \alpha_1) \cosh(\pi \alpha_1) Z_1^2 \delta_{l_1 l_2} \delta_{m_1 m_2} \delta(\alpha_2 - \alpha_1)
\end{displaymath}
where $\alpha \neq 0$ as in the massive case. Thus, our normalization constant is found to be
\begin{equation}
Z_{\alpha} = \frac{1}{\sqrt{8\sinh(\pi \alpha)\cosh(\pi \alpha)}}
\end{equation}
and our massless real scalar field solutions to the Klein-Gordon extension for the theory of inertial centers are
\begin{equation} \label{massless}
\phi^{\pm}_{\alpha, l, m} = \frac{1}{\sqrt{8\sinh(\pi \alpha)\cosh(\pi \alpha)}} \cdot \sqrt{1-\eta^2} \cdot P_{l}^{-2i\alpha}(\eta) \cdot \frac{(r^2)^{\pm i \alpha}}{r} \cdot Y_l^m(\theta, \phi)
\end{equation}
Notice the similarity between the forms for $h_{\alpha}$ in the massless case (\ref{massless}) versus the massive given in equation (64) of \cite{Feldman}.

\bibliography{Geodesic_and_massless.bib}